\documentclass[a4paper,fleqn,usenatbib,useAMS]{mnras}

\usepackage{graphicx}
\usepackage{amsmath}
\usepackage{amssymb}
\usepackage{bm}
\usepackage{pdflscape}
\usepackage{float}
\usepackage[T1]{fontenc}
\usepackage{ae,aecompl}
\usepackage{color}
\usepackage{caption}
\usepackage{mathtools}
\usepackage{upgreek}

\pdfoutput=1

\newcommand{\hatP}{{\hat P}}

\newcommand{\tildeE}{{\tilde E}}

\title[Chaotic Tides in Migrating Gas Giants]{Chaotic Tides in Migrating Gas Giants: Forming Hot and Transient Warm Jupiters via High-Eccentricity Migration}

\author[M. Vick, D. Lai, and K. Anderson]{Michelle Vick$^{1}$, Dong Lai$^{1}$, Kassandra R. Anderson$^{1}$\\$^{1}$Cornell Center for Astrophysics and Planetary Science, Department of Astronomy, Cornell University, Ithaca, NY 14853, USA}

\pubyear{2018}

\begin{document}


\label{firstpage}
\pagerange{\pageref{firstpage}--\pageref{lastpage}}
\maketitle

\begin{abstract}
	High-eccentricity migration is an important channel for the formation
	of hot Jupiters (HJs). In particular, Lidov-Kozai (LK) oscillations of
	orbital eccentricity/inclination induced by a distant planetary or
	stellar companion, combined with tidal friction, have been shown to
	produce HJs on Gyr timescales, provided that efficient tidal
	dissipation operates in the planet.  We re-examine this scenario with
	the inclusion of dynamical tides. When the planet's orbit is in a
	high-eccentricity phase, the tidal force from the star excites
	oscillatory f-modes and r-modes in the planet. For sufficiently large
	eccentricity and small pericentre distance, the mode can grow
	chaotically over multiple pericentre passages and eventually dissipate
	non-linearly, drawing energy from the orbit and rapidly shrinking the
	semi-major axis. 
	We study the effect of such chaotic tides on the planet's orbital
	evolution. We find that this pathway produces very eccentric
	($e\gtrsim 0.9$) warm Jupiters (WJs)
	on short timescales (a few to 100 Myrs). These WJs efficiently circularize to 
	become HJs due to their persistently small pericentre distances.
	Chaotic tides can also save some planets from tidal disruption by 
	truncating the LK eccentricity oscillations, significantly 
	increasing the HJ formation fraction for a range of planet masses and radii.
	Using a population synthesis calculation, we determine the characteristics
	of WJs and HJs produced in this scenario, including the final period distribution,
	orbital inclinations and stellar obliquities. Chaotic tides
	endow LK migration with several favorable features to explain observations of HJs.
	We expect that chaotic tides are also important in other flavours of high-$e$
	migration.
\end{abstract}


\section{Introduction}\label{sec:Introduction}

Despite over two decades of observations, the origins of hot Jupiters
(HJs, giant planets with orbital periods $\lesssim 10$ days) remain
puzzling \citep[see][for a review]{Dawson18}.
The materials and conditions necessary to form such massive planets are
not thought to exist so close to a protostar. Some have suggested that
HJs could form in-situ when a protoplanet migrates inward and
accumulates a gas envelope \citep{Boley16,Batygin16}. However, most
theories propose that a fully formed gas giant migrated toward its
host star either via interaction with the protoplanetary disk \citep[e.g.][]{Lin96, Kley12} 
or through tidal decay and circularization of a high-eccentricity
orbit, a process termed ``high-eccentricity migration."
Equally puzzling is the origin of warm Jupiters (WJs), giant planets
with periods between about 10 days and 200 days. Although these close-in
giant planets (HJs and WJs) represent a relatively small population of
exoplanetary systems compared to the more abundant super-earths, their
dynamical history can potentially shape the architecture of planetary
systems.

High-eccentricity migration is an appealing avenue for HJ
formation. In this scenario, a gas giant is excited into a highly
eccentric orbit via interactions with other planets or with a distant 
stellar companion.
Strong planet-planet scattering \citep{Rasio96,Chatterjee08,Tremaine08}, 
various forms of secular interactions \citep{Wu11,Hamers16,Petrovich15b}, 
or a combination of both \citep{Nagasawa08,Beauge12}, 
can produce very eccentric gas giants. 
An highly inclined stellar or planetary companion can excite
``Lidov-Kozai'' (LK) oscillations \citep[][for a review]{Lidov62,Kozai62,Naoz16},
pushing the eccentricity of the giant planet to near unity
\citep{Wu03,Fabrycky07,Naoz12,Correia12,Petrovich15,Anderson16}. 
In all cases, the orbit circularizes and decays to a period of few days due
to the tidal dissipation within the planet.

High-eccentricity migration can account for several puzzling
characteristics of observed HJs. For instance, tidal orbital decay
produces the observed pile-up of HJs at an orbital period of $\sim 3$
days \citep[e.g.][]{Santerne16}, corresponding to semi-major axes of
a few times the Roche radii of the planets. Lidov-Kozai migration
driven by an inclined stellar/planetary companion can naturally
generate significant ``spin-orbit'' misalignments between the rotation
axis of the host star and the orbital angular momentum axis of the
planet, as observed in many HJ systems
\citep{Hebrard08,Narita09,Winn09,Triaud10,Albrecht12,Winn15}; the
chaotic evolution of the stellar spin axis driven by the changing 
planetary orbit plays a dominant role in producing the observed stellar
oblqiuities \citep{Storch14,Anderson16,Storch17}.  Other flavors of
high-$e$ migration can also produce appreciable spin-orbit misalignments
\citep[e.g.][]{Lithwick14,Petrovich15b,Teyssandier18}.
In addition, direct RV and AO-imaging searches have shown that a large fraction of HJs 
have external massive planet companions at 5-20 au or distant stellar companions
(50-2000 au) \citep{Knutson14,Ngo15,Wang15,Bryan16}, suggesting that dynamical interactions play a role in HJ formation. \citet{Huang16} showed that most HJs do not have any detectable neighbors
(while half of WJs are closely flanked by small neighbors), again suggesting 
that a significant fraction of HJs may have formed through high-$e$ migration.

A significant uncertainty in any high-$e$ migration scenario
is tidal dissipation in the planet. Regardless of how the planet attains
its high eccentricity, efficient tidal dissipation is necessary in order to
circularize the planet's orbit and to bring it from a semi-major axis of
several au's to $\lesssim 0.05$ au.  So far, almost all studies of
high-$e$ migration have relied on the parameterized weak-friction theory 
of equilibrium tides \citep{Alexander73,Hut81}.
In order achieve tidal circularization within $\sim 5$~Gyrs in
high-$e$ migration scenarios, the giant planet generally should be
more dissipative than Jupiter by more than an order of magnitude
\citep[e.g.][]{Socrates12,Petrovich15,Anderson16}. 

In general, the response of a fluid body (such as a
giant planet) to tidal perturbation involves the excitation of
internal waves and modes at different frequencies \citep[see, e.g.,][]{Ogilvie14}. 
For a body in a sufficiently eccentric orbit,
oscillatory modes excited at pericentre by the star's tidal potential
can grow chaotically in amplitude, rapidly draining energy from the
orbit over multiple close passages \citep{Mardling95a, Mardling95b}. 
This effect is important for high-eccentricity migration because the
same conditions that allow a planet to migrate quickly -- small
pericentre distance and large eccentricity -- are those that result in
chaotic tidal behaviour.  \citet{IP04} first studied the possibility
that chaotic dynamical tides could speed up the circularization of
eccentric gas giants. They developed an iterative algebraic map to
simply follow mode evolution over many orbits from one pericentre
passage to the next.  More recently, \citet{Vick18} used a similar map
(generalized to include linear mode dampling) to quantify the various
dynamical behaviours of the ``mode + eccentric orbit'' system (see
also Section \ref{sec:Chaos} below); they suggested that chaotic tides could quickly
produce eccentric warm Jupiters (WJs), with a semi-major axis between
0.1 and 1~au, that efficiently circularize to HJs.  \cite{Wu18} also
studied the orbital evolution of a planet experiencing chaotic tides,
and uncovered some of the key features of LK high-$e$ migration with chaotic tides
-- we will examine these features in more details in the later sections of this paper.

In this paper, we present a comprehensive study of 
high-eccentricity migration of giant planets, incorporating a detailed
model of chaotic tidal evolution.  We focus on migration via the LK
effect with a stellar companion. Although this flavour of high-$e$
migration may only account for a fraction of the HJ population, it is
most suitable for systematic study since the initial conditions (such
as the orbital properties of the stellar perturber) are reasonably well-justified,
and it serves as a benchmark for other high-$e$ migration
scenarios. Indeed, several aspects of our results presented in this
paper can be applied to other flavours of high-$e$ migration \citep[see][which studies dynamical tides in the
``secular chaos'' scenario]{Teyssandier18}.  Comparing to our previous study of LK
migration with static tides \citep{Anderson16,Munoz16},
we show that including chaotic tides leads to a number of ``favorable''
features for the formation of close-in giant planets: e.g., it naturally
produces eccentric WJs on short timescales ($\lesssim 10-100$~Myrs)
and speeds up the formation of HJs; it generates a wider HJ period distribution; and it prevents some gas giants
from undergoing tidal disruption, thereby increase the HJ formation fraction.
Overall, chaotic tides make high-$e$ migration a more promising 
mechanism for producing HJs  

The structure of the paper is as follows. In Section~\ref{sec:Chaos}
we revisit the derivation of the iterative map for the evolution of
planetary oscillation modes (dynamical tides), starting from
hydrodynamical equations and including the effect of planetray
rotation; we examine the different dynamical behaviours of the ``modes
+ eccentric orbit'' system and discuss the conditions for various
modes in the planet to become chaotic. 
In Section~\ref{sec:TidesandKozai} we describe our model for coupling LK
oscillations with chaotic tides.  
Section~\ref{sec:Features} discusses
the key features and signposts of LK migration with chaotic tides
and includes a number of analytical results that characterize chaotic tidal migration.  
In Section~\ref{sec:popSynthesis} we present our population
synthesis study for planets undergoing LK oscillations with chaotic tides.
In Section~\ref{sec:Rates} we give analytical calculations
for the HJ formation and tidal disruption fractions from
this migration mechanism and compare them with our numerical results.
We conclude in Section 7 with a summary of
results and discussion of their implications.

\section{Chaotic Dynamical Tides in Giant Planets in Eccentric Orbits} \label{sec:Chaos}
	Consider a planet (mass $M_p$ and radius $R_p$) in a highly eccentric orbit around a star (mass $M_*$). At each pericentre passage, the time-varying gravitational potential of the star excites oscillatory modes (e.g. f-modes and r-modes) in the planet. Over multiple pericentre passages, the orbit and the modes exchange energy and angular momentum. In general, this exchange can occur in either direction depending on the phase of the modes \citep{Kochanek92,Mardling95a}. Provided the oscillation amplitudes are sufficiently small, one can determine the dynamical behaviour of the system using linear hydrodynamics \citep[e.g.][]{Mardling95a, Mardling95b, Lai96, Kumar96, Lai97}. However, following the evolution of the system over many orbits can be challenging because of the short timescale associated with the pericentre passage. Fortunately, when the orbit is eccentric enough, such that the mode and the orbital evolution mainly occur at pericentre, the full hydrodynamic solution of the ``eccentric orbit + oscillation modes" system can be expressed as an iterative map \citep{IP04, Vick18}. 		

\subsection{Hydrodynamics of Tidally Forced Oscillations \& Iterative Map} \label{sec:Mapping}
	In \citet{Vick18}, we presented a simple iterative map to describe the long-term evolution of an ``eccentric orbit + oscillation modes" system. Our map neglected the rotation of the planet and assumed that the change in the mode amplitude during each pericentre passage is constant throughout the orbital evolution. Giant planets can have substantial rotations, which affect their mode properties and introduce a few subtleties to the mapping. Also, during the long-term evolution of the planet's orbit driven by an external companion (the Lidov-Kozai effect), the pericentre distance can change appreciably, which affects mode excitation. Here we present a general derivation of the iterative map from linear hydrodynamics, allowing for the planet to rotate and for the pericentre distance (and thus the change in mode amplitude at pericentre) to vary between passages. 
	
	The stellar gravitational potential that excites oscillations in the planet is given by
	\begin{equation}
	U({\bf r},t)=-GM_*\sum_{lm}{W_{lm}r^l\over D^{l+1}}\,\, e^{-im\Phi(t)}
	Y_{lm}(\theta,\phi_i), \label{eq:defQuadPotential}
	\end{equation}
	where ${\bf r}=(r,\theta,\phi_i=\phi+\Omega_s t)$ is the position vector (in spherical
	coordinates) relative to the centre of mass of the planet (the azimuthal angle $\phi$ is measured in the rotating frame of the planet, and $\Omega_s$ is the rotation rate of the planet\footnote{We assume the spin axis of the planet is aligned with the orbital angular momentum axis throughout this paper.}). In equation~(\ref{eq:defQuadPotential}), $D(t)$ is the time-varying separation between the star and planet, $\Phi(t)$ is the orbital true anomaly, and $W_{lm}$ is a constant defined in \citet{Press77}. The dominant (quadrupole) terms have $l=|m|=2$ with $W_{2\pm2} = \sqrt{3\uppi/10}$. 
	
	The response of the planet to tidal forcing is described by the Lagrangian displacement, $\boldsymbol{\xi}(\boldsymbol{r},t)$. A free oscillatory mode of frequency $\omega_\alpha$ (in the rotating frame) has the form $\boldsymbol{\xi}_\alpha(\boldsymbol{r},t)=\boldsymbol{\xi}_\alpha(\boldsymbol{r})\,\text{e}^{-\text{i}\omega_\alpha t}\propto
	\text{e}^{\text{i}m\phi-i\omega_\alpha t}$, where $\alpha$ is the mode index. We expand $\boldsymbol{\xi}(\boldsymbol{r},t)$ in terms of the eigenmodes in the phase space \citep{Schenk02}:
	\begin{equation}
	\left[\begin{array}{c}
	\bxi\\
	{\partial\bxi/\partial t}
	\end{array}\right]
	=\sum_\alpha b_\alpha(t)
	\left[\begin{array}{c}
	\bxi_\alpha(\boldsymbol{r})\\
	-i\omega_\alpha\bxi_\alpha(\boldsymbol{r})
	\end{array}\right].
	\end{equation}
	The linear fluid dynamics equations then reduce to a set of first-order differential equations \citep{Lai06},
	\begin{equation}
	\dot{b}_\alpha + i\omega_\alpha b_\alpha = \frac{i M_*W_{lm}Q_\alpha}{2\epsilon_\alpha D^{l+1}}
	\, \text{e}^{-im\Phi(t)+im\Omega_s t}, \label{eq:cdoteq}
	\end{equation}
	where 
	\begin{equation}
	Q_\alpha= \int d^3x\,\rho \bxi_\alpha^\star\cdot\nabla (r^lY_{lm}) \label{eq:defQ}
	\end{equation} 
	is the dimensionless tidal overlap integral (in units where $G=M_p=R_p=1$), and
	\begin{equation}
	\epsilon_\alpha = \omega_\alpha + \int d^3x\,\rho \bxi_\alpha^\star\cdot(\text{i} \boldsymbol{\Omega_s} \times \boldsymbol{\xi_\alpha})\label{eq:epsdef}.
	\end{equation}
	Note that in the slow rotation limit, $\epsilon_\alpha$ is simply the mode frequency for a non-rotating planet, $\omega_\alpha(0)$. When $\epsilon_\alpha \gg \Omega_s$, the first-order correction to the mode frequency due to rotation is $\omega_\alpha - \omega_\alpha(0) \approx - m C_{\alpha}\Omega_s = -\int d^3x\,\rho \bxi_\alpha^\star\cdot(\text{i} \boldsymbol{\Omega_s} \times \boldsymbol{\xi_\alpha})$ \citep[e.g.][]{Unno89}. In equations~(\ref{eq:cdoteq})-(\ref{eq:epsdef}), the eigenmode is normalized according to 
	\begin{equation}
	\int d^3x \rho(\boldsymbol{r}) |\bxi_\alpha(\boldsymbol{r})|^2=1.
	\end{equation}
	The general solution to equation~(\ref{eq:cdoteq}) is
	\begin{equation}
	b_\alpha(t) = \text{e}^{-i\omega_\alpha t} \int_{t_0}^t
	\frac{iM_*W_{lm}Q_\alpha}{2\epsilon_\alpha  D(t')^{l+1}}\,\text{e}^{i\sigma_\alpha t'-im\Phi(t')}\,dt' + b_\alpha(t_0), \label{eq:cSolution}
	\end{equation}
	where 
	\begin{equation}
	\sigma_\alpha = \omega_\alpha + m \Omega_s
	\end{equation}
	is the mode frequency in the inertial frame. 
	
	For a highly eccentric orbit, we can assume that any energy transfer between a mode and the orbit occurs at pericentre. We can then manipulate equation~(\ref{eq:cSolution}) into a discrete form by defining $t_k$ as the time at apocentre after the $k$-th pericentre passage, i.e.
	\begin{equation}
	t_k = t_{k-1} + \frac{1}{2}\left(P_{k-1} + P_k\right),
	\end{equation}
	where $P_k$ is the orbital period after the $k$-th pericentre passage. We also define
	\begin{equation}
	\Delta b_{\alpha,k} = \int_{-P_{k-1}/2}^{P_k/2} \frac{iM_*W_{lm}Q_\alpha}{2\epsilon_\alpha D(t')^{l+1}}
	\,\text{e}^{i\sigma_\alpha t'-im\Phi(t')}\;dt'\label{eq:Deltacdef}.
	\end{equation}
	Physically, this is the real change in mode amplitude during the $k$-th pericentre passage. Equation~(\ref{eq:cSolution}) becomes 
	\begin{equation}
	b_{\alpha} = b_{\alpha,0} + \text{e}^{-\text{i}\omega_\alpha t_k} \sum_{j=1}^k \text{e}^{\text{i}\sigma_\alpha (t_{j-1}+P_{j-1}/2)}\Delta b_{\alpha,j} \label{eq:bkdiscrete}.
	\end{equation}
	We assume the initial condition $b_{\alpha,0} = 0.$ Equation~(\ref{eq:bkdiscrete}) can be rewritten in an iterative form:
	\begin{equation}
	b_{\alpha,k} = b_{\alpha,k-1}\text{e}^{-\text{i}\omega_\alpha(P_{k-1}+P_k)/2} + \Delta b_{\alpha,k}\text{e}^{\text{i}(m \Omega_s t_k - \sigma_\alpha P_k/2)}.\label{eq:bkiterative}
	\end{equation}
	We now define the mode amplitude in the inertial frame and shift the index $k$ to count pericentre passages using
	\begin{equation}
	c_{\alpha,k} \equiv b_{\alpha,k} \text{e}^{-\text{i}(m\Omega_s t_k+\sigma_\alpha P_k/2)}.
	\end{equation}
	Physically, $c_{\alpha,k}$ is the mode amplitude just before the $(k+1)$-th pericentre passage. Equation~(\ref{eq:bkiterative}) then becomes
	\begin{equation}
	c_{\alpha,k} = ( c_{\alpha,k-1} + \Delta c_{\alpha,k})\text{e}^{-\text{i}\sigma_\alpha P_{k}},\label{eq:Map}
	\end{equation}
	where $\Delta c_{\alpha,k} = \Delta b_{\alpha,k}$. 
	Equation~(\ref{eq:Map}) has a straight-forward physical interpretation: when the planet passes through pericentre (the $k$-th passage), the mode amplitude changes by $\Delta c_{\alpha,k}$, and the orbital period changes to $P_k$; as the planet completes its orbit, the phase of the mode evolves due to the passage of time, and the complex mode amplitude gains a factor $\text{e}^{-\text{i}\sigma_\alpha P_k}$. Note that in general $\Delta c_{\alpha,k}$ depends on the ``current" parameters of the system. If the pericentre distance and the shape of the orbit near pericentre remain unchanged over many passages, $\Delta c_{\alpha,k}$ is nearly constant from one passage to the next. In this case, equation~(\ref{eq:Map}) reduces to the result from \citet{Vick18} but without mode damping and with the phase evolution determined by the mode frequency in the inertial frame \citep[see also][]{IP04}.  
	 
	The mode amplitude is directly related to the mode energy via 
	\begin{equation}
	E_{\alpha,k} = 2 \sigma_\alpha \epsilon_\alpha|c_{\alpha,k}|^2, \label{eq:tildecdef}
	\end{equation} 
	where the factor of 2 arises because a mode with given $m\;(>0)$, and $\sigma_\alpha$ is physically identical to that with $(-m)$ and $ (-\sigma_\alpha)$ --- we treat them as the same mode.
	We define the dimensionless mode energy, $\tildeE_{\alpha,k}$, in units of the initial binary (planetary) orbital energy, $|E_{B,0}| = GM_*M_p/(2a_0)$ (where $a_0$ is the semi-major axis), and define the re-normalized mode amplitude $\tilde{c}_{\alpha,k}$ such that
	\begin{equation}
	\tildeE_{\alpha,k}\equiv|\tilde{c}_{\alpha,k}|^2\equiv\frac{E_{\alpha,k}}{|E_{B,0}|}.
	\end{equation}
	The total energy transfer to modes in the $k$-th passage is therefore
	\begin{equation}
	\Delta\tildeE_{k}= \sum_{\alpha} \Delta \tildeE_{\alpha,k} =\sum_{\alpha}(|\tilde{c}_{\alpha, k-1}+\Delta \tilde{c}_{\alpha}|^2 - |\tilde{c}_{\alpha,k-1}|^2). \label{eq:DeltaE}
	\end{equation}
	From energy conservation, the orbital energy ($\tildeE_{B,k}$) immediately after the $k$-th passage is given by
	\begin{equation}
	\tildeE_{B,k} = \tildeE_{B,k-1} - \Delta\tildeE_{k} = \tilde{E}_{B,k-1} - \sum_{\alpha}\tilde{E}_{\alpha,k}, \label{eq:Eb}
	\end{equation}
	and the corresponding orbital period is
	\begin{equation}  \label{eq:Kepler}
	{P_k\over P_0}=\left({\tildeE_{B,0}\over\tildeE_{B,k}}\right)^{3/2},
	\end{equation}
	where $\tildeE_{B,0}=-1$, and $P_0$ is the initial period.
	
	Equations~(\ref{eq:Map}) and (\ref{eq:DeltaE})-(\ref{eq:Kepler}) complete the map. The required inputs are  the initial mode energy, $\tildeE_{\alpha,0} = |\tilde{c}_{\alpha,0}|^2$, and $\Delta \tilde{c}_{\alpha,k}$, the change in mode amplitude during each pericentre passage. This map accurately determines the evolution of the orbit and mode energies provided that the orbit is highly eccentric and the modes remain linear (i.e., the sum of the mode energies is much less than the binding energy of the planet).
	
	The dimensionless change in mode amplitude during a pericentre passage (suppressing the subscript `k'), $\Delta \tilde{c}_\alpha$, depends on the orbital parameters (mainly the pericentre distance $r_{\rm p}$) and the mode properties. It is related to the energy transfer $\Delta E_\alpha$ by 
	\begin{equation}
	|\Delta \tilde{c}_{\alpha}|^2 = \frac{\Delta E_\alpha}{|E_{\rm B, 0}|} \label{eq:Deltatildecdef}.
	\end{equation}
	Note that in general the energy transfer to mode $\alpha$ in a pericentre passage depends on the amplitude and phase of ``pre-existing" oscillations [see equation~(\ref{eq:DeltaE})]. In equation~(\ref{eq:Deltatildecdef}), $\Delta E_\alpha$ refers to the energy transfer when the mode has zero amplitude prior to the pericentre passage --- in \citet{Vick18}, we term this ``the first passage." The energy transfer $\Delta E_\alpha$ for a parabolic encounter ($e\to 1$) was first derived by \citet{Press77}. This can be generalized to eccentric orbits and modified to include the effect of rotation \citep{Lai97,Fuller12a}.
	For a pericentre distance $r_{\rm p}$, $\Delta E_\alpha$ can be written as (keeping only the $l=2$ terms)
	\begin{align}
	\Delta E_\alpha 
	&= \frac{GM_*^2}{r_{\rm p}^6}R_p^5 T(\eta, \sigma_\alpha/\Omega_p, e) \nonumber \\ &= \frac{GM_p^2}{R_p}\left(\frac{M_*}{M_{\rm tot}}\right)^{2} \eta^{-6} T(\eta, \sigma_\alpha/\Omega_p, e), \label{eq:DelE}
	\end{align} 
	where $\eta$ is the ratio of $r_{\rm p}$ and the tidal radius $r_{\rm tide}$, i.e.
	\begin{align}
	\eta \equiv \frac{r_{\rm p}}{r_{\rm tide}}, && r_{\rm tide} \equiv R_p \left(\frac{M_{*}}{M_p}\right)^{1/3}, \label{eq:defeta}
	\end{align}
	with $M_{\rm tot} = M_*+M_p \simeq M_*$, and $\Omega_{\rm p}$ is the pericentre orbital frequency
	\begin{equation}
	\Omega_p \equiv \left(\frac{G M_{\rm tot}}{r_{\rm p}^3}\right)^{1/2}= \left(\frac{GM_p}{R_p^3}\right)^{1/2}\eta^{-3/2}.
	\end{equation}
	Note that $\sigma_\alpha/\Omega_{p} = \bar{\sigma}_\alpha\eta^{3/2}$, with $\bar{\sigma}_\alpha \equiv \sigma_\alpha (R_p^3/GM_p)^{1/2}$. To avoid tidal disruption of the planet, we require \citep{Guillochon11} 
	\begin{align}
	r_{\rm p} \ge r_{\rm p,dis} = 2.7 r_{\rm tide}, &&\text{or} && \eta \ge 2.7. \label{eq:dis}
	\end{align}
	The dimensionless function $T$ is given by
	\begin{equation}
	T = 2\pi^2 \frac{\sigma_\alpha}{\epsilon_\alpha}(Q_\alpha K_{lm})^2, \label{eq:defT}
	\end{equation}
	where $Q_\alpha$ is an overlap integral (see equation \ref{eq:defQ}), and
	\begin{equation}
	K_{lm} = \frac{W_{lm}}{2 \pi} \int_{-P/2}^{P/2}dt \; \left(\frac{r_{\rm p}}{D(t)}\right)^{l+1} \text{e}^{\text{i}\sigma_\alpha t - \text{i} m \Phi(t)}. \label{eq:Klmdef}
	\end{equation}
	In equation~(\ref{eq:Klmdef}), all quantities are in units where $G=M_p=R_p=1$ so that $K_{lm}$ is dimensionless.
	
	To follow the evolution of the system over many orbits, it is useful to be able to compute $K_{lm}$ (and thus $\Delta E_\alpha$) efficiently. For the dominant $l=m=2$ prograde mode ($\sigma_\alpha > 0$), \citet{Lai97} derived an approximation for $K_{22}$ that is accurate to within 2\% for $(1-e)\ll 1$ and $z\equiv \sqrt{2} \sigma_\alpha/\Omega_p \gtrsim$ a few: 
	\begin{equation}
	K_{22} \simeq \frac{2 z^{3/2}\exp(-2z/3)}{\sqrt{15}}\left(1- \frac{\sqrt{\uppi}}{4 \sqrt{z}}\right) \eta^{3/2}. \label{eq:Kapprox}
	\end{equation}
	
	As shown in \citet{Vick18}, when the mode has no initial energy, i.e. $\tilde{E}_{\alpha,0}=0$, the important quantity that determines the dynamical behaviour of the system is $\Delta \hat{P}_\alpha = \sigma_\alpha \Delta P$, where $\Delta P = P_1 - P_0 \simeq - (3/2) P_0 (\Delta E_\alpha/|E_{B,0}|)$ is the change in the initial orbital period ($P_0$) due to the energy transfer $\Delta E_\alpha$. From equation~(\ref{eq:DelE}), we have 
	\begin{align}
	|\Delta \hat{P}_\alpha|= \sigma_\alpha \Delta P &\simeq \frac{3}{2} \sigma_\alpha P_0 \frac{\Delta E_\alpha}{|E_{B,0}|}\nonumber\\ &\simeq \frac{6\pi \sigma_\alpha/\Omega_{p}}{(1-e)^{5/2}} \left(\frac{M_p}{M_*}\right)^{2/3} \eta^{-5} T(\eta,\sigma_\alpha/\Omega_p,e). \label{eq:defDelPhat}
	\end{align} 
	Physically, $|\Delta \hat{P}_\alpha|$ is the phase change in the mode due to energy transfer at pericentre when there is no ``pre-existing" mode energy.
\subsection{Planetary Oscillation Modes}\label{sec:ModeTypes}
	
	\begin{table*}
		\caption{Properties of $l=m=2$ f-modes and inertial modes for a $\gamma=2$ polytrope and a $\gamma = 5/3$ polytrope. The mode frequencies $\omega_\alpha$ (in the rotating frame), $\epsilon_\alpha$ (equation \ref{eq:epsdef}), and $\sigma_\alpha$ (in the inertial frame), and the planetary rotation rate $\Omega_s$ are in units of $(GM_p/R_p^3)^{1/2}$, i.e. $\omega_\alpha = \bar{\omega}_\alpha (GM_p/R_p^3)^{1/2}$. The tidal overlap integral $Q_\alpha$ is defined by equation~(\ref{eq:defQ}).}
		\begin{tabular}{l|l|l|l|l|l}
			Model Type & Mode & $\bar{\omega}_\alpha$ & $\bar{\epsilon}_\alpha$ & $\bar{\sigma}_\alpha$ & $Q_\alpha$\\
			\hline
			$\gamma=2$ polytrope & f-mode & $1.22-\bar{\Omega}_s$ & $1.22$ & $1.22 + \bar{\Omega}_s$ & 0.56 \\
			&i-mode 1 ($j=3$) & 0.56 $\bar{\Omega}_s$ & 0.28 $\bar{\Omega}_s$ &  2.56 $\bar{\Omega}_s$ & $0.015 \bar{\Omega}_s^2$\\
			& i-mode 2 ($j=3$) & -1.1 $\bar{\Omega}_s$ & -0.55$\bar{\Omega}_s$ &0.9 $\bar{\Omega}_s$ & 0.01 $\bar{\Omega}_s^2$\\
			$\gamma = 5/3$ polytrope & f-mode & $1.46-\bar{\Omega}_s$ & $1.46$ & $1.46 + \bar{\Omega}_s$ & 0.49\\
			\hline
		\end{tabular}
		\label{tab:ModeProperties}
	\end{table*}
	
	\begin{figure*}
		\begin{center}
			\includegraphics[width=5.5 in]{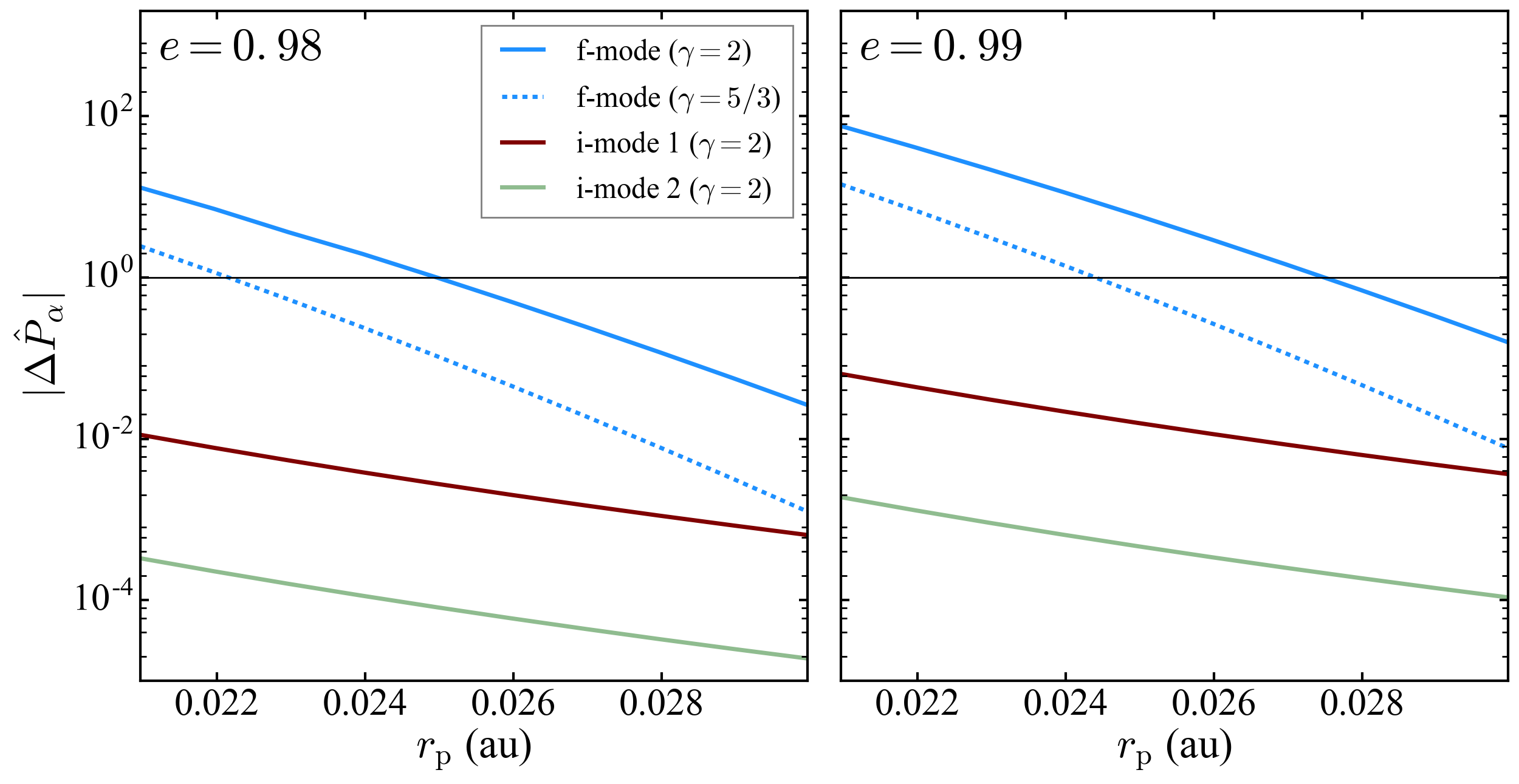}
			\caption{The dimensionless quantity $|\Delta \hatP_{\alpha}|$ (see equation~\ref{eq:defDelPhat}) as a function of the pericentre distance $r_{\rm p}$ for various planetary oscillation modes: the $l=m=2$ f-mode and two i-modes of a $\gamma=2$ polytrope, and the $l=m=2$ f-mode of a $\gamma=5/3$ polytrope. The properties of these modes are provided in Table~\ref{tab:ModeProperties}. The left and right panels show the results for two values of the orbital eccentricity ($e=0.98,0.99$). For all models, we use $R_p= 1.6~R_J$, $q=M_*/M_p = 10^3$. The black line marks $\Delta \hatP_\alpha=1$.}
			\label{fig:ModeComparison}
		\end{center}
	\end{figure*}
		
	The gravitational potential of a star can excite many types of oscillation modes in the eccentric orbiting planet. Some of the modes will be more strongly excited than others and therefore more influential in long-term orbital evolution. The dominant modes are quadrupole, with $l = |m| = 2$. The $m=0$ modes are generally much less important because $K_{l0}$ (equation \ref{eq:Klmdef}) is small. In this paper, we adopt the convention $m>0$, so a mode with $\sigma_\alpha > 0$ ($\sigma_\alpha < 0$) is prograde (retrograde) with respect to the rotation of the planet in the inertial frame. For the rest of this paper, we only consider modes with $l=m=2$. 
	
	We adopt a simple giant planet model composed of a neutrally stratified fluid characterized by a $\gamma=2$ polytrope equation of state ($P \propto \rho^2$). Of the acoustic modes (with the restoring force being pressure), the f-mode (fundamental mode) has a frequency closest to the frequency of tidal forcing near pericentre, and the largest tidal overlap integral $Q_\alpha$. Other acoustic modes (p-modes) have higher frequencies and smaller tidal overlaps, and are much less important than the f-mode for energy transfer. For a rotating planet, the stellar potential can also excite i-modes [i.e. inertial modes, with the restoring force being the Coriolis force; see \citet{Xu18} and references therein]. In a gas giant that has stratification, g-modes (driven by buoyancy) may also be excited \citep{Fuller14}. The presence of a large solid core also supports elastic modes which can ``mix" with the f-mode \citep[e.g.][]{Fuller13}. We do not consider such planet models in this paper because there are considerable uncertainties in the stratification and core size of giant planets \citep[e.g.][]{Guillot05}.
	
	Table \ref{tab:ModeProperties} lists the key properties of the $l=m=2$ f-modes and the first two inertial modes for the $\gamma=2$ polytropic planet model. For comparison, the f-mode properties of the $\gamma = 5/3$ polytrope are also given. The f-modes are calculated using the slow-rotation approximation, which gives $\omega_\alpha = \omega_\alpha(0) - m C_\alpha \Omega_s$, and we find $C_\alpha \simeq 0.5$ for both the $\gamma=2$ and $\gamma=5/3$ models. The i-mode result is from \citet{Xu18} (see their Table IV), based on calculations using a non-perturbative spectral code. Note that for $m=2$, the mode frequencies in the rotating frame ($\omega_\alpha$) and in the inertial frame ($\sigma_\alpha$) are related by $\sigma_\alpha = \omega_\alpha + 2\Omega_s$. 
	
	To determine which of these modes produces the strongest dynamical tides, we calculate $|\Delta \hatP_{\alpha}|$ with equation~(\ref{eq:defDelPhat}) as a function of $r_{\rm p}$ for each of the modes (see Fig.~\ref{fig:ModeComparison}). For definiteness, we assume that the planet rotates at the pseudosynchronous rate $\Omega_{\rm ps}$ given by the weak friction theory of equilibrium tides \citep[e.g.][]{Hut81}, i.e.
	\begin{equation}
	\Omega_{\rm s} = \Omega_{\rm ps} \equiv \frac{f_2(e)}{(1-e^2)^{3/2}f_5(e)}n, \label{eq:Pseudosynchronous}
	\end{equation}
	where $n= (GM_{\rm tot}/a^3)^{1/2}$ is the mean motion of the planet and
	\begin{align}
	f_2(e) &= 1+\frac{15}{2}e^2 + \frac{45}{8} e^4 + \frac{5}{16}e^6, \label{eq:deff2}\\
	f_5(e) &= 1+ 3 e^2 + \frac{3}{8}e^4. \label{eq:deff5}
	\end{align}
	For $(1-e) \ll 1$, $\bar{\Omega}_{\rm ps}= \Omega_{\rm ps} (GM_p/R_p^3)^{-1/2} \simeq 1.17/\eta^{3/2}$, and the mode frequency in the inertial frame, $\sigma_\alpha$, is nearly independent of the orbital eccentricity. In this limit, $T(\eta, \sigma_\alpha/\Omega_p, e)$ also has negligible eccentricity dependence, and $|\Delta \hatP_\alpha| \propto (1-e)^{-5/2}$ [see equation~(\ref{eq:defDelPhat})]. The two panels of Fig.~\ref{fig:ModeComparison} shows that $|\Delta \hatP_\alpha|$ indeed increases with $e$ for all modes. We are most interested in systems where $r_{\rm p}$ is sufficiently small such that  $|\Delta \hatP_\alpha| \gtrsim 1 $ (the criterion for chaotic mode growth; see Section~\ref{sec:ChaosConditions}). For such systems, it is clear that the $l=m=2$ f-mode is most strongly excited at pericentre passages. For the remainder of the paper, we focus on the prograde $l=2$ f-mode and neglect the contributions to energy transfer from other modes. 

\subsection{Conditions for Chaotic Tides}\label{sec:ChaosConditions}
	\begin{figure}
		\begin{center}
			\includegraphics[width=\columnwidth]{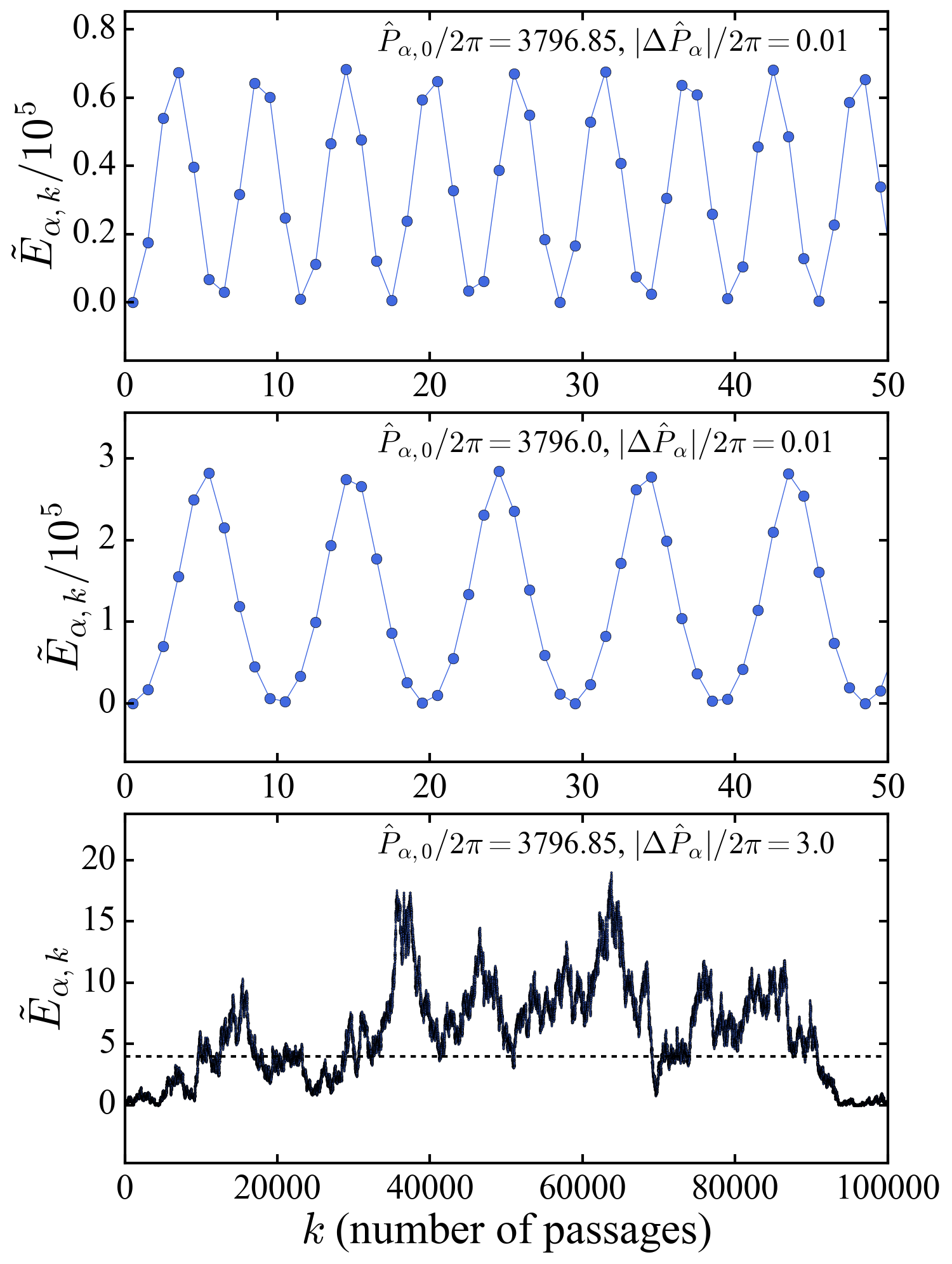}
			\caption{The evolution of the f-mode energy (scaled to the initial orbital energy $|E_{\rm B,0}|$) over multiple pericentre passages for star-planet systems with different values of $\hatP_{\alpha,0}$ and $|\Delta \hatP_{\alpha}|$. Note that the top two panels have a different $y$-scale and $x$-axis range from the bottom panel. Top: An example of low-amplitude oscillations (with $|\Delta \hat{P}_\alpha| \lesssim 1$). Middle: An example of resonant mode evolution (with $\hat{P}_0/2\pi =$ integer). Bottom: An example of chaotic mode growth (with $|\Delta \hat{P}_\alpha| \gtrsim 1$). For a system with $M_*= 1 M_\odot$, $M_p=1M_J$, $R_p = 1.6 R_J$, and the f-mode properties of a $\gamma=2$ polytrope, the three panels correspond to $a \approx 1.5$ au and $e\approx0.98$ in the top two panels and $e\approx0.985$ in the bottom panel. This calculation does not include dissipation of the mode energy. In the bottom panel, the planet binding energy is shown with a dashed line; physically, the mode must dissipate energy well before $\tildeE_{\alpha,k}$ climbs to this value.}
			\label{fig:3Behaviors}
		\end{center}
	\end{figure}
	When the planetary orbit is sufficiently eccentric, the change in mode amplitude in a pericentre passage, $\Delta c_{\alpha}$ (or equivalently the energy transfer in the ``first" passage, $\Delta E_\alpha$, see equation \ref{eq:DelE}), is constant over many pericentre passages.\footnote{This requires that the pericentre distance is constant, which in turn requires that the fractional change in the orbital angular momentum, $\Delta L/L$, remains small throughout orbital evolution. See equation~(13) of \citet{Vick18}.} In this case, if the mode has negligible dissipation, the dynamical behaviour of the iterative map depends on $\hatP_{\alpha,0} = \sigma_\alpha P_0$, the initial mode energy $E_{\alpha,0}$, and $|\Delta \hatP_\alpha|$ (see equation \ref{eq:defDelPhat}), assuming that the initial mode energy $E_{\alpha,0}$ is negligible (see Section~\ref{sec:extendChaosCondition}). In particular, a system with $|\Delta \hatP_\alpha|$ greater than a critical value, $\Delta \hat{P}_{\rm crit}$, behaves very differently from systems with $|\Delta \hatP_\alpha| \lesssim \Delta \hat{P}_{\rm crit}$. The exact value of $\Delta \hat{P}_{\rm crit}$ depends on the dimensionless orbital period, $\hatP_{\alpha,0}$. On average, $\Delta \hat{P}_{\rm crit} \sim 1$. The map exhibits three classes of behaviour \citep{Vick18}: 
	
	(i) When $|\Delta \hatP_\alpha| \lesssim \Delta \hatP_{\rm crit} \sim 1$, the mode energy undergoes low-amplitude oscillations. A small amount of energy (of order $\Delta E_\alpha$) is transferred back and forth between the mode and the orbit over multiple pericentre passages. An example of this behaviour is shown in the top panel of Fig.~\ref{fig:3Behaviors}. 
	
	(ii) When $|\Delta \hatP_\alpha| \lesssim \Delta \hatP_{\rm crit}$ and $\hat{P}_{\alpha,0}/(2\pi)$ is close to an integer (i.e., when $\sigma_\alpha$ is an integer multiple of the orbital frequency, $2\pi/P_0$), the mode energy exhibits larger-amplitude oscillations with a mean mode energy $\gg \Delta E_\alpha$ (see the middle panel of Fig.~\ref{fig:3Behaviors}.) 
	
	(iii) When $|\Delta \hatP_\alpha| \gtrsim \Delta \hatP_{\rm crit}$, the mode amplitude evolves chaotically and can grow to very large values of order the initial orbital energy $|E_{B,0}|$; meanwhile, the orbit experiences significant decay in the semi-major axis. The bottom panel of Fig.~\ref{fig:3Behaviors} provides an example. Note that in this example the binding energy of the planet is of order the initial orbital energy ($GM_p^2/R_p = 2.6\;|E_{\rm B,0}|$ for an initial semi-major axis $a_0= 1$~au). For such chaotic systems, the mode energy will eventually grow to values comparable to the binding energy of the planet, at which point the linear treatment is no longer appropriate and the mode will dissipate energy due to non-linear effects. We will discuss the implications of non-linear dissipation in later sections.
	
	These three different behaviours can be characterized by the maximum mode energy that is attained over many pericentre passages.  
	The examples in Fig.~\ref{fig:3Behaviors} demonstrate that the mode energy remains a small fraction of the initial orbital energy for non-chaotic evolution.  However, when the system behaves chaotically, the maximum mode energy, $\rm{max}(E_{\alpha,k})$, can be significant compared to $|E_{\rm B,0}|$. 
	
	\begin{figure*}
		\begin{center}
			\includegraphics[width= 7in]{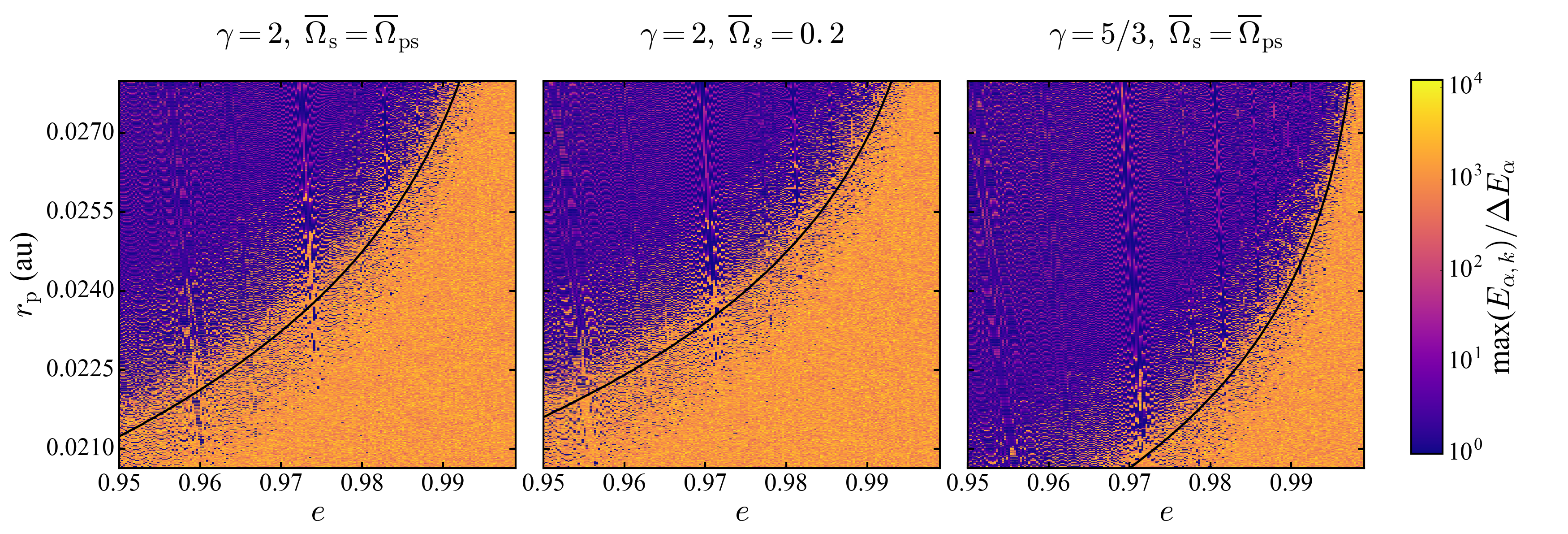}
			\caption{The maximum planet f-mode energy, $\max(E_{\alpha,k})$ (in units of $\Delta E_{\alpha}$), after $10^3$ orbits for a system with $M_* = 1 M_\odot$, $M_p = 1 M_J$, $R_p =1.6~R_J$ and a polytropic planet model with either $\gamma=2$ or $\gamma=5/3$ as labelled. Planets below the minimum $r_{\rm p} = $0.0207~au for the assumed star/planet parameters are expected to tidally disrupt; see equation~(\ref{eq:dis}). In the left and right panels, the planet is assumed to be rotating pseudo-synchronously (see equation \ref{eq:Pseudosynchronous}). In the middle panel, the planet rotates at the rate $\Omega_s = 0.2 (GM_p/R_p^3)^{1/2}$. Systems in the blue region display low-amplitude oscillations, while those in the orange region exhibit chaotic mode growth. The light purple ridges correspond to resonances between the mode frequency and the orbital frequency. The solid black lines correspond to $|\Delta \hat{P}_\alpha| = 1$ (equations \ref{eq:defDelPhat} and \ref{eq:echaos}).}
			\label{fig:HJParams}
		\end{center}
	\end{figure*}
	
	In Fig.~\ref{fig:HJParams} we explore the boundary for chaotic behaviour in the $r_{\rm p} - e$ plane by recording $\text{max}(E_{\alpha,k})$ over $10^3$ orbits for different planet models with $M_p = 1M_J$ and $R_p=1.6~R_J$ (the stellar mass is fixed at $M_*=1M_\odot$). The left and right panels of Fig.~\ref{fig:HJParams} assume that $\Omega_s$ is the pseudosynchronous rate from equation~(\ref{eq:Pseudosynchronous}) while the middle panel uses a constant value of $\Omega_s = 0.2(GM_p/R_p^3)^{1/2}$. Small changes in $\Omega_s$ do not have a large effect on the chaotic boundary. The right panel of Fig.~\ref{fig:HJParams} uses a $\gamma=5/3$ polytropic planet model rather than our standard choice of $\gamma=2$. A comparison between the right panel and others suggests that changes in the structure of the planet can significantly shift the boundary for chaotic behaviour. 
	
	Figure~\ref{fig:HJParams} demonstrates that the condition $|\Delta \hat{P}_\alpha| \gtrsim 1$ generally captures the boundary for chaotic behaviour. This boundary is somewhat fuzzy because the precise value of $\Delta \hatP_{\rm crit}$ spans a wide range, from $<0.1$ to $>1$, depending on the value of $\hatP_{\alpha,0}$ \citep[see Fig.~1 of][]{Vick18}. In particular, $\Delta \hatP_{\rm crit} \ll 1$ when $\hatP_{\alpha,0}/(2\pi)$ is close to an integer (resonance). Figure~\ref{fig:HJParams} confirms that $\Delta \hatP_{\rm crit} \sim 1$ on average. When $|\Delta \hat{P}_\alpha| \gtrsim 1,$ the change in the mode phase is nearly a random number mod $2\uppi$, and each pericentre ``kick" to the mode occurs at a random phase. Under this condition, the mode energy grows in a diffusive manner \citep{Mardling95a,IP04,Wu18}. Note that this ``diffusion" is approximate; as explained in \citet{Vick18} (see their Fig.~3), there exists an upper limit to the mode energy even in the absence of dissipation. For this reason, we prefer to call the mode growth and dissipation in the regime $|\Delta \hat{P}_\alpha| \gtrsim 1$ ``chaotic tides." 
	
	From equation~(\ref{eq:defDelPhat}), we see that $|\Delta \hatP_\alpha|$ is larger for more eccentric orbits, and smaller $\eta$ (i.e. smaller $r_{\rm p}$ and larger $R_p$). A young gas giant that is pumped into a highly eccentric orbit can satisfy the condition for chaotic tidal evolution (see Fig.~\ref{fig:HJParams}) and rapidly transfer orbital energy to the f-mode. This behaviour is most likely to occur for $e \gtrsim 0.95$.
	The critical condition for chaotic tides, $|\Delta \hatP_\alpha|\simeq \Delta \hatP_{\rm crit} \sim 1$, can be written (using equation \ref{eq:defDelPhat}) as
	\begin{equation}
	(1-e_{\rm crit}) \simeq 0.11 \bar{\sigma}_\alpha^{2/5} \left(\frac{10^3M_p}{M_*}\right)^{4/15} \left(\frac{\eta}{3}\right)^{-7/5}(\Delta \hatP_{\rm crit}T)^{2/5}. \label{eq:echaos}
	\end{equation}
	Note that $T$ depends on $\eta$ and $\sigma_\alpha/\Omega_{\rm p} = \bar{\sigma}_\alpha \eta^{3/2},$ and is nearly independent of $e$ for large eccentricities. Equation (\ref{eq:echaos}) gives the critical eccentricity, for a given $\sigma_\alpha$, above which chaotic tides operate as a function of the dimensionless pericentre distance $\eta$ (equation \ref{eq:defeta}).\footnote{The mode frequency $\sigma_\alpha$ depends on the planet rotation rate $\Omega_s$, which may depend on $e$ if the equilibrium rotation rate depends on $e$. For pseudosynchronous rotation, this dependence is very weak when $(1-e)\ll 1$ (for which $\Omega_{s} \simeq 1.17 \Omega_{p}$); see equation~(\ref{eq:Pseudosynchronous}).} If we make further simplification for the function $T$ by replacing $K_{22}$ in equation~(\ref{eq:defT}) with the approximate fitting formula $K_{2,2} \approx 1.79 \times 10^{4} z^{-6} \eta^{3/2}$ \citep{Wu18}, equation~(\ref{eq:echaos}) can be inverted to yield 
	\begin{align}
	\eta_{\rm crit} \simeq& 4.25 \;\bar{\sigma}_\alpha^{-0.59}Q_\alpha^{0.11} \left(\frac{1-e}{0.02}\right)^{-0.135} \left(\frac{10^3 M_p}{M_*}\right)^{0.036}\nonumber\\ &\times \left(\frac{\sigma_\alpha}{\epsilon_\alpha\Delta \hat{P}_{\rm crit}}\right)^{0.054}, \label{eq:etaBoundary}
	\end{align}
	or equivalently,
	\begin{align}
	r_{\rm p,crit}\simeq&\; (0.0206\;\rm{au})\;\bar{\sigma}_\alpha^{-0.59}Q_\alpha^{0.11} \left(\frac{1-e}{0.02}\right)^{-0.135} \left(\frac{R_p}{R_J}\right)\nonumber\\ &\times \left(\frac{10^3 M_p}{M_*}\right)^{-0.297} \left(\frac{\sigma_\alpha}{\epsilon_\alpha\Delta \hat{P}_{\rm crit}}\right)^{0.054}. \label{eq:BoundaryApprox}
	\end{align}
	Equation (\ref{eq:BoundaryApprox}) gives the critical pericentre distance for chaotic tides to operate as a function of eccentricity.
	
\subsection{Conditions for Chaotic Tides When $E_{\alpha,0} > 0$}\label{sec:extendChaosCondition} 

\begin{figure*}
	\begin{center}
		\includegraphics[width= 7in]{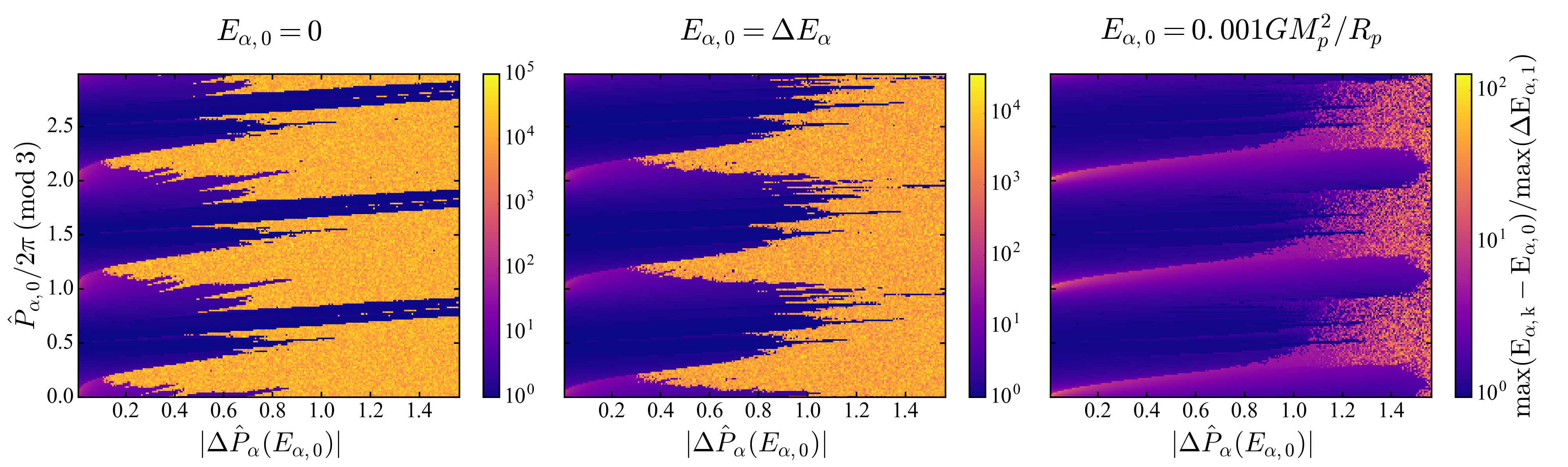}
		\caption{The maximum mode energy, shifted by the initial mode energy $E_{\alpha,0}$ and scaled to $\text{max}(\Delta E_{\alpha,1})$ (see equation \ref{eq:defmaxEtilde}) for a range of $\hatP_{\alpha,0}$ and $\Delta \hatP_{\alpha}(E_{\alpha,0})$. The dark purple regions exhibit low-amplitude oscillations, the pink regions exhibit resonant behaviour, and the orange regions correspond to chaotic mode growth. Each panel shows calculations for a different choice of $E_{\alpha,0}$. The left panel corresponds to no initial energy in the mode. The middle panel corresponds to a mode that is already undergoing low-amplitude oscillations with $ E_{\alpha,0}=\Delta E_\alpha$. The right panel shows a mode with a significant amount of initial energy such that $E_{\alpha,0} \gg \Delta E_\alpha$. For the plotted range of $\hat{P}_{\alpha,0}$, $\Delta \tildeE_{\alpha} \simeq 2.5 \times 10^{-4}$ in the right panel. In general, the boundary for chaotic behaviour varies with $\hat{P}_{\alpha,0}$ and shifts to larger $|\Delta \hatP_{\alpha}(E_{\alpha,0})|$ for larger $E_{\alpha,0}$.}
		\label{fig:PhaseSpaceEinit}
	\end{center}
\end{figure*}

\begin{figure*}
	\begin{center}
		\includegraphics[width= 7in]{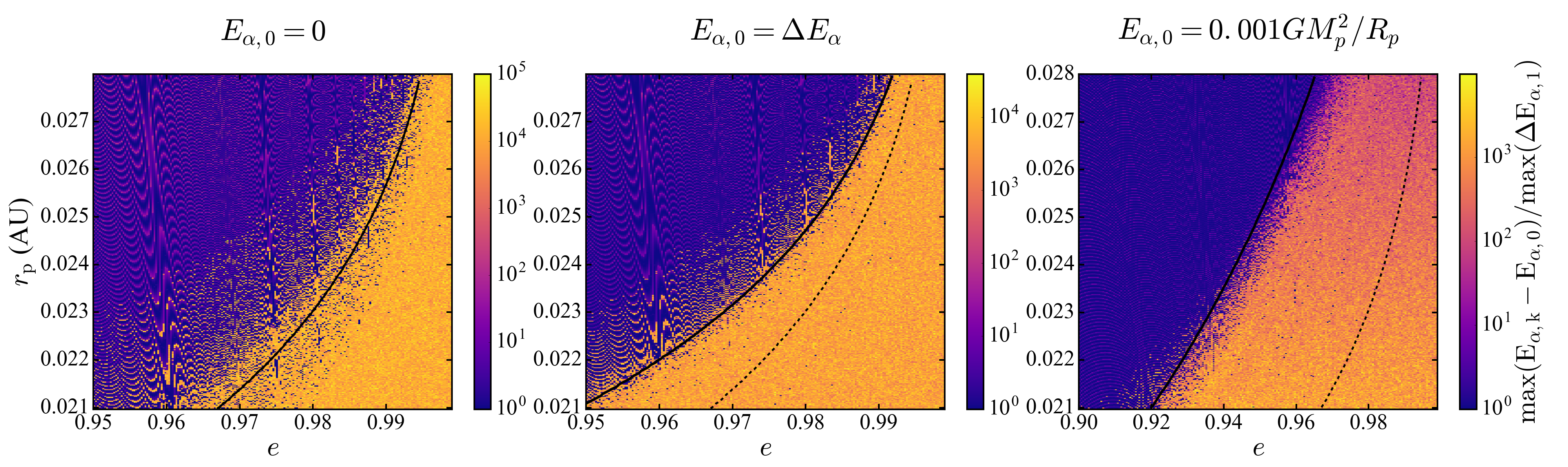}
		\caption{Similar to Fig.~\ref{fig:HJParams}, but showing the effect of finite initial mode energy $E_{\alpha,0}$ (see also Fig.~\ref{fig:PhaseSpaceEinit}). Each panel corresponds to a $\gamma=2$ polytrope and assumes that the planet spin, $\Omega_s$, is pseudosynchronous (see equation \ref{eq:Pseudosynchronous}). In general the boundary for chaotic behaviour shifts to larger $r_{\rm p}$ and lower $e$ for larger $E_{\alpha,0}$. The solid black lines show $|\Delta \hatP_{\alpha}(E_{\alpha,0})|=1$ while the dashed lines show $|\Delta \hatP_\alpha|=|\Delta \hatP_{\alpha}(E_{\alpha,0}=0)|=1$; in the left panel, the two lines coincide.}
		\label{fig:HJParamsEinit}
	\end{center}
\end{figure*}

In Section~\ref{sec:ChaosConditions}, we have discussed the condition for the onset of chaotic mode growth for an ``eccentric orbit + oscillatory modes" system where there is no initial energy in the mode ($E_{\alpha,0} = 0$). When dynamical tides are the only influence on the planet's orbital evolution, and when the total energy of the system is conserved (i.e. the modes do not dissipate), $\hatP_{\alpha,0}$ and $|\Delta \hatP_{\alpha}|$ completely determine the behaviour of the system. For instance, a system with $\Delta \hatP_{\alpha} \lesssim \Delta \hatP_{\rm crit}\sim 1$ will never undergo chaotic mode growth, and one with $\Delta \hatP_{\alpha} \gtrsim \Delta \hatP_{\rm crit} $ will always be chaotic. However, when non-tidal effects alter the planet's orbit (e.g. LK eccentricity oscillations driven by an external companion), a system can transition in and out of the chaotic regime as $r_{\rm p}$ is driven below and above $r_{\rm p,crit}$ [see equations~(\ref{eq:echaos})-(\ref{eq:BoundaryApprox}) and Fig.~\ref{fig:HJParams}]. Therefore, it is useful to know whether a system is ``currently chaotic" when there is ``pre-existing" energy in the mode ($E_{\alpha,0} > 0$). Here, we generalize the criterion for the onset of chaotic tides to account for the ``pre-existing" mode energy \citep[see][]{Mardling95a,Wu18}. 

As discussed in Section~\ref{sec:ChaosConditions}, the mode amplitude evolves chaotically when the mode phase at pericentre is nearly random from one passage to the next. This occurs when the phase-shift from pericentre energy transfer is of order unity. For a mode with no ``pre-existing" energy, the pericentre phase shift is solely due to $\Delta E_\alpha$. When the mode already has energy $E_{\alpha,0}$ prior to the first passage, the pericentre phase shift can reach a maximum value of
\begin{equation}
|\Delta \hatP_{\alpha}(E_{\alpha,0})| \equiv \sigma_\alpha \text{max}(|P_{1}-P_{0}|) \simeq \frac{3}{2} \hatP_{\alpha,0}\;  \frac{\text{max}(\Delta E_{\alpha,1})}{|E_{B,0}|}, \label{eq:DelPhatmax}
\end{equation}
where $\text{max}(\Delta E_{\alpha,1})$ is the maximum possible energy transfer in the first pericentre passage given the initial mode energy $E_{\alpha,0}$, and is given by
\begin{equation}
\text{max}(\Delta E_{\alpha,1}) = \Delta E_\alpha + 2\sqrt{\Delta E_\alpha E_{\alpha,0}}. \label{eq:defmaxEtilde}
\end{equation}
The general condition for chaotic behaviour is that the maximum pericentre phase shift is sufficiently large that that the mode phase is nearly random at pericentre, i.e. 
\begin{equation}
|\Delta \hatP_{\alpha}(E_{\alpha,0})| \gtrsim \Delta \hatP_{\rm crit} \sim 1. \label{eq:GeneralChaosCondition}
\end{equation}
Note that $|\Delta \hatP_{\alpha}(E_{\alpha,0})|$ can be many times larger than $|\Delta \hatP_{\alpha}| = |\Delta \hatP_{\alpha}(0)|$ if  $E_{\alpha,0} \gg \Delta E_\alpha$; in this case equation~(\ref{eq:GeneralChaosCondition}) reduces to equation~(18) of \citet{Wu18}.\footnote{\citet{Wu18} adopted $\Delta \hat{P}_{\rm crit} = 2/3$. In reality, $\Delta \hat{P}_{\rm crit}$ depends strongly on $\hat{P}_0$; see Fig.~\ref{fig:PhaseSpaceEinit}.}

Figure~\ref{fig:PhaseSpaceEinit} shows how the dynamical behaviour of the system depends on $\hatP_{\alpha,0}$ and $|\Delta \hatP_{\alpha}(E_{\alpha,0})|$ for $E_{\alpha,0} = 0, \Delta E_\alpha$, and $0.001 GM_p^2/R_p$. The left panel is similar to Fig.~1 from \citet{Vick18}. The middle panel corresponds to a system where the planet has an initial energy $E_{\alpha,0} = \Delta E_{\alpha}$. This is likely the case the time LK oscillations drive the planet's orbital eccentricity to the regime for chaotic mode growth. In the right panel, $E_{\alpha,0} \gg \Delta E_\alpha$, which is relevant for determining whether a system undergoing chaotic mode growth will continue to behave chaotically (see Section~\ref{sec:FullModel}). As in Fig.~\ref{fig:HJParams}, the orange regions indicate chaotic tides. While the exact value of $\Delta \hatP_{\rm crit}$ depends on whether the mode frequency is near a resonance with the orbital frequency, in general a system with $|\Delta \hatP_{\alpha}(E_{\alpha,0})| \gtrsim 1$ is more likely than not to experience chaotic tides. This criterion works best for $\tilde{E}_{\alpha,0} \equiv  E_{\alpha,0}/|E_{B,0}| \ll 1$; for larger $\tilde{E}_{\alpha,0}$, $\Delta \hat{P}_{\rm crit}$ gradually moves to larger values, as suggested by the right panel of Fig.~\ref{fig:PhaseSpaceEinit}. 

We can also test the generalized criterion for chaotic tides over the range of orbital parameters relevant to migrating gas giants. Figure~\ref{fig:HJParamsEinit} shows the similar results as Fig.~\ref{fig:HJParams}, but for multiple values of $E_{\alpha,0}$. We see that the condition $|\Delta \hatP_{\alpha}(E_{\alpha,0})| = 1$ (the solid black line) matches well with the boundary for chaotic tidal behaviour. The critical pericentre distance for chaotic tides (for a given eccentricity $e$), $r_{\rm p, crit}$, increases with $E_{\alpha,0}$ while $e_{\rm crit}$ (for given $r_{\rm p}$) decreases.

In summary, chaotic tides are easier to achieve when planetary modes are already oscillating. The larger the mode energy, the more relaxed the conditions for chaotic tidal behaviour.

\subsection{Conditions for Continued Chaotic Behaviour}\label{sec:DissipativeChaosCondition} 
As mentioned in Section~\ref{sec:ChaosConditions}, we expect the f-mode of a planet experiencing chaotic tides to dissipate a significant amount of energy when the mode amplitudes become non-linear. As a mode drains energy from the orbit, the orbit becomes more tightly bound and a larger amount of energy transfer is required to significantly alter the mode phase at pericentre. Eventually, a system undergoing chaotic tides can dissipate enough energy that the planetary mode enters a quasi-steady state. For dissipative systems, it is useful to know when the orbit is too tightly bound for the mode to continue chaotic behaviour. 

To determine whether the system is chaotic after the $k$-th pericentre passage, we ask whether the current maximum pericentre phase shift, $|\Delta \hatP_{\alpha,k}(E_{\alpha,k-1})|$ is sufficiently large ($\gtrsim 1$). A natural way to think of this condition is to rescale $\Delta \hatP_\alpha(\tildeE_{\alpha,0})$ to the current orbital energy and period, $E_{B,k}$ and $P_k$. Similar to equation~(\ref{eq:DelPhatmax}), we have
\begin{align}
|\Delta \hatP_{\alpha,k}(E_{\alpha,k-1})| &= \sigma_{\alpha,k-1} \text{max} (|P_k-P_{k-1}|) \nonumber \\ &\simeq \frac{3}{2} \sigma_{\alpha,k-1} P_{k-1} \frac{\text{max}(\Delta E_{\alpha,k})}{|E_{B,k-1}|}, \label{eq:DelPkmaxFull}
\end{align}
where 
\begin{equation}
\text{max}(\Delta E_{\alpha,k}) = \Delta E_{\alpha} + 2\sqrt{\Delta E_\alpha E_{\alpha,k-1}}. \label{eq:defmaxEkm1}
\end{equation}
When evaluating equations~(\ref{eq:DelPkmaxFull}) and (\ref{eq:defmaxEkm1}), one must use the ``current" parameters of the system, e.g. the orbital period $P_{k-1}$ and mode frequency $\sigma_{\alpha,k-1}$ just before the $k$-th pericentre passage. Note that the mode frequency may change as the system evolves because it depends on the planet's spin.

Equations (\ref{eq:DelPkmaxFull}) and (\ref{eq:defmaxEkm1}) suggest that a system can escape the influence of chaotic tides in several ways. If a non-tidal effect is dominating the orbital evolution and $r_{\rm p}$ increases or $e$ decreases, $\Delta E_\alpha$ and $\text{max}(\Delta E_{\alpha,k})$ can become too small for chaotic tides to continue operating; if the orbital evolution is dominated by chaotic tides, $r_{\rm p}$ and therefore $\Delta E_\alpha$ are nearly constant, but the orbit can become too tightly bound (i.e. $|E_{B,k-1}|$ is too large and $P_{k-1}$ is too small), for the system to satisfy $|\Delta \hatP_{\alpha,k}(E_{\alpha,k-1})|\gtrsim 1$. Chaotic behaviour can also be suppressed if the mode suddenly dissipates a large amount of energy such that $E_{\alpha,k} \ll E_{\alpha,k-1}$.

\section{Lidov-Kozai migration with chaotic tides}\label{sec:TidesandKozai}
	\begin{figure*}
		\begin{center}				
			\includegraphics[width=7 in]{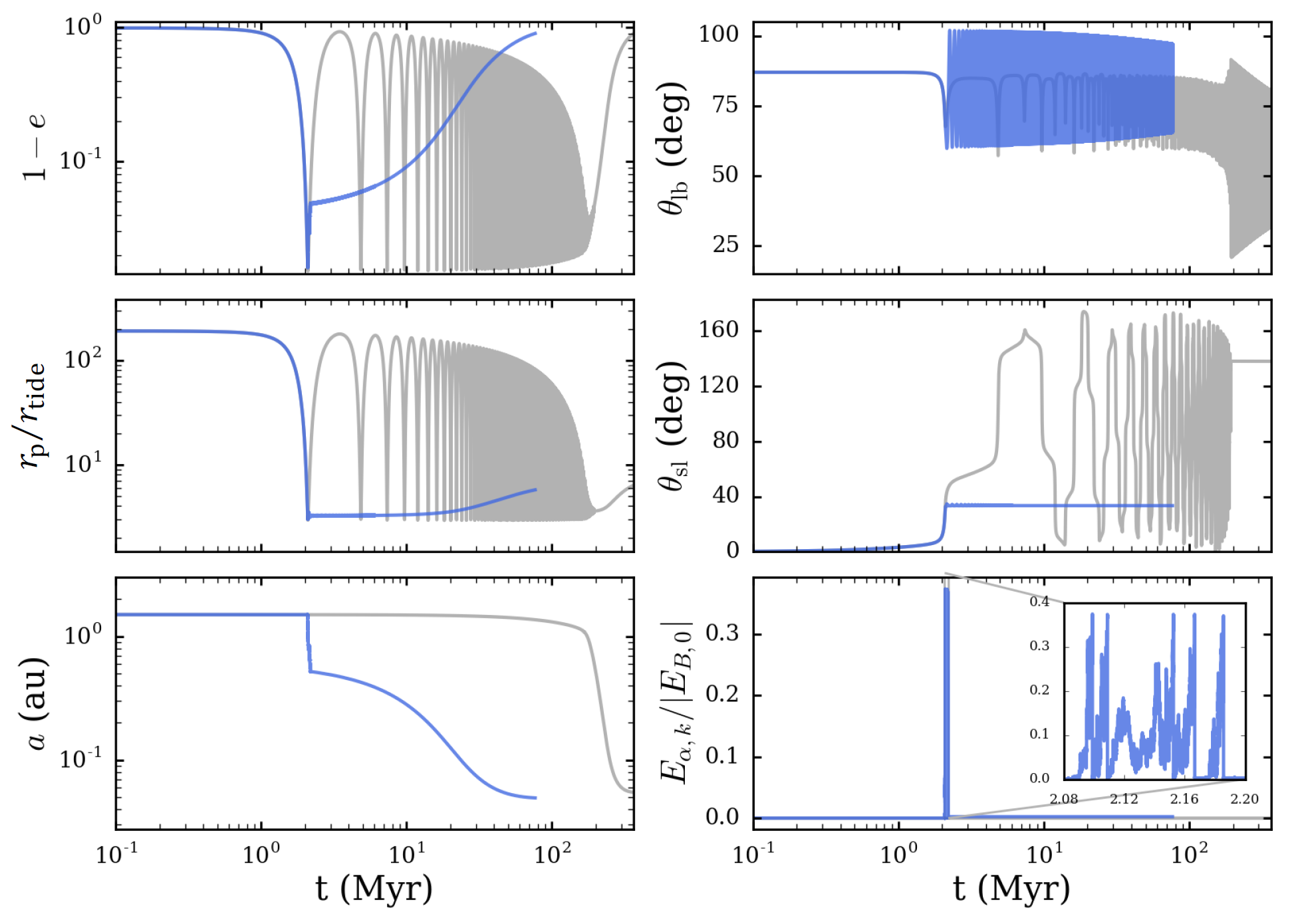}
			\caption{An example of LK migration of giant planets driven by binary companions, with and without chaotic tides. The parameters are $a_{\rm b} = 200$~au, $e_{\rm b}=0$, $M_p = M_J, M_*=M_b=M_\odot$, and $R_p = 1.6~R_J$. The initial semi-major axis and eccentricity for the planet are $a_0 = 1.5$~au and $e_0=0.01$, inclination $\theta_{\rm lb,0} = 87^\circ$, and the initial rotation period of the star is $2.3$ days. The left panels show the orbital eccentricity, pericentre distance (in units of $r_{\rm tide} \equiv R_p(M_{\rm tot}/M_p)^{1/3}$), and the semi-major axis. The right panels show $\theta_{\rm lb}$ (the inclination between the orbits of the planet and the binary companion), $\theta_{\rm sl}$ (the angle between the stellar spin vector and the planet's orbital angular momentum vector), and the mode energy $E_{\alpha,k}$. The grey lines show results without chaotic tides. The blue lines display the results that include chaotic tides for the same system, with an upper limit for the mode energy of $E_{\rm max} = 0.1 GM_p^2/R_p$ and a residual energy after non-linear dissipation of $E_{\rm resid} = 0.01 E_{\rm max}$. In both models, long-term orbital decay is due to weak tidal friction, with $\Delta t_{\rm L}=1$ s. The inset in the bottom right panel shows a zoom-in of the high-eccentricity phase where chaotic tides operate.}
			\label{fig:LongTermEvolution}
		\end{center}
	\end{figure*} 
	
	We now discuss our implementation of dynamical tides in giant planet migration driven by the Lidov-Kozai (LK) mechanism. In this scenario, an external stellar or planetary companion induces quasi-periodic oscillations in the eccentricity of the giant planet's orbit; as the eccentricity attains a large value $(\sim 1)$, tidal dissipation in the planet at pericentre reduces the orbital energy, leading to circularization and decay of the planet's orbit. As noted in Section~\ref{sec:Introduction}, previous works on LK migration adopted the (parametrized) weak friction theory of static tides (ST). The grey curves of Fig.~\ref{fig:LongTermEvolution} give an example of this ``standard" LK migration scenario.
	
	The theory of LK migration with static tides involves two important timescales. The first is the timescale for quadrupole LK eccentricity oscillations, $t_{\rm LK}$, given by 
	\begin{align}
	t_{\rm LK} =& \left(\frac{10^6}{2\pi}\text{yr}\right)\left(\frac{M_b}{M_\odot}\right)^{-1}\left(\frac{M_*}{M_\odot}\right)^{1/2}\left(\frac{a_0}{1~\text{au}}\right)^{-3/2}\nonumber\\
	&\times \left(\frac{a_{\rm b,eff}}{100~\text{au}}\right)^{3} \label{eq:tKozai},
	\end{align}
	where $a_0$ is the initial semi-major axis of the planet's orbit, $M_b$ is the mass of the (stellar) companion, and 
	\begin{equation}
	a_{\rm b,eff} \equiv a_{\rm b} (1-e_{\rm b}^2)^{1/2} \label{eq:defabeff},
	\end{equation}
	with $a_{\rm b}$ and $e_{\rm b}$ the semi-major axis and eccentricity of the companion's orbit. The time that the planet spends near the maximum eccentricity, $e_{\rm max}$, is of order \citep[e.g.][]{Anderson16}
	\begin{equation}
		\Delta t(e_{\rm max}) \sim (1-e_{\rm max}^2)^{1/2} t_{\rm LK}. \label{eq:deltat}
	\end{equation} 
	The second important timescale is that of orbital decay due to static tides given by \citep{Alexander73,Hut81}.
	\begin{align}
	t^{-1}_{\rm ST} =& \left\lvert\frac{\dot{a}}{a}\right\rvert_{\rm ST} = (6 k_{2p} \Delta t_{L}) \frac{M_*}{M_p} \left(\frac{R_p}{a}\right)^{5}\frac{n^{2}}{(1-e^2)^{15/2}} \nonumber\\ & \times \left[f_1(e)\ - \frac{f_2^2(e)}{f_5(e)}\right], \label{eq:deftst}
	\end{align}
	where $k_{2p}$ is the tidal Love number of the planet, $\Delta t_L$ is the lag time, $f_1(e) = 1 + 31e^2/2 + 255e^4/8 + 185e^6/16 + 25e^8/64$, and $f_2(e)$ and $f_5(e)$ are given in equations~(\ref{eq:deff2}) and (\ref{eq:deff5}). (We have assumed that the planet has a pseudosynchronous spin rate.) Tidal dissipation is most efficient near the maximum eccentricity. Since the planet only spends a fraction ($\sim \sqrt{1-e_{\rm max}^2}$) of the time near $e_{\rm max}$, the effective orbital decay rate during LK migration is 
	\begin{align}
	t^{-1}_{\rm ST,LK} =& \left(\left\lvert\frac{\dot{a}}{a}\right\rvert_{\rm ST} \sqrt{1-e^2}\right)_{e_{\rm max}} \nonumber \\
	\approx& \frac{1.27}{{\rm Gyr}} \left(\frac{k_{\rm 2p}}{0.37}\right) \left(\frac{\Delta t_{\rm L}}{1~{\rm s}}\right)\left(\frac{M_p}{M_{\odot}}\right)^2 \left(\frac{M_p}{M_J}\right)^{-1} \nonumber \\
	& \times \; \left(\frac{R_p}{R_J}\right)^5\left(\frac{a_0}{1~{\rm au}}\right)^{-1} \left(\frac{r_{\rm p, min}}{0.025~{\rm au}}\right)^{-7} \label{eq:tst,lk}
	\end{align}
	where $r_{\rm p,min} = a_0(1-e_{\rm max})$ is the minimum pericentre distance (in the second of the above equalities, we have used $e_{\rm max} = 0.96$). Successful migration within a few Gyrs requires $r_{\rm p,min} \lesssim 0.025$ au, corresponding to a final (circularized) planet semi-major axis $a_{\rm F} \lesssim 0.05$ au. A planet would need to be more dissipative than Jupiter (larger $\Delta t_{\rm L}$)\footnote{For Jupiter, $\Delta t_L\simeq0.1$~s, corresponding to $k_{2p}/Q_{p}\simeq 10^{-5}$ at a tidal forcing period of $0.5$~hrs.} to become a HJ with a larger $a_{\rm F}$.  

	As we shall see, chaotic tides can change the standard LK migration scenario in several important aspects (see Fig.~\ref{fig:LongTermEvolution}). Although the planet spends only a small fraction of time in the high-eccentricity phase of the LK oscillations, this time covers enough orbits to allow chaotic mode growth. Consider the example depicted in Fig.~\ref{fig:LongTermEvolution}: the critical eccentricity (for $a_0 = 1.5$~au) for chaotic mode growth to operate is $e_{\rm crit} \simeq 0.985$ (see Fig.~\ref{fig:HJParams} and equation \ref{eq:echaos}). So the time the planet spends around $e \gtrsim e_{\rm crit}$ is $\sim \sqrt{1 - e_{\rm crit}^2} t_{\rm LK} \approx 0.17 t_{\rm LK} \approx 0.23$~Myr, which is more than $1.2 \times 10^5$ orbital periods. The planet's oscillation mode can grow to a large amplitude within this time frame, stealing orbital energy in the process and allowing the orbit to decay within one or a few LK cycles. 
	
\subsection{Description of the Model}\label{sec:FullModel}

	We now describe our method for coupling the evolution of the oscillation mode (the $l=m=2$ f-mode) of a gas giant with the evolution of its orbit driven by an external binary companion (LK oscillations). The planet starts out with eccentricity $e_0 \simeq 0$ and zero mode amplitude ($E_{\alpha,0} = 0$). The other planet and stellar properties as well as the initial values of $e_{\rm b}$, $a_0$, $a_{\rm b}$, $\theta_{\rm lb,0}$ (the mutual inclination of the orbits) and $\Omega_0$ (the longitude of the ascending node of the planet's orbit) are taken as input parameters. Apart from dynamical tides, we evolve the orbital angular momentum vector $\boldsymbol{L}$ and eccentricity vector $\boldsymbol{e}$ of the planet, and the spin angular momentum vector, $\boldsymbol{S_*}$ of the host star in the same way as in \citet{Anderson16} (hereafter ASL16), i.e.
	\begin{align}
	\frac{d \boldsymbol{L}}{dt} &= \left.\frac{d \boldsymbol{L}}{dt}\right\rvert_{\rm LK} +  \left.\frac{d \boldsymbol{L}}{dt}\right\rvert_{\rm SL} +  \left.\frac{d \boldsymbol{L}}{dt}\right\rvert_{\rm ST}\label{eq:dLdt},\\
	\frac{d \boldsymbol{e}}{dt} &= \left.\frac{d \boldsymbol{e}}{dt}\right\rvert_{\rm LK} + \left.\frac{d \boldsymbol{e}}{dt}\right\rvert_{\rm SL} + \left.\frac{d \boldsymbol{e}}{dt}\right\rvert_{\rm SRF} + \left.\frac{d \boldsymbol{e}}{dt}\right\rvert_{\rm ST}\label{eq:dedt},\\
	\frac{d \boldsymbol{S_*}}{dt} &= \left.\frac{d \boldsymbol{S_*}}{dt}\right\rvert_{\rm SL} +  \left.\frac{d \boldsymbol{S_*}}{dt}\right\rvert_{\rm MB}.\label{eq:dSdt}
	\end{align}
	Here the LK terms are contributions (to the octupole order) from the binary companion that give rise to LK oscillations [equations A1-A2 of ASL16; from \citet{Liu15}]; the SL terms arise from the spin-orbit coupling between the host star spin, $\boldsymbol{S_*}$, and the planet's orbital angular momentum (equations 60-61 of ASL16); and the short-range force (SRF) terms account for periastron precession of the planet's orbit due to general relativity (GR) and tidal/rotational distortions of the planet (sections A3 and A4 of ASL16). The stellar spin evolution includes the spin-down torque due to magnetic braking (MS), i.e.
	\begin{equation}
	\left.\frac{d\boldsymbol{S_*}}{dt}\right\rvert_{\rm MB} = - \alpha_{\rm MB}I_*\Omega_*^3 \hat{\boldsymbol{S_*}},\label{eq:Skumanich}
	\end{equation}
	where $\alpha_{\rm MB}$ is taken to be $1.5 \times 10^{-14}$ yr to model a solar type star, $I_*$ is the moment of inertia of the star, and $\Omega_*$ is its rotation rate. We assume an initial spin period of 2.3 days.
	We also evolve the angular momentum and eccentricity vectors of the outer orbit ($\boldsymbol{L}_b$ and $\boldsymbol{e}_b$) according to the octupole LK effect \citep[equations 19 and 20 of][]{Liu15}; these changes in the outer orbit are small because $M_p \ll M_b$. Throughout evolution, we take the planet's rotation rate to be the pseudosynchronous rotation rate given by equation~(\ref{eq:Pseudosynchronous}). Equations (\ref{eq:dLdt})-(\ref{eq:dedt}) include contributions from the dissipation of static tides (ST); these are parameterized by the tidal lag time $\Delta t_{\rm L}$, and are distinct from the effects of dynamical tides. 
	
	In the presence of dynamical tides, we must supplement equations~(\ref{eq:dLdt})-(\ref{eq:dSdt}) with the evolution of the mode amplitude. This is based on the iterative map discussed in Section~\ref{sec:Mapping}. Here, we provide a summary of the implementation of this procedure. At a certain time in the orbital evolution, just before the $k$-th pericentre passage, the orbital parameters are $a_{k-1}$ and $e_{k-1}$, the planet spin rate is $\Omega_{s,k-1}$, and the mode amplitude is $c_{\alpha, k-1}$. To advance to the next orbit, we carry out the following steps: 
	
	(i) Calculate $\Delta E_\alpha$ from equations~(\ref{eq:DelE}), (\ref{eq:defT}), and (\ref{eq:Kapprox}) using mode parameters from Table \ref{tab:ModeProperties} and the current orbital/planet parameters (i.e. $a_{k-1}$, $e_{k-1}$, $\Omega_{s,k-1}$, etc.) and obtain $\Delta c_\alpha \propto \sqrt{ \Delta E_\alpha}$. In practice, we normalize mode energy by $|E_{B,0}|$ (the initial orbital energy), and thus $\Delta \tilde{c}_\alpha = \sqrt{\Delta \tildeE_\alpha} = \sqrt{\Delta E_\alpha/|E_{B,0}|}$. 
	
	(ii) Calculate the energy transfer in the $k$-th passage using $\Delta E_{\alpha,k} = |E_{B,0}|(|\tilde{c}_{\alpha,k-1} + \Delta \tilde{c}_\alpha|^2 - |\tilde{c}_{\alpha,k-1}|)$ (see equation \ref{eq:DeltaE}) and obtain the new orbital energy $E_{B,k} = E_{B,k-1} - \Delta E_{\alpha,k}$. 
	
	(iii) Obtain the new orbital semi-major axis and eccentricity (after the $k$-th pericentre passage) according to 
			\begin{align}
			a_k &= \frac{E_{B,k-1}}{E_{B,k}}a_{k-1}\\
			e_k &= \left[1-\frac{E_{B,k}}{E_{B,k-1}}(1-e_{k-1}^2)\right]^{1/2} \label{eq:orbEvolution},
			\end{align}
	where we have assumed that the orbital angular momentum is conserved during the passage.
	
	(iv) Evolve the (complex) mode amplitude over one orbit to obtain its value just before the $(k+1)$-th passage using $c_{\alpha, k} = (c_{\alpha,k-1} + \Delta c_\alpha)\text{e}^{-\text{i}\sigma_{\alpha,k}P_k}$ (see equation \ref{eq:Map}), where $P_k$ is the orbital period corresponding to $a_k$ and $\sigma_{\alpha,k}$ is the mode frequency after the $k$-th passage (assuming a pseudosynchronous rotation rate).
	
	\begin{figure}
		\begin{center}
			\includegraphics[width=\columnwidth]{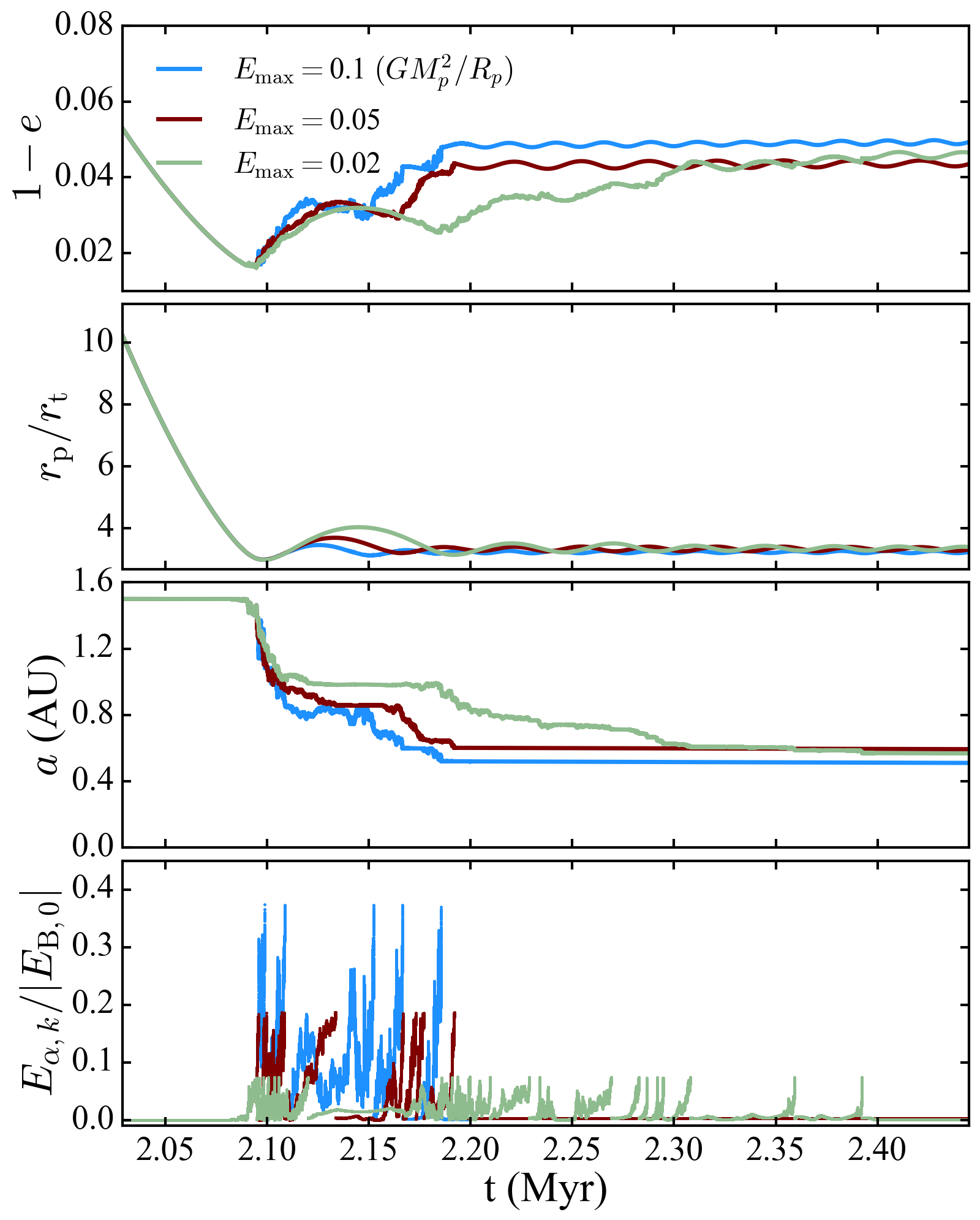}
			\caption{A zoom-in of the chaotic tidal evolution of the system depicted in Fig.~\ref{fig:LongTermEvolution} showing the effect of different choices of $E_{\rm max}$ (the maximum mode energy at which non-linear dissipation occurs). In all three cases, the residual mode energy after non-linear dissipation is set to $E_{\rm resid} = 0.001 GM_p^2/R_p.$ In general, the value of $E_{\rm max}$ does not have a large effect on orbital parameters of the planet that result from chaotic tidal evolution.}
			\label{fig:EmaxComparison}
		\end{center}
	\end{figure}
	\begin{figure}
		\begin{center}
			\includegraphics[width=\columnwidth]{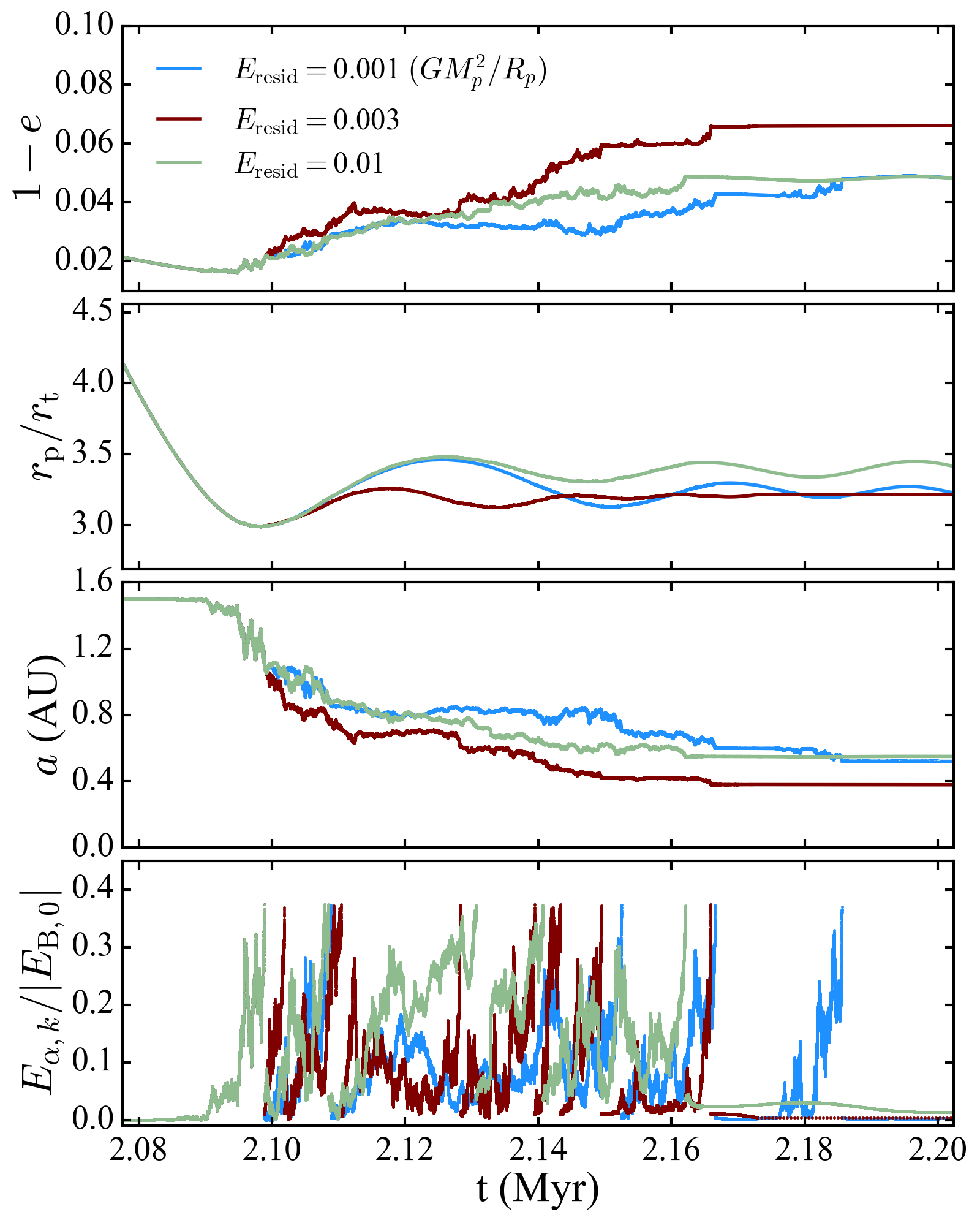}
			\caption{The same as Fig.~\ref{fig:EmaxComparison} but showing the effect of different choices of $E_{\rm resid}$. In all cases, $E_{\rm max}$ is set to $0.1 GM_p^2/R_p$.}
			\label{fig:EresComparison}
		\end{center}
	\end{figure}
	
	Left unchecked, the mode energy in this model can grow to un-physically large values (as in the bottom panel of Fig.~\ref{fig:3Behaviors}). We therefore assume that, when the mode energy reaches a significant fraction of the planet's binding energy, non-linear effects dissipate nearly all of the mode energy within an orbital period. We parametrize the maximum energy that the mode can reach before non-linear effects become important, $E_{\rm max}$, and the residual energy in the mode after an episode of nonlinear mode dissipation and tidal heating, $E_{\rm resid}$. Thus, when $|\tilde{c}_{\alpha,k}|^2 > E_{\rm max}/|E_{B,0}|$, the mode amplitude is immediately changed to $|\tilde{c}_{\alpha,k}|^2 = E_{\rm resid}/|E_{B,0}|$. This method for handling dissipation was used in \citet{Wu18}. The parameter $E_{\rm max}$ does not change the qualitative features of evolution due to chaotic tides (see Fig.~\ref{fig:EmaxComparison}), but can alter the details. The quantity $E_{\rm resid}$ is more important because chaotic behaviour is easier to excite when the f-mode already has some energy, as discussed in Section~\ref{sec:extendChaosCondition}, but still does not change the overall behaviour of the model (see Fig. \ref{fig:EresComparison}). Unless stated otherwise, we use $E_{\rm max}=0.1 (GM_p^2/R_p)$ and $E_{\rm resid}=0.01 E_{\rm max}=10^{-3} (GM_p^2/R_p)$ in the results presented in this paper. 
	
	Since dynamical tides must be implemented on an orbit by orbit basis, it is not practical to evolve the whole system over many Myrs. Because dynamical tides are effective only during the high-e phase of the LK cycle, when chaotic mode growth occurs, we only need to evolve the modes when the eccentricity is sufficiently high and the pericentre distance is sufficiently small. As discussed in Section~\ref{sec:DissipativeChaosCondition}, chaotic tides operate when $|\Delta \hat{P}_{\alpha,k}(E_{\alpha,k-1})|\gtrsim \Delta \hat{P}_{\rm crit}$ (see equation \ref{eq:DelPkmaxFull}). In our implementation, we turn on dynamical tides only when $|\Delta \hat{P}_{\alpha,k}(E_{\alpha,k-1})|\gtrsim 0.01$. As we see from Fig.~\ref{fig:PhaseSpaceEinit}, this is a conservative choice, allowing us to capture the onset of chaotic behaviour for all reasonable parameters regardless of resonances and the value of ``pre-existing" mode energy. 

	After the orbit has decayed through chaotic tides, the system will eventually satisfy $|\Delta \hat{P}_{\alpha,k}(E_{\alpha,k-1})|<\Delta \hat{P}_{\rm crit}\sim 1$ (see the right panel of Fig. \ref{fig:PhaseSpaceEinit}) while the orbital eccentricity is still large. This typically occurs just after the mode energy is dissipated non-linearly from $E_{\rm max}$ to $E_{\rm resid}$. Thus, when the system satisfies
	\begin{equation}
	|\Delta \hat{P}_{\alpha,k}(E_{\alpha,k-1})| \simeq \frac{3 \sqrt{\Delta E_\alpha E_{\alpha,k-1}}}{|E_{B,k}|}\sigma_{\alpha,k}P_k \lesssim 1, \label{eq:stop}
	\end{equation} 
	(where we have assumed $E_{\alpha,k-1} \sim E_{\rm resid}\gg \Delta E_{\alpha}$) the f-mode will no longer behave chaotically and influence the orbital evolution. Instead, static tides drive the gradual decay and circularization of the planet's orbit. To save computation time, when equation~(\ref{eq:stop}) is satisfied while the orbital eccentricity is large, we take note. If equation~(\ref{eq:stop}) is still satisfied in 30,000 orbits, we stop evolving the dynamical tides; if equation~(\ref{eq:stop}) is not satisfied after 30,000 orbits, we check again in another 30,000 orbits. The multiple checks are to account for the fact that $\Delta \hatP_{\rm crit}$ varies significantly depending on the orbital period and can be less than 1.
		
	\begin{figure*}
		\begin{center}
			\includegraphics[width=7 in]{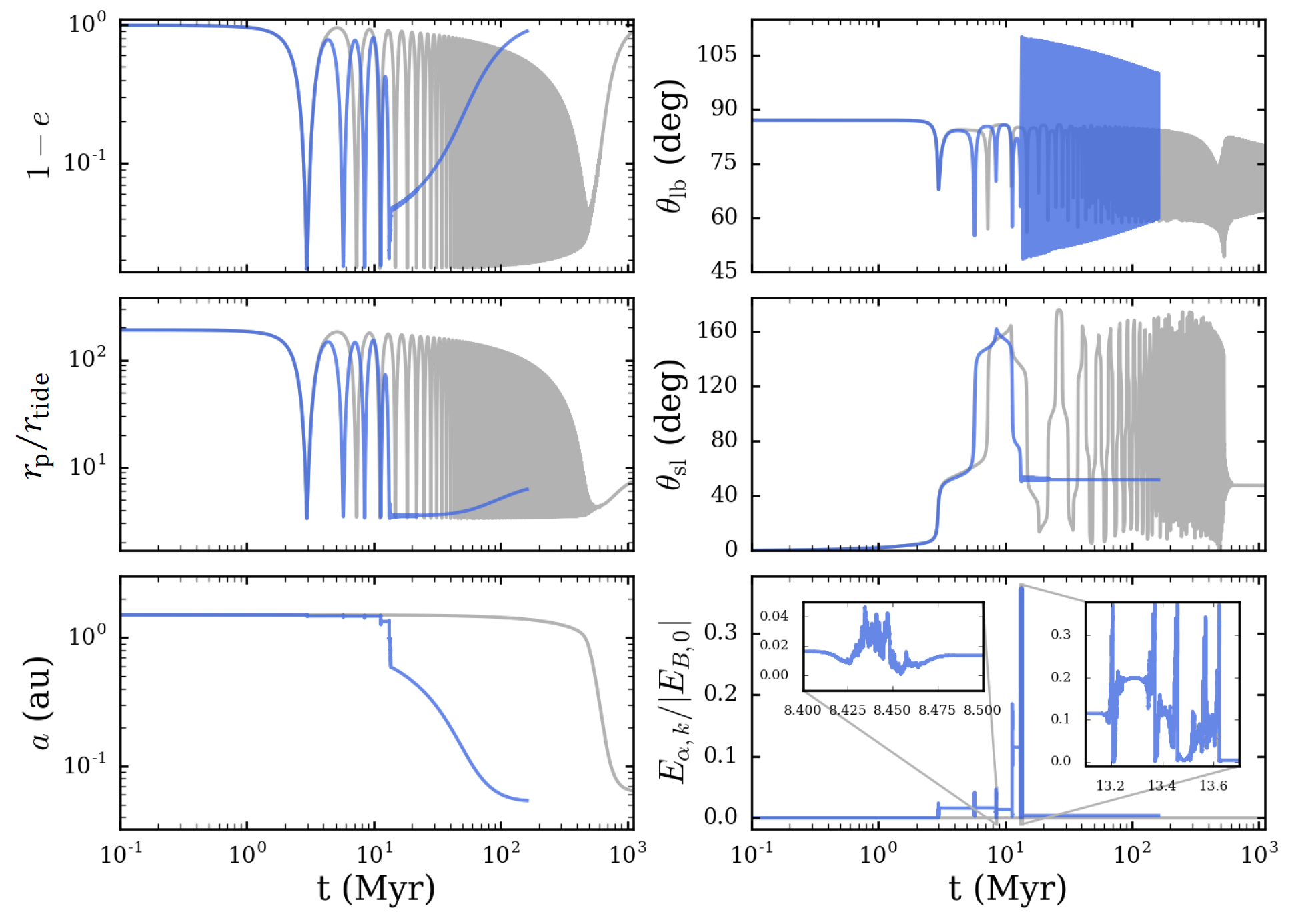}
			\caption{Same as Fig.~\ref{fig:LongTermEvolution}, except for $a_{\rm b} = 225$~au. The insets in the bottom right panel show a closer look at two high-eccentricity phases. In the first inset, the mode energy does not dissipate. In the second inset, the planet undergoes multiple episodes of tidal dissipation.}
			\label{fig:LongTermEvolutionLargerEta}
		\end{center}
	\end{figure*}
	
\subsection{Examples of the Model}\label{sec:Results}
	LK migration with chaotic tides proceeds in three steps (see the blue curves in Fig.~\ref{fig:LongTermEvolution}). First, the external companion drives the planet's orbital eccentricity beyond $e_{\rm crit}$ (see equation \ref{eq:echaos}), where chaotic tides set in and begin to take over orbital evolution, freezing $r_{\rm p}$ due to angular momentum conservation. The blue lines in Figs.~\ref{fig:EmaxComparison} and \ref{fig:EresComparison} show a zoom-in view. We see that $r_{\rm p}$ decreases due to the LK effect until settling to an approximately constant value around 2.1 Myr while $e$ and $a$ start to wander chaotically. In the second stage, chaotic tides dominate orbital evolution and quickly shrink the orbit until it is too bound for such tides to continue operating. In Figs.~\ref{fig:EmaxComparison} and \ref{fig:EresComparison}, it is clear that the planet undergoes multiple episodes of non-linear dissipation (where the mode energy rapidly dissipates to a small residual value of $E_{\rm resid}$ after climbing to maximum energy $E_{\rm max}$) while subject to chaotic tides. The result is a very eccentric WJ. At this new semi-major axis, the planet is decoupled from the influence of the stellar companion, and no longer experiences significant eccentricity oscillations. Finally, weak tidal friction circularizes and shrinks the orbit to form a HJ on a Gyr timescale (depending on the dissipation rate of static tides in the planet).
	
	In some cases, a system will undergo multiple high-eccentricity phases before chaotic tides reach the threshold $E_{\rm max}$ and the f-mode dissipates energy. An example is shown in the blue lines of Fig.~\ref{fig:LongTermEvolutionLargerEta}, where the system has the same parameters and initial conditions as in Fig.~\ref{fig:LongTermEvolution} but with a slightly larger companion semi-major axis ($a_{\rm b} = 225$ au). We see that the planet's eccentricity undergoes five LK oscillations before the f-mode energy climbs to $E_{\rm max}$. Before an episode of tidal dissipation, any energy transferred from the orbit to the mode can easily pass back to the orbit. For example, in the third high-eccentricity phase of Fig.~\ref{fig:LongTermEvolutionLargerEta} (shown in an inset of the bottom right panel) the f-mode loses a small amount of energy and the semi-major axis increases slightly. Once the mode energy reaches $E_{\rm max}$ and is dissipated, the orbit shrinks irreversibly and continues to rapidly decay until the orbit is too bound for chaotic tides to operate. 
	
	We can gain some insight into when a system requires multiple high-eccentricity phases to reach $E_{\rm max}$ by studying the rate of chaotic orbital decay predicted by the iterative map from Section~\ref{sec:Mapping}. While $|\Delta \hat{P}_\alpha| > 1$ (for $E_{\alpha,0} = 0$), the mode amplitude grows in an approximately diffusive manner [Fig.~5 of \citet{Vick18}; see also \citet{Mardling95a,IP04,Wu18}], with the averaged mode energy
	\begin{equation}
	<E_{\alpha,k}> \simeq \Delta E_\alpha k \label{eq:stochastic}
	\end{equation}
	For a sufficiently eccentric orbit, $\Delta E_\alpha$  is roughly constant over many pericentre passages. We can define a timescale for orbital decay from the energy transfer and dissipation,
	\begin{equation}
	t_{\rm decay} \equiv \frac{P_0|E_{\rm B,0}|}{\Delta E_\alpha}. \label{eq:tchaos}
	\end{equation}
	If we use the largest value of $e$ attained by the system in Fig.~\ref{fig:LongTermEvolutionLargerEta}, we find that $t_{\rm decay} = 0.31$ Myr and the timescale for the system to remain at high eccentricity is $\Delta t(e=0.983) = 0.13$ Myr (see equation \ref{eq:deltat}). Because $t_{\rm decay} < \Delta t(e=0.983)$, the system must undergo multiple LK cycles for the mode energy to climb to $E_{\rm max}$.
	In contrast, the system in Fig.~\ref{fig:LongTermEvolution} reaches $e=0.989$, for which $t_{\rm decay} = 3.9 \times 10^{-2}$ Myr and $\Delta t(e=0.989)=0.15$~Myr. For this system, where the timescale for chaotic tides to alter the orbit is very short, the planet's orbit decays within one high-eccentricity phase.
	
\section{Key Features of Chaotic Tidal Migration: Analytical Understanding}\label{sec:Features}

	LK migration with chaotic tides has a few characteristic features. First, it can prevent some gas giants from undergoing tidal disruption. Second, it naturally produces very eccentric WJs. Lastly, this population of WJs circularizes relatively quickly, due to weak tidal friction because the planets are ``detached" from the companions' perturbations. In the following, we discuss the necessary conditions for LK chaotic tidal migration, hallmarks of this process, and predictions for the stellar obliquities of HJs formed via this mechanism.
	
	\subsection{Condition for Chaotic Tidal Migration}
	As discussed in Section~\ref{sec:ChaosConditions}, a planet must have $e\gtrsim e_{\rm crit}$ to initiate chaotic tides, where $e_{\rm crit}$ as a function of $a_0$ (the initial semi-major axis of the planet) is implicitly given by equation~(\ref{eq:echaos}). To reach such a high eccentricity, a system must have sufficiently large initial mutual inclination, $\theta_{\rm lb,0}$. In the idealized case of quadrupole LK oscillations, the maximum eccentricity is 
	$e_{\rm max} = \left[1 - (5/3) \cos^2 \theta_{\rm lb,0}\right]^{1/2}.$
	Accordingly, $e_{\rm max} \sim 1$ can be achieved only for $\theta_{\rm lb,0}\approx 90^\circ$. Including octupole terms complicates the relationship between the initial mutual inclination and the maximum attainable eccentricity and broadens the inclination window for extreme eccentricity excitation. A realistic treatment of the LK effect must also account for the short-range forces (SRFs) that cause the planet's pericentre to precess. These limit the maximum attainable eccentricity in LK cycles to a value $e_{\rm lim}$, where $e_{\rm lim}$ is the maximum eccentricity achieved when $\theta_{\rm lb,0} = 90^\circ$ in the quadrupole limit. In the presence of the octupole potential, the eccentricity still cannot exceed $e_{\rm lim}$, but this eccentricity can be realized for a wider range of initial inclinations \citep{Liu15}. The relevant SRF's for this problem are GR and the effect of static tides raised on the planet. When $(1-e_{\rm lim}) \ll 1$, the limiting eccentricity is given by
		\begin{equation}
		\frac{\epsilon_{\rm GR}}{(1-e_{\rm lim}^2)^{1/2}}+\frac{7}{24}\frac{\epsilon_{\rm Tide}}{(1-e_{\rm lim}^2)^{9/2}} \simeq \frac{9}{8}, \label{eq:elim}
		\end{equation}
		where $\epsilon_{\rm GR}$ and $\epsilon_{\rm Tide}$ measure the strengths of precession due to GR and the planet's tidal bulge relative to the quadrupole LK effect, respectively:
		\begin{align}
		\epsilon_{\rm GR} &\equiv \frac{3 G M_*^2 a_{\rm b,eff}^3}{a_0^4c^2M_b},\\
		\epsilon_{\rm Tide} &\equiv \frac{15 M_*^2a_{\rm b,eff}^3 k_{2p}R_p^5}{a_0^8M_pM_b}.
		\end{align}
		For our ``standard" system (shown in Fig.~\ref{fig:LongTermEvolution}) with $a_0=1.5$ au, $a_{\rm b}=200$ au, and $e_{\rm b} = 0$, we find $e_{\rm lim} \approx 0.998$ and $e_{\rm crit} \approx 0.985$, so the planet can reach large enough eccentricities to experience chaotic tides, provided the initial mutual inclination angle, $\theta_{\rm lb,0}$, is sufficiently large.
		
\begin{figure}
			\begin{center}
				\includegraphics[width=\columnwidth]{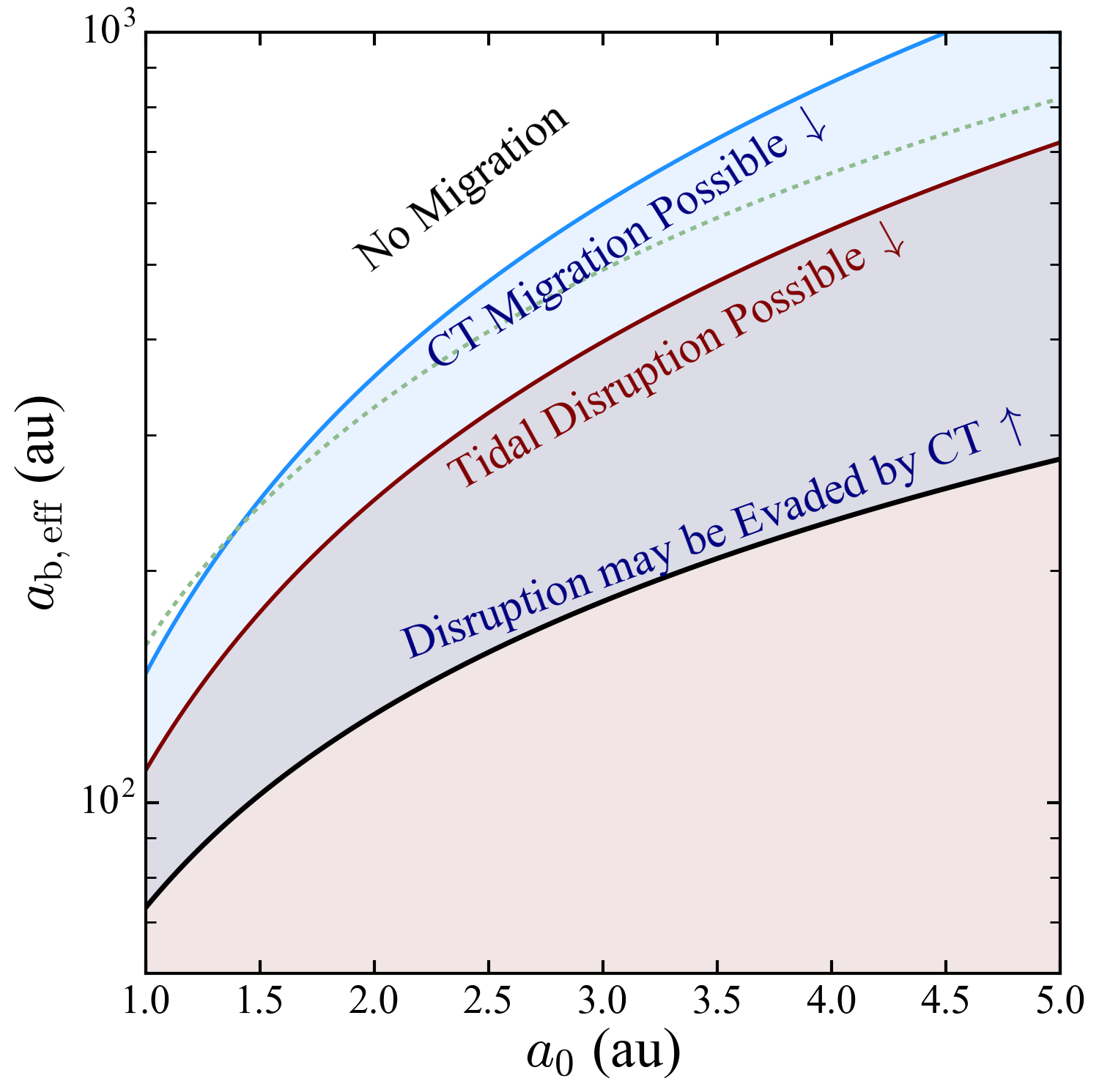}
				\caption{The effective outer companion semi-major axis $a_{\rm b,eff}$ (equation \ref{eq:defabeff}) vs. the initial semi-major axis of the planet, for $M_p = 1 M_J$, $R_p = 1.6 R_J$ and $M_* = M_\odot$. LK migration via chaotic tides (CT) is possible below the blue line, equation~(\ref{eq:abeffMax}); tidal disruption becomes possible below the red line, equation~(\ref{eq:abeffMax}) with $r_{\rm p,dis}$ in place of $r_{\rm p,crit}$; tidal disruption may be evaded by chaotic tides above the black line, equation~(\ref{eq:abeffMin}). The dashed line corresponds to the maximum $a_{\rm b,eff}$ for standard LK migration (with static tides) to operate within $10^{9}$ yrs, assuming $\Delta t_{\rm L} = 1$~s, see equation~(\ref{eq:rpmig}).} \label{fig:SavedSystems}
			\end{center}
		\end{figure}

		In general, the necessary condition for chaotic tidal migration is
		\begin{equation}
		e_{\rm lim} \ge e_{\rm crit},
		\end{equation} 
		or equivalently $r_{\rm p, lim} \equiv a_0(1-e_{\rm lim}) \le r_{\rm p,crit}$, where $e_{\rm crit}$ and $r_{\rm p,crit}$ are given by equation~(\ref{eq:echaos}) (or more approximately by equation \ref{eq:BoundaryApprox}). Because $e_{\rm lim}$, given by equation~(\ref{eq:elim}), depends on $a_{\rm b, eff}$, the condition on $e_{\rm lim}$ translates to an upper limit on $a_{\rm b,eff}$ for chaotic migration to occur:
		\begin{align}
		a_{\rm b,eff} &\le \frac{3^{1/3}}{2^{5/6}} \frac{a_0^{7/6}}{R_p^{1/6}}\left(\frac{M_bM_p}{M_*^2}\right)^{1/3}\left[\frac{GM_p}{R_p c^2}\left(\frac{R_p}{r_{\rm p,crit}}\right)^{1/2}\right. \nonumber\\&+ \left. \frac{35}{384}\frac{k_{2p}}{2^{9/2}}\left(\frac{R_p}{r_{\rm p,crit}}\right)^{9/2} \right]^{-1/3},\label{eq:abeffMax}
		\end{align}
		The upper limit in equation~(\ref{eq:abeffMax}) is shown as a blue line in Fig.~\ref{fig:SavedSystems}.
		
		We can also derive the maximum $r_{\rm p}$ that allows for standard LK migration with static tides within a stellar lifetime. Using equation~(\ref{eq:tst,lk}) and requiring that the planet migrate within $t_{\rm mig}$ (i.e. $t_{\rm ST,LK} \lesssim t_{\rm mig}$), we find
		\begin{align}
		r_{\rm p} \lesssim r_{\rm p, ST}  \equiv&  (0.021~\text{au}) 
		\left(\frac{t_{\rm mig}}{1~\text{Gyr}}\right)^{1/7}
		\left(\frac{R_p}{R_J}\right)^{5/7} \left(\frac{M_p}{M_J}\right)^{-1/7} \nonumber\\  
		\times& \left(\frac{a_0}{1~\text{au}}\right)^{-1/7}
		\left(\frac{k_{2p}}{0.37}\right)^{1/7}\left(\frac{M_*}{1 M_\odot}\right)^{2/7}\left(\frac{\Delta t_{L}}{1~\text{s}}\right)^{1/7},\label{eq:rpmig}
		\end{align}
		[$r_{\rm p,ST}$ is called $r_{\rm p, mig}$ in \citet{Munoz16}]. Substituting $r_{\rm p,ST}$ for $r_{\rm p, crit}$ in equation~(\ref{eq:abeffMax}) yields an approximate maximum $a_{\rm b,eff}$ for standard LK migration (without chaotic tides). The result is shown as the dashed line in Fig.~\ref{fig:SavedSystems}. For a $1 M_J$, $1.6R_J$ gas giant, the conditions for chaotic tidal migration are typically more generous than for standard LK migration. This suggests that LK migration with chaotic tides yields more HJs than LK migration with static tides.
		
		\subsection{Evading Tidal Disruption} \label{sec:Saved}
		The planet can be tidally disrupted if its pericentre distance, $r_{\rm p}$, is less than the tidal disruption limit from equation~(\ref{eq:dis}),
		\begin{equation}
		r_{\rm p,dis} = (0.013~\text{au}) \left(\frac{R_p}{R_J}\right)\left(\frac{M_*}{10^3 M_p}\right)^{1/3}. \label{eq:defrpdis}
		\end{equation}
		Therefore, when $r_{\rm p,lim} < r_{\rm p,dis}$, the companion can induce tidal disruption of the planet if $\theta_{\rm lb,0}$ is sufficiently large. Substituting $r_{\rm p, dis}$ for $r_{\rm p, crit}$ in equation~(\ref{eq:abeffMax}) yields the maximum $a_{\rm b,eff}$ for tidal disruption to be possible. This is shown as the red line in Fig.~\ref{fig:SavedSystems}.
		
		In the standard LK migration scenario (with static tides), a large fraction of migrated giant planets are tidally disrupted because $r_{\rm p, dis}$ can be quite close to $r_{\rm p,ST}$; moreover, Saturn-mass planets can rarely migrate successfully via LK oscillations and static tides because $r_{\rm p, dis} \gtrsim r_{\rm p,ST}$ \citep[ASL16,][]{Munoz16}. In contrast, with chaotic tides, there is always a region of parameter space where chaotic tidal migration is possible without disruption ($r_{\rm p,crit} > r_{\rm p, dis}$; see Fig.~\ref{fig:HJParams}) for any reasonable values of $M_p$ and $R_p$ for giant planets.
		
\begin{figure}
				\begin{center}
					\includegraphics[width = \columnwidth]{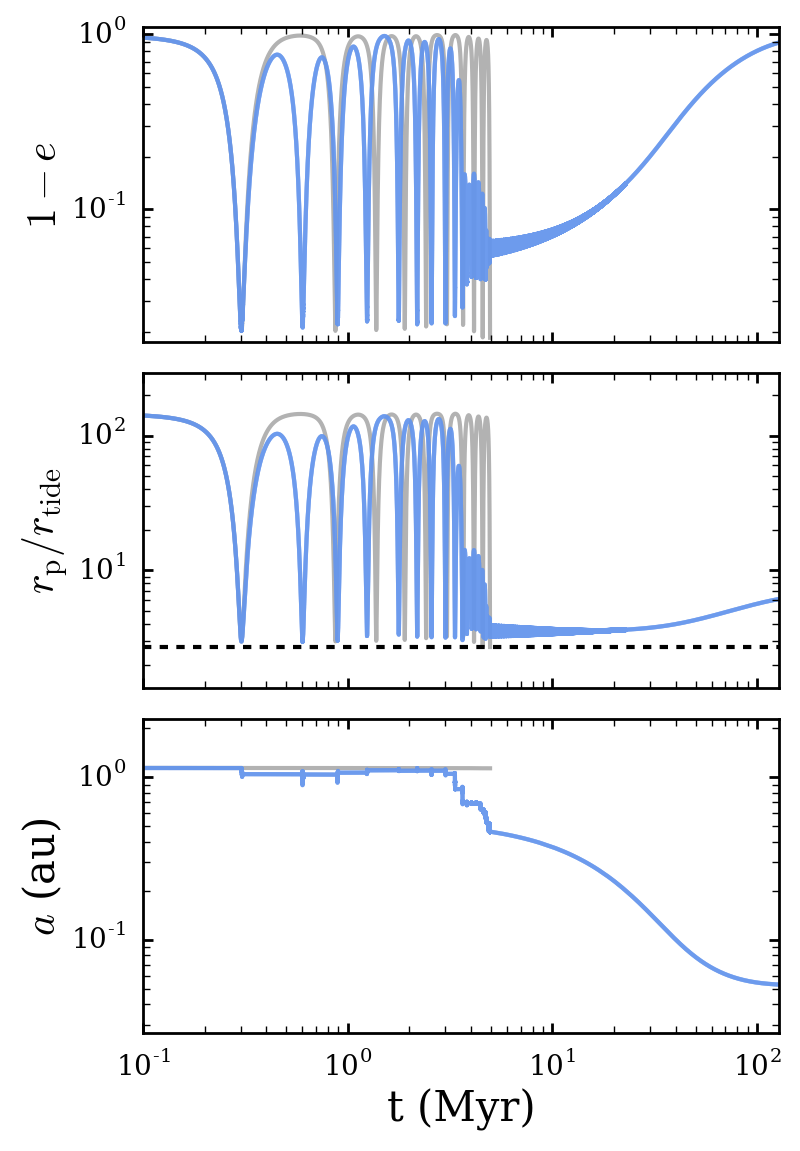}
					\caption{Evolution of a system that is spared disruption by chaotic tides. The blue lines includes chaotic tides, and the grey lines do not. The initial conditions are $a_0 = 1.14$~au, $a_{\rm b} = 102$~au, $\theta_{\rm lb,0} = 82^\circ$, $e_{\rm b} = 0.47$, and $\Omega_0 = 179^\circ$ (where $\Omega_0$ is the longitude of the ascending node of the planet). The physical properties of the planet and stars are $M_p = M_J, M_*=M_b=M_\odot$, and $R_p = 1.6~R_J$. The black dashed line is $r_{\rm p, dis}$ from equation~(\ref{eq:defrpdis}).} \label{fig:SavedSystemExample}
				\end{center}
			\end{figure}

		In some cases, chaotic tides can shepherd to safety the planets that are otherwise destined for tidal disruption by acting to rapidly decrease $e$ and increase $r_{\rm p}$. This possibility was suggested in \citet{Wu18}. Figure~\ref{fig:SavedSystemExample} shows one such example. Before the planet is pushed to extreme eccentricity and tidal disruption by the octupole potential, it is frozen into a high-eccentricity orbit with a small semi-major axis and becomes decoupled from the companion. Without further interference from the LK effect, the eccentric WJ circularizes in orbit over a Gyr timescale to eventually become a HJ.
		
		We can understand the condition for chaotic tides to save a planet from disruption using a simple time-scale argument. Tidal disruption occurs when $r_{\rm p} \le r_{\rm p,dis}$. At high eccentricity, the timescale for a planet to remain below a given $r_{\rm p}$ is $ \Delta t(<r_{\rm p})\sim t_{\rm LK}\sqrt{1-e^2} \sim \sqrt{2}t_{\rm LK}(r_{\rm p}/a_0)^{1/2}$ (see equation \ref{eq:deltat}), while the timescale for chaotic tides to decay the orbit is $t_{\rm decay}$, given by equation~(\ref{eq:tchaos}). The planet reaches a minimum $r_{\rm p}$ when these timescales are roughly equal. For planets that are ``just saved" from disruption, this minimum is nearly $r_{\rm p,dis}$. A system that can be saved from disruption must satisfy $t_{\rm decay} \lesssim \Delta t(<r_{\rm p,dis})$, i.e.
		\begin{equation}
		\frac{P_0 |E_{\rm B,0}|}{\Delta E_\alpha} 
		\lesssim \sqrt{2} t_{\rm LK} \left(\frac{r_{\rm p,dis}}{a_0}\right)^{1/2}, \label{eq:saveCondition}
		\end{equation}
		where $\Delta E_{\alpha}$ is evaluated at $r_{\rm p} = r_{\rm p, dis}$ (see equation \ref{eq:DelE}). Recall that $\Delta E_\alpha$ depends on $T(\eta, \sigma_\alpha/\Omega_p, e)$. When $(1- e) \ll 1$, $T$ and the pseudosynchronous spin rate, $\Omega_{ps}$, have negligible dependence on $e$. We can then evaluate $T(\eta)$ by using equation~(\ref{eq:defT}) with the approximate $K_{22}$ from equation~(\ref{eq:Kapprox}) and taking $\bar{\sigma}_\alpha \simeq \bar{\omega}_\alpha + 1.17/\eta^{3/2}$ for the mode frequency in the inertial frame. At $r_{\rm p,dis}$, $T(\eta=2.7) = 2.5 \times 10^{-2}$.

		By rearranging  equation~(\ref{eq:saveCondition}), we can find the minimum $a_{\rm b,eff}$ needed for chaotic tidal migration: 
		\begin{equation}
		a_{\rm b,eff}\gtrsim 27.7\;a_0^{5/6}r_{\rm tide}^{1/6} \left(\frac{M_*}{M_p}\right)^{2/9}\left(\frac{M_b}{M_*}\right)^{1/3}, \label{eq:abeffMin}
		\end{equation}
		with $r_{\rm p,dis} = 2.7 r_{\rm tide}$. 
		The limit on $a_{\rm b,eff}$ is shown as a black line in Fig.~\ref{fig:SavedSystems}.
		
	\subsection{Eccentric Warm Jupiter Formation}\label{sec:eccWJ}
	
	\begin{figure}
		\begin{center}
			\includegraphics[width=\columnwidth]{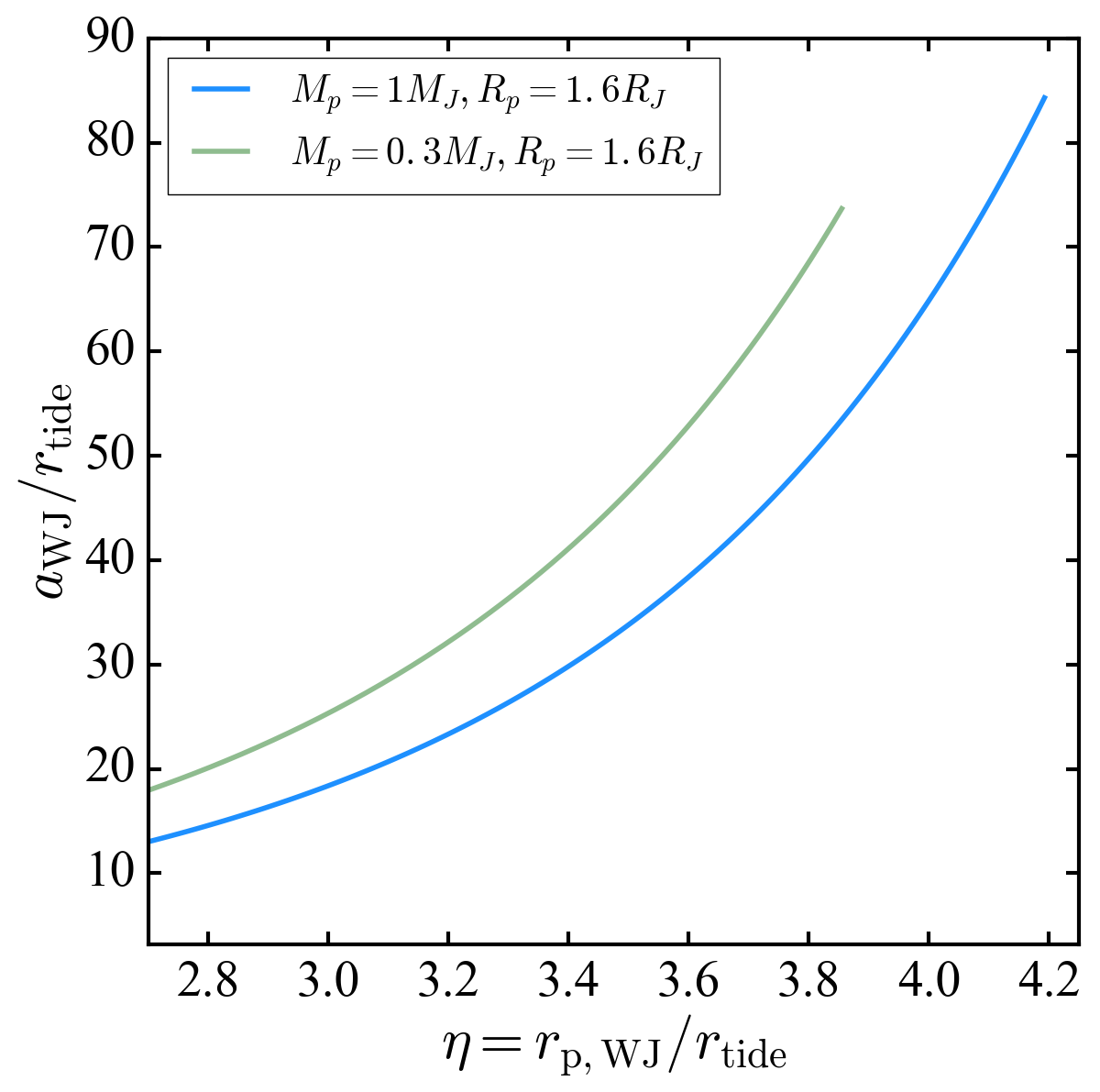}
			\caption{The predicted semi-major axis (see equation \ref{eq:aWJ}) of WJs that have undergone chaotic tidal migration as a function of the pericentre distance $r_{\rm p, WJ}$ for two different planet models. The minimum value of $\eta$ is set by the criterion for tidal disruption (equation \ref{eq:defrpdis}). The maximum value of $\eta$ is set by the chaotic tides boundary [equation~(\ref{eq:etaBoundary}) with $a_0 = 5$ au].}
			\label{fig:aWJvseta}
		\end{center}
	\end{figure}

	A giant planet that undergoes chaotic orbital decay becomes an eccentric WJ after reaching a semi-major axis where both chaotic tides and the LK effect have little influence on the continued evolution of the planet's orbit. For example, in Figs.~\ref{fig:LongTermEvolution} - \ref{fig:EresComparison}, all of the integrations produce a planet with a semi-major axis between $0.35$ and $0.65$ au and an eccentricity between 0.93 and 0.97. We can understand why this occurs by re-examining the criterion for chaotic tides to stop affecting orbital evolution (see equation \ref{eq:stop}). Typically, chaotic tides end just after a dissipation episode, where $E_{\alpha,k-1}\sim E_{\rm resid}$. Setting $|\Delta \hat{P}_{\alpha,k}(E_{\rm resid})|\sim 1$ with $P_k\propto a_k^{3/2}$ and $|E_{B,k}|\propto a_k^{-1}$, we find 
	\begin{equation}
	a_{\rm WJ} \simeq a_0 \left[(3 \sigma_\alpha P_0)^2\frac{\Delta E_\alpha E_{\rm resid}}{|E_{\rm B,0}|^2}\right]^{-1/5}, \label{eq:aWJ1}
	\end{equation}
	where $\Delta E_{\alpha}$ and $\sigma_\alpha$ are evaluated at $r_{\rm p} = r_{\rm p, WJ}$, the pericentre distance of the planet when chaotic tides stop operating. The value of $r_{\rm p, WJ}$ varies erratically for small changes of initial conditions, but must fall between $r_{\rm p, dis}$ ($\eta = 2.7$) and $r_{\rm p,crit}$ (very roughly $\eta \sim 4$) for a system with $a_0< 5$ au. Equation (\ref{eq:aWJ1}) gives the relationship between $a$ and $r_{\rm p}$ for WJs formed by chaotic tides. Parametrizing the residual mode energy by $E_{\rm resid} \equiv f G M_p^2/R_p$, and using equation~(\ref{eq:DelE}), we can write $a_{\rm WJ}$ in terms of the planet properties and $\eta$:
	 \begin{align}
	 a_{\rm WJ} \simeq  r_{\rm tide} \left(\frac{M_*}{M_p}\right)^{4/15}\left[ \left(6 \pi \bar{\sigma}_\alpha\right)^2f\frac{4T(\eta)}{\eta^6}\right]^{-1/5}. \label{eq:aWJ}
	 \end{align} 
	 Equation~(\ref{eq:aWJ}) is plotted in Fig.~\ref{fig:aWJvseta} for two different planet models. For values of $\eta$ between $2.7$ and $3.6$, equation~(\ref{eq:aWJ}) yields semi-major axes in the range of $0.3-1.0$ au (for $R_p=1.6 R_J$ and $M_*/M_p=1000$). We can also find the eccentricity of the WJs via
	 \begin{align}
	 (1-e_{\rm WJ}) &= \frac{\eta R_p}{a_{\rm WJ}}\left(\frac{M_*}{M_p}\right)^{1/3} \nonumber\\
	 &\simeq \left(\frac{M_p}{M_*}\right)^{4/15}\left[(6 \pi \bar{\sigma}_\alpha)^2f\frac{4T(\eta)}{\eta}\right]^{1/5}.\label{eq:eWJ}
	 \end{align}
	 For $\eta$ between $2.7$ and $3.6$, $(1-e_{\rm WJ})$ is between 0.027 and 0.060 (again for $R_p=1.6 R_J$ and  $M_*/M_p=1000$).
	
	Figures~\ref{fig:LongTermEvolution}, \ref{fig:LongTermEvolutionLargerEta}, and \ref{fig:SavedSystemExample} show that after the WJ forms through chaotic tides, the eccentricity freezes at a high value and the planet's orbit decouples from the companion. In general, LK eccentricity oscillations freeze when $\dot{\omega}_{\rm SRF}t_{\rm LK}\sqrt{1-e^2} \gtrsim 1,$ where $\dot{\omega}_{\rm SRF}$ is the rate of precession due to SRFs. This is equivalent to $\epsilon_{\rm GR}/\sqrt{1-e^2} \gtrsim 1$ or $\epsilon_{\rm Tide}/(1-e^2)^{9/2} \gg 1$. We find that tidal effects play a more important role than GR in the ``freezing" of eccentricity oscillations of WJs.  

\subsection{Fast Long-Term Orbital Decay and Hot Jupiter Formation}\label{sec:StaticTides}
	When the planet's orbit is no longer strongly influenced by chaotic f-mode evolution and the LK effect, the orbit decays and eventually circularizes due to tidal friction from static tides raised on the planet. The final result is a HJ. This is the same mechanism that has been used to explain HJ formation in previous studies of migration via the LK effect \citep[ASL16;][]{Petrovich15}. Here, the process occurs more quickly because the orbit is frozen into a small pericentre value rather than oscillating between high and low eccentricities. Figures~\ref{fig:LongTermEvolution} and \ref{fig:LongTermEvolutionLargerEta} compare the long-term orbital evolution predicted by our model (in blue) with the standard calculation that does not include chaotic tides (in grey). In both cases, the orbit circularizes much more quickly when chaotic tides are included.
	
\subsection{Spin-Orbit Misalignment and Final Planet-Binary Inclination}\label{sec:Spin}
		\begin{figure}
			\includegraphics[width =\columnwidth]{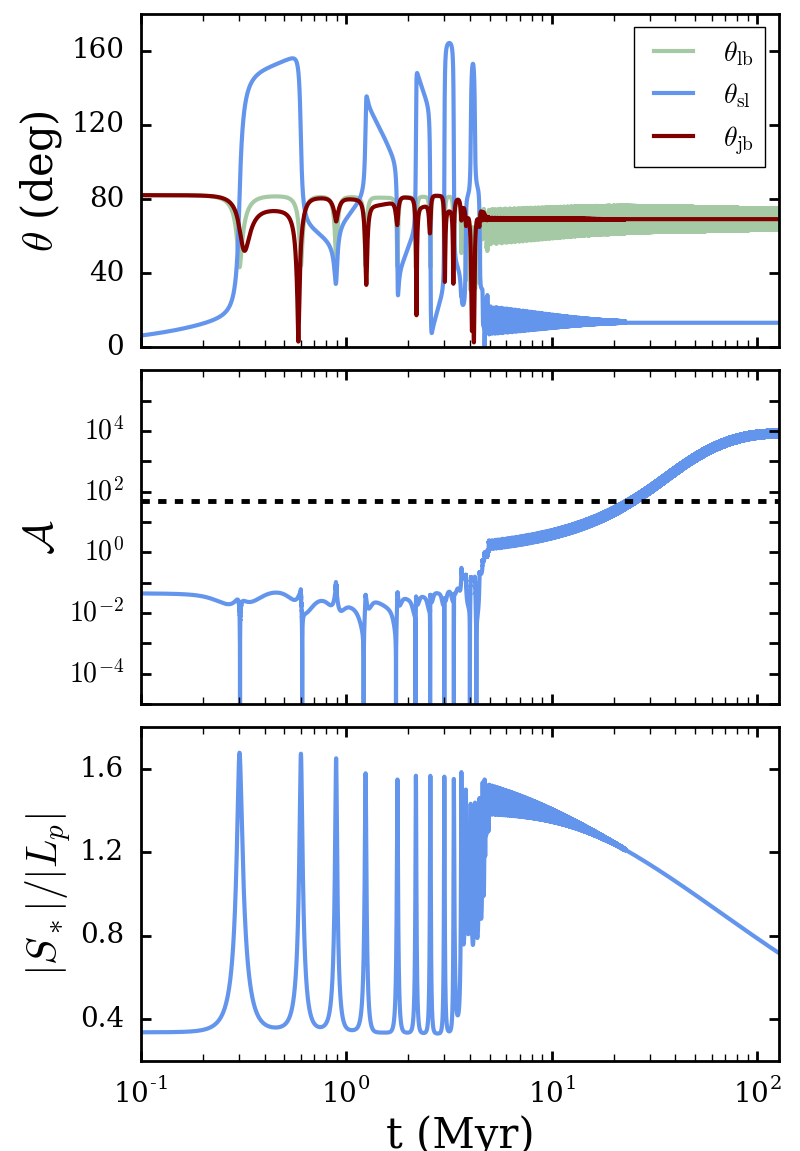}
			\caption{The same system as in Fig.~\ref{fig:SavedSystemExample} (with chaotic tides), but showing the evolution of various misalignment/inclination angles, the adiabaticity parameter $\mathcal{A}$ (equation~\ref{eq:defA}), and the ratio of the spin angular momentum of the star to the planet's orbital angular momentum. The stellar obliquity, $\theta_{\rm sl}$, freezes when $\mathcal{A} \gg 1$ (the dotted line in the second panel denotes $\mathcal{A} = 50$, the criterion used in our population synthesis).}
			\label{fig:Angles}
		\end{figure}	

		Recent studies \cite[ASL16][]{Storch14, Storch17} have shown that a giant planet undergoing LK migration can induce complex dynamics in the spin axis of the oblate host star before the spin-orbit misalignment angle (stellar obliquity), $\theta_{\rm sl}$, becomes frozen. This freezing occurs when the precession rate, $\Omega_{\rm p*}$, of the stellar spin vector ($\boldsymbol{S_*}$) around $\hat{\boldsymbol{L}}$ (the unit orbital angular momentum vector of the planet) becomes much faster than the precession rate ($\Omega_{L}$) of $\hat{\boldsymbol{L}}$ around the binary angular momentum axis $\boldsymbol{\hat{L}_b}$. The ratio of precession rates is
		\begin{equation}
		\mathcal{A} \equiv \left|\frac{\Omega_{\rm p*}}{\Omega_{\rm L}}\right|, \label{eq:defA}
		\end{equation}
		where
		\begin{equation}
		\Omega_{\rm L} \equiv \left|\frac{d\hat{\boldsymbol{L}}}{dt}\right|_{\rm LK, quad} \simeq \frac{3(1+4e^2)}{8 t_{\rm LK}\sqrt{1-e^2}}|\sin 2 \theta_{\rm lb}|,
		\end{equation}
		and 
		\begin{equation}
		\Omega_{p*} = -\frac{3}{2}\frac{k_{q*}}{k_*}\frac{M_p}{M_*}\left(\frac{R_*}{a}\right)^3\frac{\Omega_*}{(1-e^2)^{3/2}}\cos \theta_{\rm sl},
		\end{equation}
		with $k_{*}$ as the stellar moment of inertia constant, $k_{q*}$ as the stellar rotational distortion coefficient, and $\Omega_*$ as the stellar spin rate (see ASL16). We take $k_*$ = 0.1 and $k_{q*}=0.05$. We call $\mathcal{A}$ the adiabaticity parameter. The spin-orbit angle freezes when $\mathcal{A}\gg 1$.
		
		Figure~\ref{fig:Angles} shows an example of the evolution of the spin-orbit misalignment angle $\theta_{\rm sl}$ during chaotic tidal migration. The stellar spin rate evolves according to the Skumanich law given in equation~(\ref{eq:Skumanich}). When chaotic tides stop operating, around 4.5 Myr, $\mathcal{A} \sim 1$, and $\theta_{\rm sl}$ still oscillates with an amplitude of $\sim 16^\circ$. Because $\mathcal{A}$ depends on the stellar spin rate and $\theta_{\rm lb}$, the condition for the adiabatic ``spin-orbit freeze" ($\mathcal{A} \gg 1$) is generally different from the condition for ``LK oscillations freeze" [$\epsilon_{\rm Tide}/(1-e^2)^{9/2} \gg 1$; see section \ref{sec:eccWJ}]. Many of the planets that migrate via LK chaotic tides will not have a fixed $\theta_{\rm sl}$ when they become eccentric WJs.
		
		The mutual inclination $\theta_{\rm lb}$ between the orbits of the planet and the stellar companion is also of interest. We see from Fig.~\ref{fig:Angles} that $\theta_{\rm lb}$ can oscillate even at the end of the evolution (when the planet has become a HJ). This variation of $\theta_{\rm lb}$ arises from the coupling between $\boldsymbol{L}$ and $\boldsymbol{S_*}$ (with $|\boldsymbol{S_*}|$ non-negligible compared to $|\boldsymbol{L}|$). We define
		\begin{equation}
		\boldsymbol{J} \equiv \boldsymbol{L} + \boldsymbol{S_*}.
		\end{equation}
		We find that the angle between $\boldsymbol{J}$ and $\boldsymbol{L_b}$, $\theta_{\rm jb}$, is constant when the planet's orbit is decoupled from the influence of the stellar companion. To understand this, consider the time evolution of the planet's orbital angular momentum axis $\hat{\boldsymbol{L}}$ and the stellar spin axis $\hat{\boldsymbol{S}}_*$. Both are rapidly precessing around $\boldsymbol{J}$. An external torque acting on $\boldsymbol{S}_*$ (from magnetic braking) or $\boldsymbol{L}$ (from the binary companion) also acts on $\boldsymbol{J}$. But after averaging over the fast timescale $2\pi/|\Omega_{p*}|$, the net effect is that $\boldsymbol{J}$ precesses around $\boldsymbol{L}_B$. As a result, $\theta_{\rm jb}$ is constant and $\theta_{\rm lb}$ oscillates around $\theta_{\rm jb}$ with amplitude $\theta_{\rm lj}$, where
		\begin{equation}
		\tan{\theta_{\rm jl}} = \frac{|\boldsymbol{S_*}|}{|\boldsymbol{L}|} \frac{\sin \theta_{\rm sl}}{(1 + |\boldsymbol{S_*}|/|\boldsymbol{L}| \cos \theta_{\rm sl})}.
		\end{equation}
		The top panel of Fig.~\ref{fig:Angles} illustrates this behaviour. As the star gradually spins down due to magnetic braking, $\theta_{\rm lb}$ asymptotes to $\theta_{\rm jb}$.

\section{Population Synthesis}\label{sec:popSynthesis}
		In this section we carry out a population synthesis study of giant planets undergoing LK migration with chaotic tides. Our goal is to determine the production efficiencies of eccentric WJs and HJs, as well as their general properties as predicted by this scenario.
		
		\begin{table*}
			\caption{Outcomes of LK migration with chaotic tides for three planet models. The fraction of systems that undergo chaotic evolution is denoted by $\mathcal{F}_{\rm mig}$; these planets are either tidally disrupted ($\mathcal{F}_{\rm dis}$) or exit chaotic tides as eccentric WJs ($\mathcal{F}_{\rm WJ} = \mathcal{F}_{\rm mig}- \mathcal{F}_{\rm dis}$). The fraction of planets that circularize to $e=0.1$ with 1 Gyr is $\mathcal{F}_{\rm HJ}$. We compare our results with semi-analytical predictions using the method described in Section~\ref{sec:Rates}.}
			\begin{tabular}{l|l|l|l|l|l|l|l|l}
				\hline
				& \multicolumn{2}{c}{1 $R_J$, 1$M_J$} && \multicolumn{2}{c}{1.6 $R_J$, 1$M_J$} && \multicolumn{2}{c}{1.6 $R_J$, 0.3$M_J$}\\
				\hline
				& Rate & Prediction && Rate & Prediction && Rate & Prediction\\
				\cline{2-3}\cline{5-6}\cline{8-9}
				\\
				$\mathcal{F}_{\rm mig}$& 14.0 \% & 13.7 \% && 15.4 \% & 14.2 \% && 16.7 \% & 15.2\%\\
				$\mathcal{F}_{\rm dis}$ & 10.1\% & $\le$11.9\% && 11.5 \% & $\le$11.9 \% && 13.2 \% & $\le$13.1 \% \\
				$\mathcal{F}_{\rm WJ}$&3.9\% & $\ge$1.8 \% && 3.9\% & $\ge$2.3 \% && 3.6 \% & $\ge$ 2.1 \%\\
				$\mathcal{F}_{\rm HJ}$&3.9\% & --- && 3.8\% & --- && 2.2 \% & --- \\ 
				\hline
			\end{tabular}
			\label{tab:PopSynthResults}
		\end{table*}	
		
				\begin{figure*}
					\includegraphics[width=6.5in]{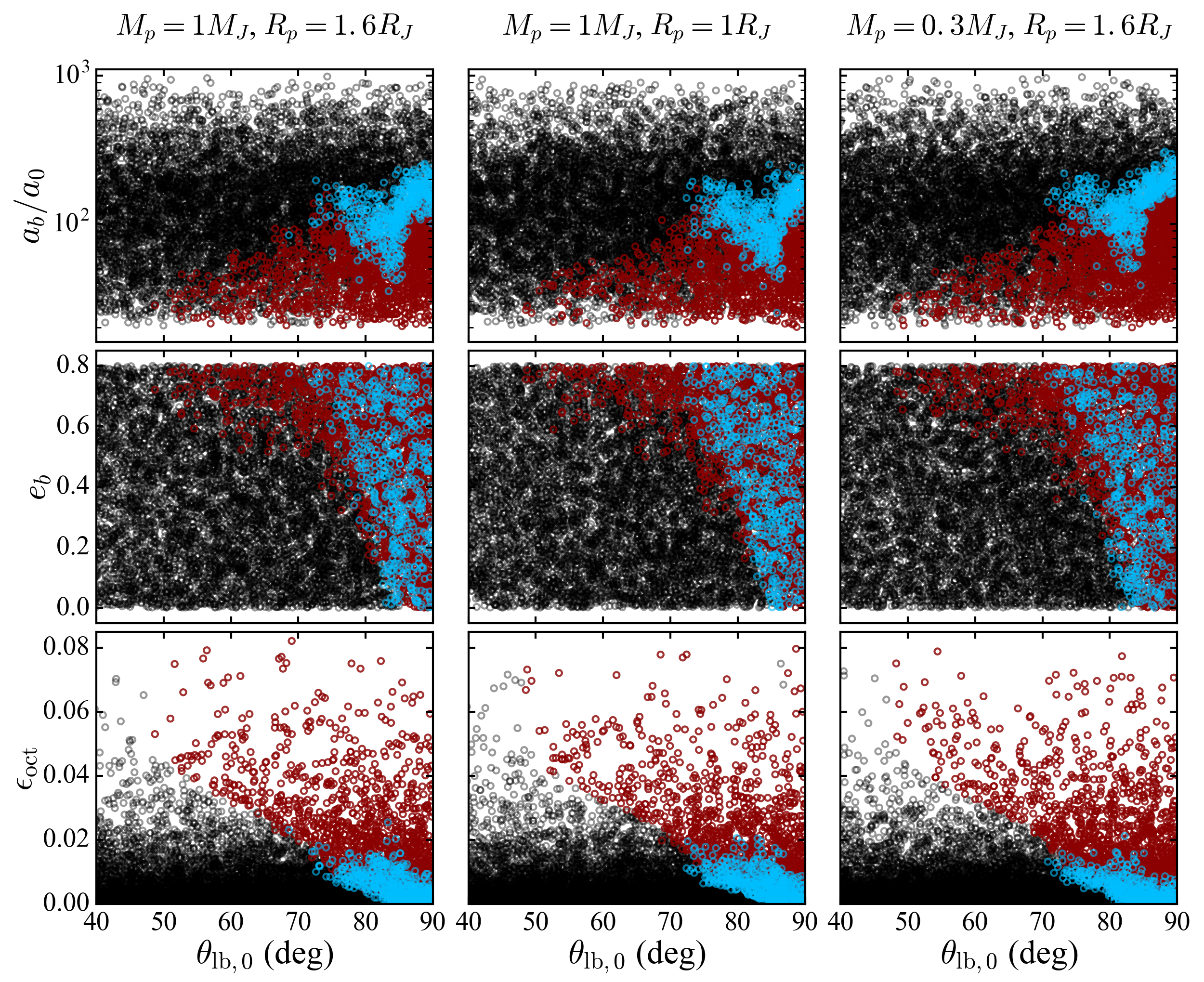}
					\caption{The parameter space that that results in WJs formed by chaotic tidal migration (blue) and tidal disruption (red). Each point represents a calculation with different initial binary inclination $\theta_{\rm lb,0}$, semi-major axes ($a_{\rm b}$ in units of $a_0$), eccentricity $e_{\rm b}$ and octupole parameter $\epsilon_{\rm oct}$. Systems with $\theta_{\rm lb,0} > 90^\circ$ are plotted as $180-\theta_{\rm lb, 0}$. For most systems, the planetary f-mode never becomes chaotic (black points). WJs are produced for a narrow range of $a_{\rm b}/a_0$ and for $\epsilon_{\rm oct} \lesssim 0.02$. The three columns show different combinations of planet mass and radius (as labelled).}
					\label{fig:phaseSpaceScan}
				\end{figure*}
		
		\subsection{Setup and Method}\label{sec:setupPopSynth}
		All of our population synthesis calculations use the same stellar and companion masses $M_*=M_b = M_\odot$. The initial rotational period of the host star is $2.3$ days, and the star spins down according to the Skumanich law (equation \ref{eq:Skumanich}). The planet model is a $\gamma=2$ polytrope. We consider two planet radii, $R_p=R_J$ and $R_p=1.6 R_J$, and two planet masses $M_p=1 M_J$ and $M_p = 0.3M_J$. 
		
		While some of our earlier examples have assumed that $e_{\rm b}=0$ (so that the octupole terms vanish), in the population synthesis we consider a broad range of $e_{b}$ and incorporate the octupole LK effect. The importance of the octupole effect (relative to the quadrupole effect) is encoded in the dimensionless parameter 
		\begin{equation}
		\epsilon_{\rm oct} = \frac{a}{a_{\rm b}}\frac{e_{\rm b}}{1-e_{\rm b}^2}.\label{eq:epsilonOct}
		\end{equation}
		We have explored a variety of initial conditions, uniformly spanning $a_0  = [1, 5]~\text{au}$, $a_{\rm b} = [10^2,10^3]$~(au) (uniformly sampled in $\log{a_{\rm b}}$), $\cos(\theta_{\rm lb,0}) = (-0.77,0.77)$, $e_{\rm b} = [0,0.8]$ and $\Omega_0 = [0,2\uppi]$ (where $\Omega_0$ is the initial longitude of the ascending node of the planet's orbit). Our choice of initial mutual inclinations, $\theta_{\rm lb,0}$, is limited to the range where the quadrupole LK effect can operate, i.e. $\cos^2\theta_{ \rm lb,0}< 3/5$. Systems that do not obey the stability criterion \citep{Mardling01},
		\begin{equation}
		\frac{a_{\rm b}}{a} > 2.8\left(1 + \frac{M_b}{M_{\rm tot}}\right)^{2/5} \frac{(1+e_{\rm b})^{2/5}}{(1-e_{\rm b})^{6/5}} \left[1-0.3\frac{\theta_{\rm lb,0}}{180^\circ}\right],
		\end{equation}
		are discarded. Note that the range of initial conditions is identical to that of ASL16 for straightforward comparison between the HJ formation fraction from LK migration with and without chaotic tides.
		
		The details of how the evolution of the planetary f-mode is calculated in combination with the LK effect are discussed in Section~\ref{sec:FullModel}. In all calculations, we use $E_{\rm max} = 0.1 GM_p^2/R_p$ and $E_{\rm resid} = 0.001 GM_p^2/R_p$ to parametrize the maximum f-mode energy and the residual energy after an episode of non-linear tidal dissipation. The planet is assumed to be rotating at the pseudo-synchronous rate (equation~\ref{eq:Pseudosynchronous}). The effect of dynamical tides on the planet's spin is not accounted for in this investigation, but could be incorporated into future studies that follow the evolution of multiple oscillatory modes in the planet.
		
		Our study is structured to efficiently determine the likelihood of three possible outcomes: ``No Chaotic Tidal Migration", ``Tidal Disruption", and `Chaotic Tidal Migration". Each calculation is stopped when the integration time reaches 1 Gyr or when one of the following conditions is met:
		
			(i) If $|\Delta \hat{P}_{\alpha,k}(E_{\alpha,k-1})|$ (equation \ref{eq:DelPkmaxFull}) has not reached 1 after min$(500 t_{\rm LK}, 5 t_{\rm LK}/\epsilon_{\rm oct})$, the calculation is terminated and the system is labelled as ``No Chaotic Tidal Migration". Note that a small percentage of such systems may migrate within 1 Gyr via standard LK migration without the assistance of chaotic tides. These are not identified as a separate outcome from planets that will not migrate at all.
			
			(ii) If the system has evolved for more than $10^7$ planetary orbits in the chaotic phase (i.e. where the orbital evolution is coupled with the f-mode evolution, as discussed in Section~\ref{sec:FullModel}) without reaching $E_{\alpha,k} = E_{\rm max}$, the integration is stopped and the system is classified as ``No Chaotic Tidal Migration". 
			
			(iii) If, at any time, the pericentre distance is within the tidal disruption radius, i.e. $r_{\rm p} < r_{\rm p,dis}$, the outcome is classified as ``Tidal Disruption", and the integration is stopped.
			
			(iv) If the planet has experienced chaotic orbital evolution and its orbit has circularized to $e=0.1$ within 1~Gyr, the integration is stopped and the system is classified as having undergone ``Chaotic Tidal Migration".  
			
		For each system, we begin by integrating the full equations of motion and incorporating the evolution of the planet f-mode (as described in Section~\ref{sec:FullModel}). However, at  different points in the giant planet's evolution, it is possible to ``turn off" various effects without losing accuracy. As discussed in Section~\ref{sec:FullModel}, the effect of dynamical tides is only accounted for when it can significantly affect orbital evolution. Additionally, when the semi-major axis is small, LK oscillations are suppressed. As the planet's orbit shrinks and circularizes on a long time-scale due to static tides [see equation~(\ref{eq:deftst}) with $\Delta t_{L} = 1$~s], the eccentricity of the planet's orbit precesses on a much shorter time-scale than that of tidal decay. Following this precession is computationally expensive and does not affect the final properties of the migrating planet. When the LK oscillations are ``frozen" [$\epsilon_{\rm Tide}/(1-e^2)^{9/2} > 30$] and the spin-orbit angle is safely adiabatic ($\mathcal{A} >50$), we continue the integration without LK and SRF terms. 
		
		\subsection{Migration and WJ/HJ Formation Fractions}	
		
		The results of our population synthesis are summarized in Table \ref{tab:PopSynthResults}. For each combination of planetary mass and radius, we ran $10^4$ calculations to determine the fraction of systems that undergo chaotic tidal migration ($\mathcal{F}_{\rm mig}$). These migrated planets are either tidally disrupted (fraction $\mathcal{F}_{\rm dis}$) or survive as eccentric WJs (fraction $\mathcal{F}_{\rm WJ} = \mathcal{F}_{\rm mig} - \mathcal{F}_{\rm dis}$) that undergo further orbital decay and circularization due to static tides, eventually becoming HJs. The fraction of systems that evolve into HJs within 1 Gyr is denoted by $\mathcal{F}_{\rm HJ}$. Note that our population synthesis did not sample initial mutual inclinations with $|\cos(\theta_{\rm lb,0})|>0.77$, as such planets do not experience large excursions in eccentricity. The migration, disruption, WJ, and HJ fractions are calculated assuming a uniform distribution in $\cos(\theta_{\rm lb,0})$, e.g. $\mathcal{F}_{\rm mig} = 0.77 N_{\rm mig}/N_{\rm run}$ with $N_{\rm mig}$ the number of systems that displayed chaotic tidal migration and $N_{\rm run}$ the total number of runs. For $1M_J$, $1R_p$ planets, the HJ formation percentage from our population synthesis is larger than the value obtained for standard LK migration by a factor of $\sim 1.6$ (3.9 \% vs 2.4 \%); see\citep[ASL16;][]{Munoz16}.
		
		\subsection{Parameter Space for WJ Formation}
\begin{figure}
				\includegraphics[width=\columnwidth]{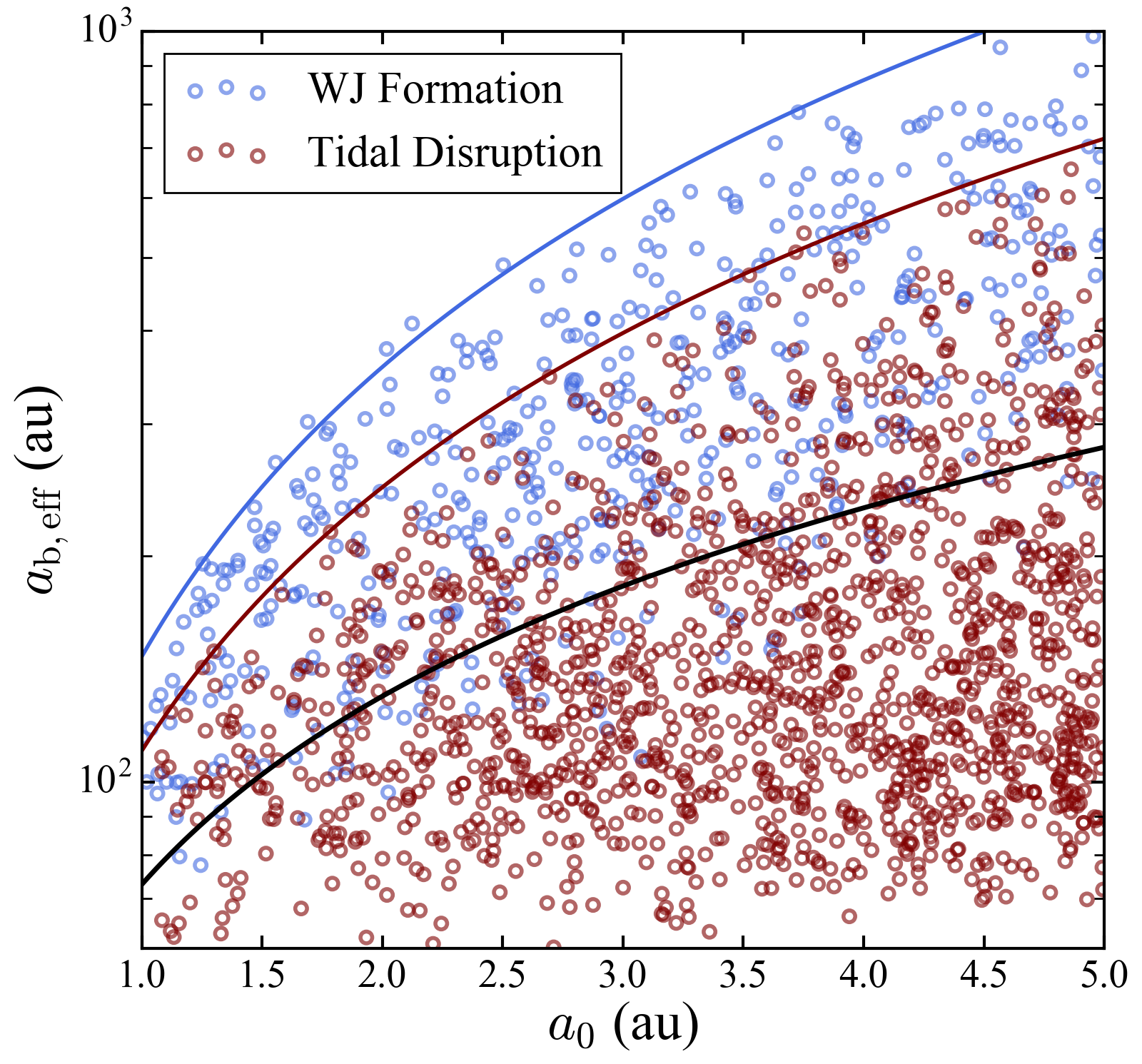}
				\caption{Initial conditions for LK migration with chaotic tides, coloured by the outcomes of the evolution, either the formation of an eccentric WJ (blue) or tidal disruption (red). The giant planet has $M_p=1M_J$ and $R_p=1.6R_J$. The blue, red, and black lines are the same as in Fig.~\ref{fig:SavedSystems}.}
				\label{fig:Saved}
			\end{figure}

		We can learn about the parameter space that leads to eccentric WJ formation by examining how the outcome of orbital evolution depends on the initial conditions and the planet properties (see Fig. \ref{fig:phaseSpaceScan}; see also Fig.~18 from ASL16 for comparison with the standard LK migration results). It is clear that smaller values of $\theta_{\rm lb,0}$ do not yield migration. This is unsurprising as the eccentricities required for chaotic tides are very large, which necessitates large initial mutual inclinations. Another pronounced feature is that the systems that produce eccentric WJs are clustered in $a_{\rm b}/a_0$. The reason for this is discussed in Section~\ref{sec:Features} (see Fig.~\ref{fig:SavedSystems}). In essence, only a narrow window in $a_{\rm b}/a_0$ can produce systems that will reach large enough eccentricities to undergo chaotic tides (see equation \ref{eq:abeffMax}), but not so large that the planets are tidally disrupted (see equation \ref{eq:abeffMin}). Figure~\ref{fig:Saved} demonstrates that the systems that produce surviving WJs indeed satisfy equations~(\ref{eq:abeffMax}) and (\ref{eq:abeffMin}). The range of $\epsilon_{\rm oct}$ that can produce eccentric WJs is capped by the limit on $a_{\rm b}/a_0$ (see equation \ref{eq:epsilonOct}). Lastly, Fig.~\ref{fig:phaseSpaceScan} shows that changing the properties of the gas giant has little effect on the parameter space that produces eccentric WJs. Unlike standard LK migration with static tides, where systems with $M_p=0.3 M_J$ produce hot Saturns at a rate of  $0.5\%$ (assuming $\Delta t_{\rm L} = 1$ s) after evolving for 5 Gyr \citep[see ASL16,][]{Munoz16}, chaotic tides allow low-mass planets to survive high-eccentricity migration without suffering tidal disruption, and produce hot Saturns at a rate of about $2.2 \%$ after only 1 Gyr. 
		
		\subsection{Properties of WJs Formed by Chaotic Tidal Evolution}\label{sec:eccentricWJs}
		\begin{figure*}
			\begin{center}
				\includegraphics[width = 6.5in]{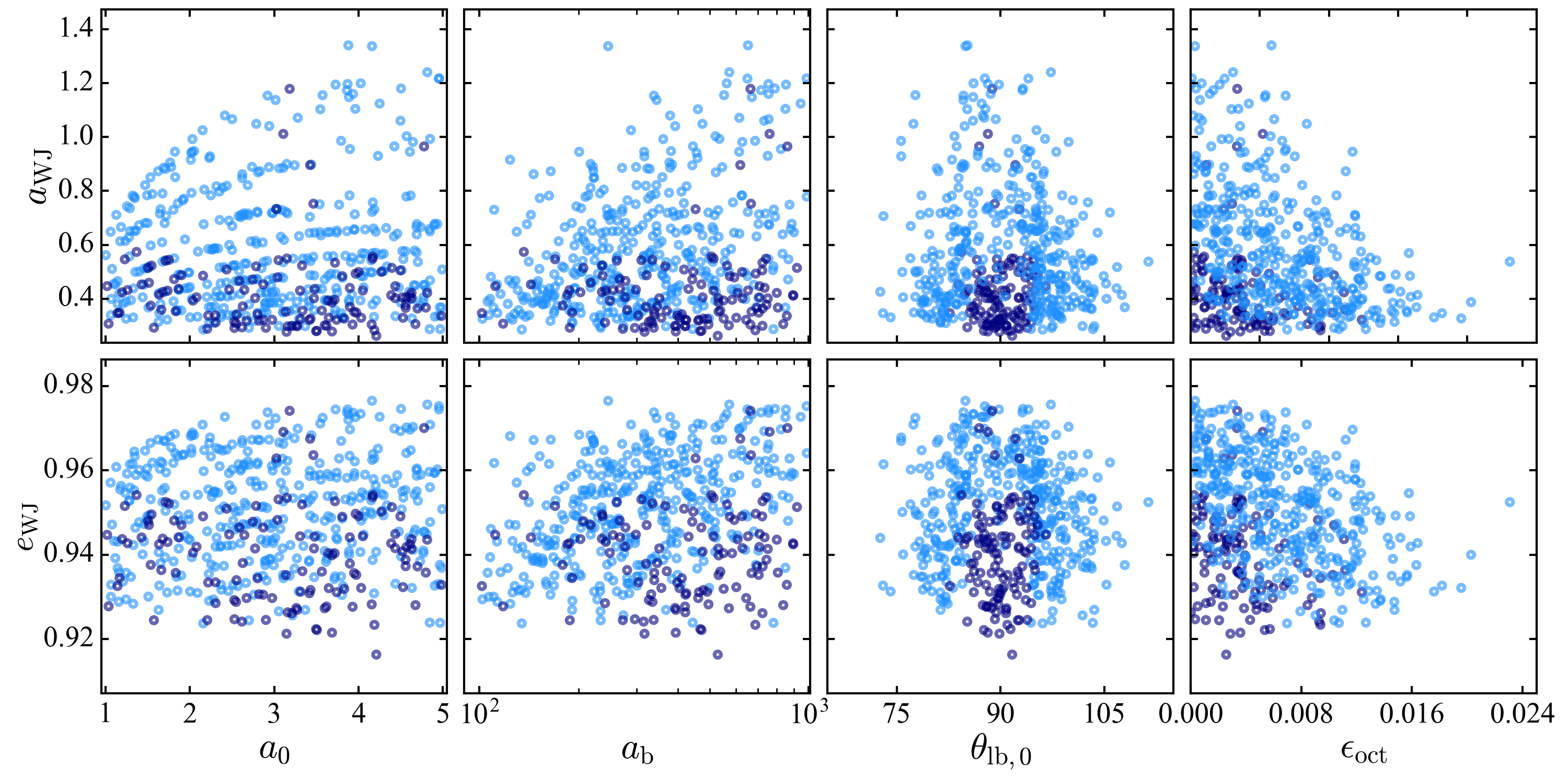}
				\caption{The orbital semi-major axis $a_{\rm WJ}$ (top row) and eccentricity $e_{\rm WJ}$ (second row) of WJs formed by chaotic tidal migration (for planets with $M_p =1M_J$, $R_p = 1.6 R_J$) as a function the initial $a_0$ of the planet, $a_{\rm b}$ of binary, the initial binary inclination, $\theta_{\rm lb,0}$, and the octupole parameter $\epsilon_{\rm oct}$ [see equation~(\ref{eq:epsilonOct})]. Systems that only experience one high-eccentricity phase are shown in dark blue, and those that undergo multiple LK cycles are shown in light blue.} \label{fig:scatterPlots}
			\end{center}
		\end{figure*}
		\begin{figure*}
			\includegraphics[width=6.5in]{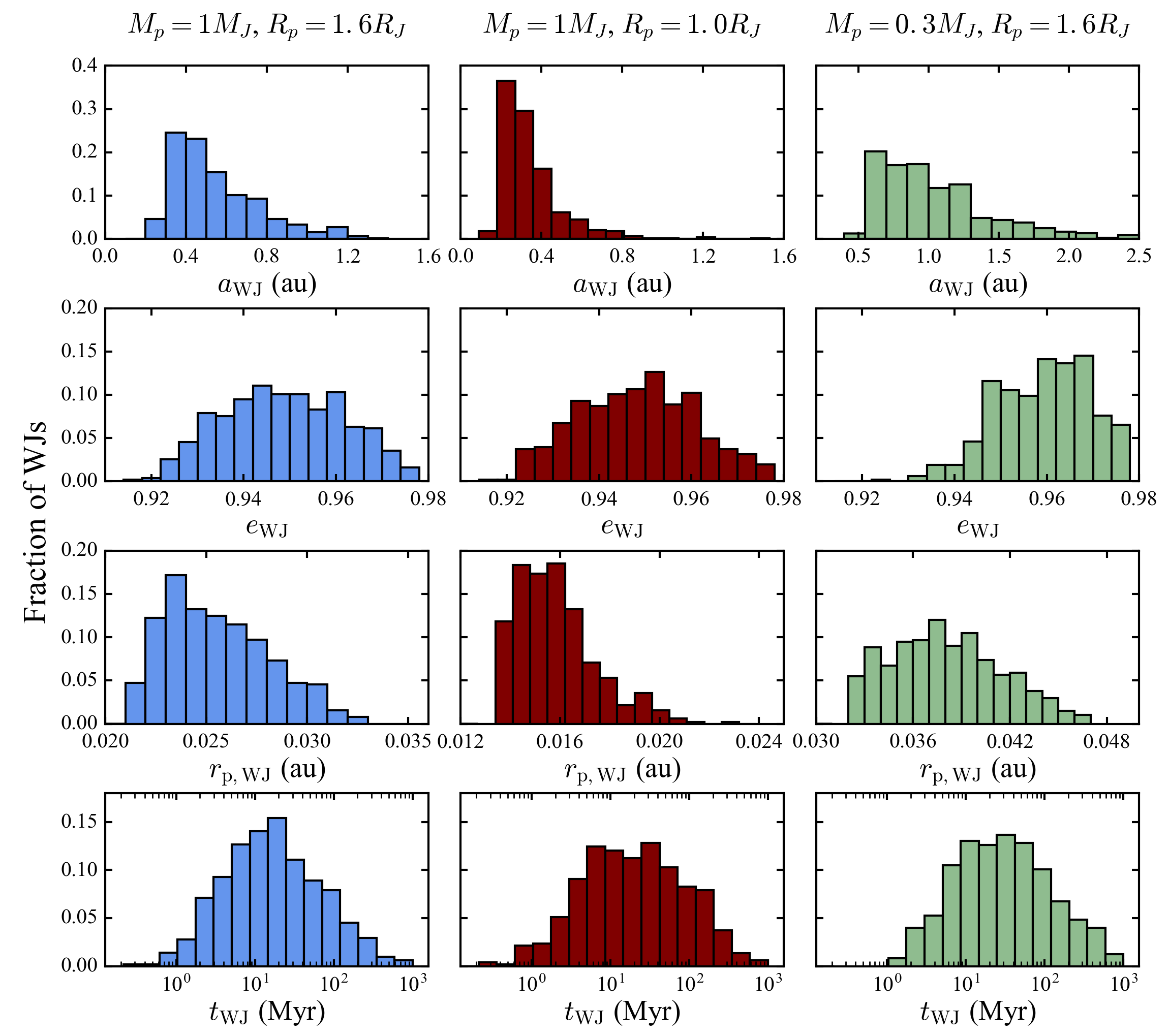}
			\caption{The distributions of $a_{\rm WJ}$, $e_{\rm WJ}$, $r_{\rm p,WJ}$, and $t_{\rm WJ}$ (the time at which chaotic evolution stops) for WJs produced by LK migration with chaotic tides for three different planet models. The ranges of $a_{\rm WJ}$ and $e_{\rm p,WJ}$ are in agreement with equations~(\ref{eq:aWJ}) and (\ref{eq:eWJ}), (see also Fig. \ref{fig:grid}). The peaks of the $r_{\rm p,WJ}$ distributions occur near $\eta \sim 3$ [see equation~(\ref{eq:defeta})] for all three planet models.} \label{fig:Histograms}
		\end{figure*}

		We now examine the orbital properties of planets produced by chaotic tides. The scatter plots in Fig.~\ref{fig:scatterPlots} show an overview of the WJ eccentricities and semi-major axes ($e_{\rm WJ}$ and $a_{\rm WJ}$) for the $M_p = 1 M_J, R_p = 1.6 R_J$ planet model after the planetary f-mode has settled into a quiescent state (but before weak tidal friction circularizes the orbit over a long time-scale). Most planets have $a_{\rm WJ}$ between $0.2$ and $1$ au and $e_{\rm WJ}$ between 0.91 and 0.97. Systems that have undergone multiple LK cycles tend to have larger $a_{\rm WJ}$ and $e_{\rm WJ}$.
		This is expected because larger values of $a_{\rm WJ}$ and $e_{\rm WJ}$ correspond to systems that do not reach a small enough value of $r_{\rm p}$ for chaotic tides to significantly alter the orbit within the time-frame of a single high-eccentricity phase. In general, $a_{\rm WJ}$ and $e_{\rm WJ}$ increase with $\eta = r_{\rm p}/r_{\rm tide}$ [see equations~(\ref{eq:aWJ}) and (\ref{eq:eWJ}) and Fig.~\ref{fig:aWJvseta}]. 
		
		In the top left panel of Fig.~\ref{fig:scatterPlots}, the values of $a_{\rm WJ}$ seem to fall on a set of $a_{\rm WJ}$ - $a_0$ curves. This feature arises from our treatment of non-linear dissipation, where the f-mode rapidly dissipates energy after reaching $E_{\rm max}$. Each curve corresponds to some number $N$ of non-linear dissipation episodes, after which the orbital energy becomes $(E_{B,0} - N E_{\rm max})$. The systems with the most distant stellar companions can produce WJs with relatively large semi-major axes ($\gtrsim$1 au) after chaotic tidal evolution. This occurs because the minimum $\eta$ required for chaotic tides to compete with the LK effect is larger for systems with larger $a_{\rm b,eff}$. In Fig.~\ref{fig:scatterPlots}, we can also see that  most systems that exhibit chaotic behaviour have initial mutual inclinations around $90^\circ$, as expected. Lastly, as seen earlier in Fig.~\ref{fig:phaseSpaceScan}, large values of $\epsilon_{\rm oct}$ do not produce surviving WJs.
		
		Figure~\ref{fig:Histograms} shows the histogram of the parameters of WJs produced by chaotic tidal migration. The distributions of planet properties are generally more sharply peaked for denser planets. The ranges of $a_{\rm WJ}$ and $e_{\rm p,WJ}$ are in agreement with equations~(\ref{eq:aWJ}) and (\ref{eq:eWJ}), which predict that $a_{\rm WJ}$ scales linearly with $R_p$ while $e_{\rm WJ}$ has no dependence on the planet radius and a weak dependence on the planet mass (for a given minimum $\eta$ achieved during the orbital evolution). Combining equations~(\ref{eq:aWJ}) and (\ref{eq:eWJ}), we can see that the distribution of $r_{\rm p, WJ} = a_{\rm WJ}(1-e_{\rm WJ})$ should scale directly with $r_{\rm tide}$. This would imply that the peak at $0.015$ au in the $r_{\rm p,WJ}$ distribution for the $M_p= 1M_J, R_p = 1\;R_J$ planet model should appear near $0.024$ au and $0.036$ au in the $M_p= 1M_J, R_p = 1.6\;R_J$ and $M_p= 0.3M_J, R_p = 1.6\;R_J$ models, respectively, as is the case in Fig.~\ref{fig:Histograms}. The timescale for chaotic evolution to shrink the orbit, $t_{\rm WJ}$, peaks near 10 Myr for all planet models.
		
\begin{figure}
			\begin{center}
				\includegraphics[width=\columnwidth]{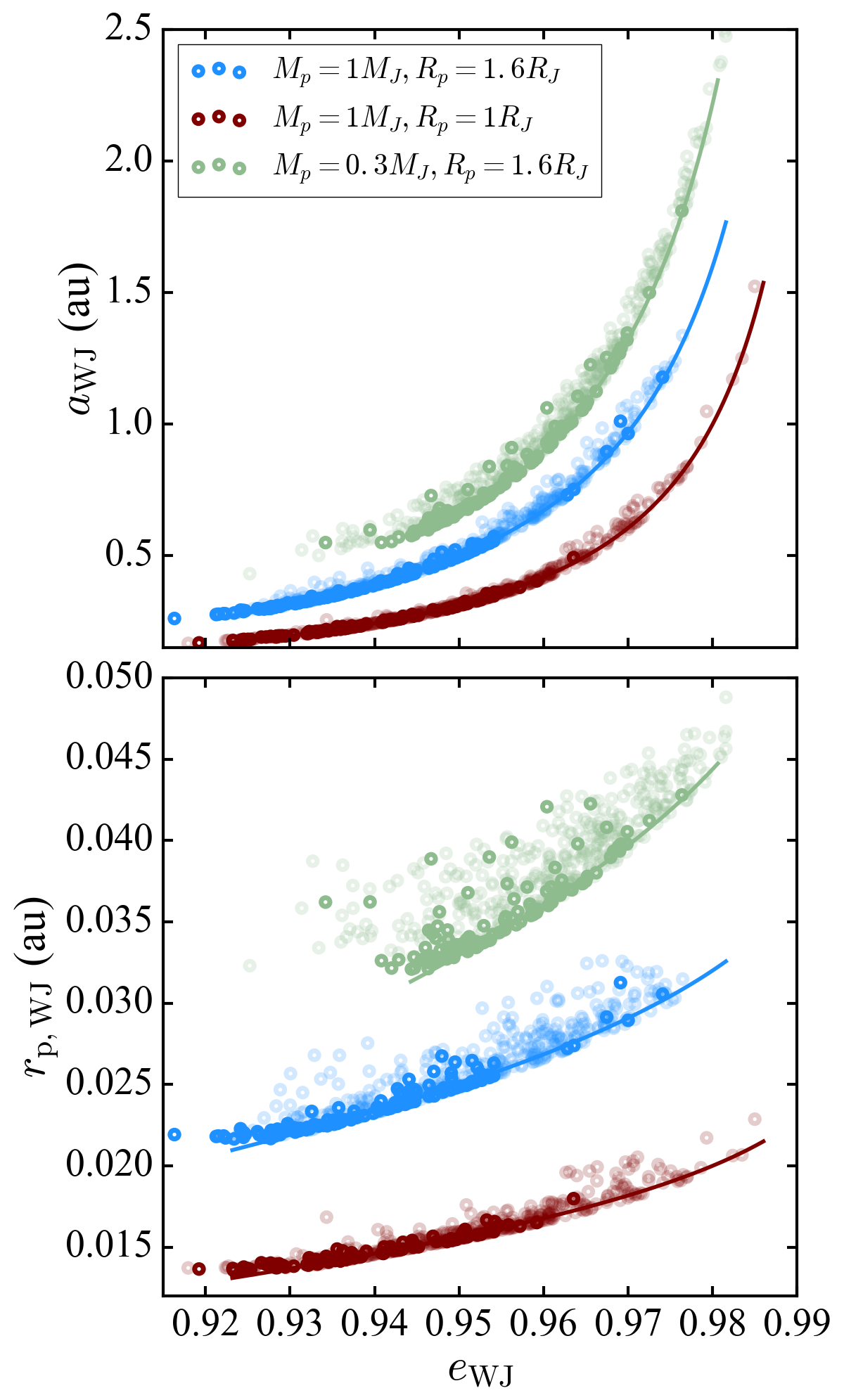}
				\caption{Semi-major axis ($a_{\rm WJ}$), pericentre distance ($r_{\rm p, WJ}$), and eccentricity ($e_{\rm WJ}$) correlations of WJs produced by chaotic tidal migration for three planet models.
				Dark points indicate systems that have only undergone one LK cycle, and light points show systems that have experienced multiple LK cycles. The solid lines denote $|\Delta \hatP_{\alpha}(\tilde{E}_{\rm resid})| = 1.0$ for different planet properties calculated from equations~(\ref{eq:aWJ}) and (\ref{eq:eWJ}).} \label{fig:grid}
			\end{center}
		\end{figure}

		The range of $r_{\rm p, WJ}$ is narrow for all three planet models. The lower edge of this distribution is determined by the tidal disruption radius. A planet cannot survive  if $\eta < 2.7$. The upper value of $r_{\rm p, WJ}$ is roughly set by the pericentre distance where the planet with the largest semi-major axis (in our simulation, $a_0=5$ au) crosses the chaos boundary, generally near $\eta \sim 4$. Figure~\ref{fig:grid} shows the relationship between $a_{\rm WJ}$, $e_{\rm WJ}$ and $r_{\rm p,WJ}$ for planets that have undergone chaotic evolution. The solid lines show equation~(\ref{eq:eWJ}) for each planet model. Recall that equation~(\ref{eq:eWJ}) yields the eccentricity where $|\Delta \hat{P}_{\alpha}(E_{\rm resid})| = 1$ for a given $r_{\rm p}$. Therefore, the restrictions on $r_{\rm p,WJ}$ determine the range in the distributions of $a_{\rm WJ}$ and $e_{\rm WJ}$. This simple calculation explains the properties of planets that survive chaotic tidal evolution very well. The spread in the results from the population synthesis arises from the fuzziness in the chaotic tides boundary, i.e. variations in $\Delta \hat{P}_{\rm crit}$. These variations are larger for smaller $E_{\rm resid}$ (see Fig.~\ref{fig:PhaseSpaceEinit}). On average, systems with smaller $M_{p}$ and larger $R_p$ have smaller $E_{\rm resid}$. Accordingly, the spread in the results is largest for the $M_p= 0.3M_J, R_p = 1.6\;R_J$ model and smallest for the $M_p= 1M_J, R_p = 1\;R_J$ model.
		
		\subsection{Hot Jupiter Properties}
\begin{figure*}
			\begin{center}
				\includegraphics[width = 6.5in]{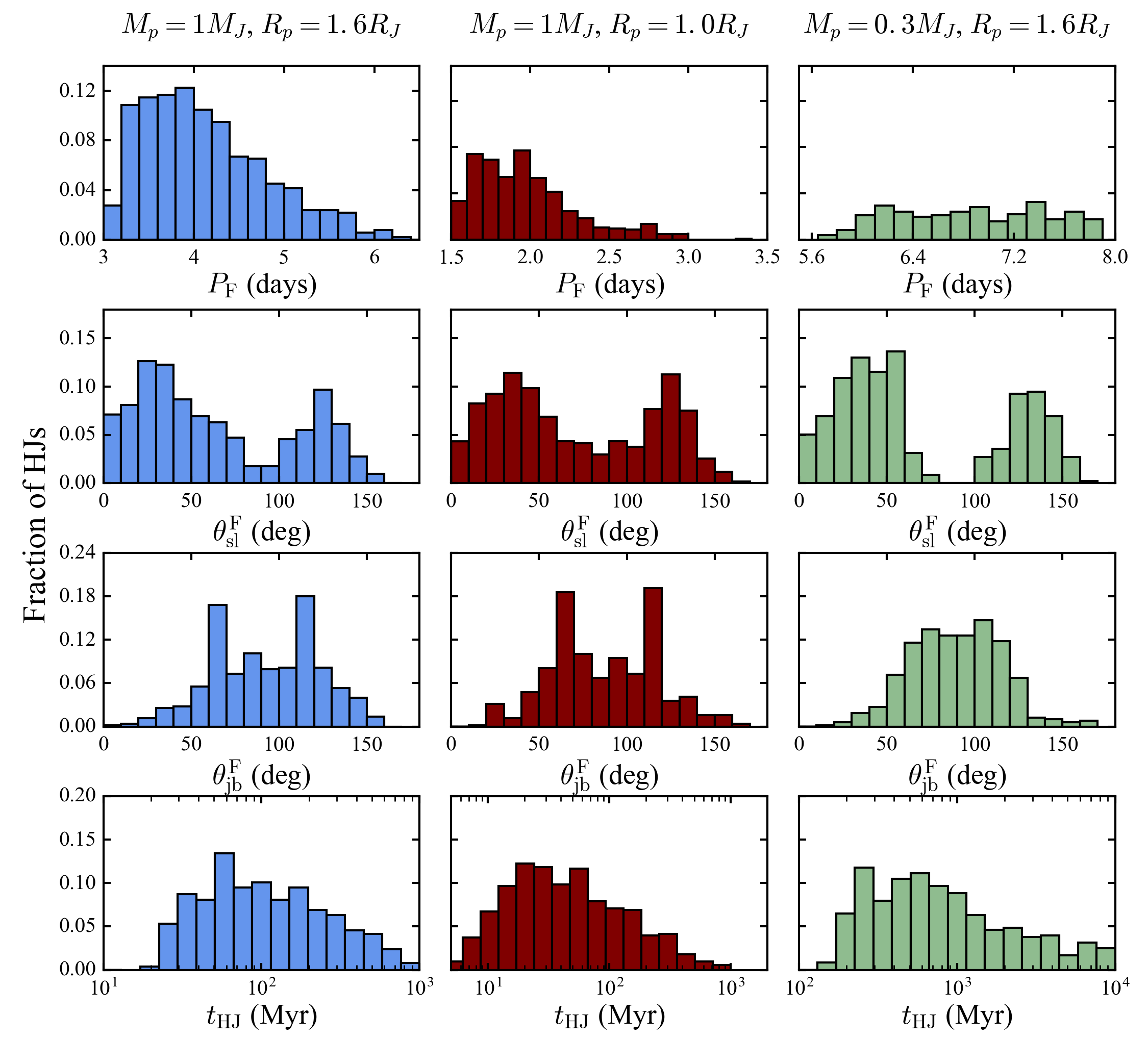}
				\caption{The distribution of the final orbital periods, spin-orbit misalignments ($\theta_{\rm sl}^{\rm F}$), asymptotic mutual orbital inclinations ($\theta_{\rm jb}^{\rm F}$; see Section~\ref{sec:Spin}), and circularization times (when $e=0.1$) for HJs that have formed in our population synthesis calculation. LK chaotic tidal migration can produce HJs with orbital periods in the range of 1.5-8 days depending on the planet mass and radius. This migration mechanism yields a bimodal distribution in $\theta_{\rm sl}^{\rm F}$. The final mutual orbital inclinations can span a large range of values from $\theta_{\rm jb}^{\rm F} = 0^{\circ}$ to $\sim 170^{\circ}$. LK chaotic tidal migration can comfortably generate HJs within 10 Gyr time-scales.} \label{fig:HJPeriods}
			\end{center}
		\end{figure*}

		The eccentric WJs formed by chaotic tidal migration continue to experience orbital decay and circularization due to static tides. As long as the static tide is sufficiently dissipative, the planet will circularize to 
		\begin{equation}
		a_{\rm F} = a_{\rm WJ}(1-e^2_{\rm WJ}) \simeq 2 r_{\rm p,WJ}
		\end{equation} 
		For our population synthesis, most WJs circularize within 1 Gyr (assuming $\Delta t_{\rm L} = 1 $~s)  to become HJs (see Table~\ref{tab:PopSynthResults}). Those that do not are almost exclusively from the low-density planet model with the largest $r_{\rm tide}$. The HJ period distributions for our population synthesis calculations are shown in Fig.~\ref{fig:HJPeriods}. The peak lies between 3.5 and 4 days for our standard model with $M_p=1M_J, R_p = 1.6 R_J$; this shifts to smaller periods for denser planets (with $M_p=1M_J$, $R_p = 1R_J$). As discussed in Section~\ref{sec:eccentricWJs}, the range of $r_{\rm p,WJ}$ is set by the pericentre distance requirements for tidal disruption and for chaotic tidal behaviour. Both of these conditions vary directly with $r_{\rm tide}$, so the peak of the HJ period distribution scales as $r_{\rm tide}^{3/2}$.
		
		The final spin-orbit misalignments ($\theta_{\rm sl}$) are also shown in Fig.~\ref{fig:HJPeriods}. For all three planet models, the distribution in $\theta_{\rm sl}$ is bimodal, peaking near $30^\circ$ and $140^\circ$. These distributions are qualitatively similar to those obtained in ASL16 for standard LK migration with static tides. In general, the percentage of retrograde configurations ($\theta_{\rm sl} > 90^\circ$) is larger for chaotic tidal migration than for standard LK migration. This may be due to the fact that, because chaotic tides dramatically speed up orbital decay, the star is rotating more rapidly when $\theta_{\rm sl}$ becomes ``frozen," and the feedback torque from the star on the orbit is larger. 
		
		We have also provided the final values of $\theta_{\rm jb}$, the angle between $\boldsymbol{J} = \boldsymbol{S_*} + \boldsymbol{L}$ and $\boldsymbol{L}_{\rm B}$. In all of our calculations, the ratio $|\boldsymbol{S_*}|/|\boldsymbol{L}|$ is still of order unity near $t_{\rm HJ}$ (the time when the planet's orbit has circularized to $e=0.1$), so the mutual inclination of the inner and outer orbits, $\theta_{\rm lb}$, is not fixed at $t_{\rm HJ}$. As the star continues to spin down, $\theta_{\rm lb}$ will approach $\theta_{\rm jb}$, as discussed in Section~\ref{sec:Spin}. The distributions of $\theta_{\rm jb}$ have two strong peaks around $65^\circ$ and $115^\circ$.
		
		Lastly, Fig.~\ref{fig:HJPeriods} shows the distribution of ``arrival times" for HJs, $t_{\rm HJ}$, when $e=0.1$. For the $M_p=1M_J, R_p = 1 R_J$ planet model, all orbits circularize to $e=0.1$ within 1 Gyr. For the $M_p=0.3M_J, R_p = 1.6 R_J$ model, only 65\% of the WJs circularize within a Gyr timeframe, yet 97\% can become HJs within 10 Gyr. In general, planets with larger $r_{\rm tide}$ finish chaotic evolution at larger $r_{\rm p, WJ}$ and take longer to circularize via static tides. Most planets that undergo chaotic tidal evolution can be expected to become HJs within the lifetimes of their host stars. As a result, the population of eccentric WJs formed by chaotic tides is transient (assuming $\Delta t_{\rm L} = 1$~s).
				
\section{Analytical Calculation of WJ Formation Fraction}\label{sec:Rates}
		
		\citet{Munoz16} developed an analytical method to calculate the HJ formation fraction in the standard LK migration (with static tides) scenario. In this section, we adapt this method to calculate the fractions of systems that experience chaotic tidal migration or tidal disruption. 
		A system will become chaotic or suffer disruption when the planet's pericentre distance is smaller than a critical value --- for chaotic behaviour, $r_{\rm p,crit}$ from equation~(\ref{eq:BoundaryApprox}), for disruption, $r_{\rm p,dis}$ from equation~(\ref{eq:defrpdis}). We define $\theta_{\rm lb,crit}$ as the minimum initial mutual inclination necessary for a system to reach $r_{\rm p,crit}$, or equivalently $e_{\rm crit}$ (see equation \ref{eq:echaos}). Assuming uniform distribution in $\cos{\theta_{\rm lb,0}}$, the fraction of systems (for a given $a_0$, $a_{\rm b}$, $e_{\rm b}$) that experience chaotic tidal behaviour is then
		\begin{equation}
		f_{\rm mig}(a_0,a_{\rm b},e_{\rm b}) = \cos{\theta_{\rm lb,crit}}, \label{eq:f}
		\end{equation}
		where $\cos{\theta_{\rm lb,crit}}=0$ when $e_{\rm lim} < e_{\rm crit}$. We refer to \citet{Munoz16} for the calculation of $\cos{\theta_{\rm lb,crit}}$ given $e_{\rm crit}$. By performing a Monte Carlo integration over $a_0 = [1,5]$ (au), $a_{\rm b} = [10^2,10^3]$ (au) (uniformly sampled in $\log{a_{\rm b}}$), and $e_{\rm b} = [0,0.8]$, we can predict the fraction of systems in our population synthesis where the planet undergoes chaotic tidal migration ($\mathcal{F}_{\rm mig}$) and fraction of systems where the planet is tidally disrupted ($\mathcal{F}_{\rm dis}$). The fraction of systems that become chaotic but are not disrupted is given by
		\begin{equation}
		\mathcal{F}_{\rm WJ}=\mathcal{F}_{\rm mig}-\mathcal{F}_{\rm dis}.
		\end{equation} 
		As discussed in Section~\ref{sec:Saved}, chaotic tides can compete with the LK effect and prevent planet tidal disruption. Our analytic calculations therefore provide an upper limit to the tidal disruption fraction ($\mathcal{F}_{\rm dis}$) and a lower bound on the fraction of WJs formed ($\mathcal{F}_{\rm WJ}$) in this scenario. Table \ref{tab:PopSynthResults} shows that our analytical results are in good agreement with the populaion synthesis results. Our analytical $\mathcal{F}_{\rm mig}$ is slightly smaller than the value obtained from population synthesis, likely because the boundary for chaotic tides is fuzzy and we have chosen a conservatively low value for $r_{\rm p,crit}$. 
		
		In addition to calculating $\mathcal{F}_{\rm mig}$ and $\mathcal{F}_{\rm dis}$, we can estimate the total migration fraction $\mathcal{F}^{\; \rm st+ct}_{\rm mig}$ where the planet either migrates via chaotic tides or migrates within 1 Gyr due to weak tidal friction (the standard LK migration scenario). For a given $a_0,a_{\rm b}$, and $e_{\rm b}$, this is
		\begin{equation}
		f^{\; \rm st+ct}_{\rm mig}(a_0,a_{\rm b},e_{\rm b}) = \max(\cos{\theta_{\rm lb,crit}},\cos{\theta_{\rm lb,st}}), \label{eq:fmig}
		\end{equation} 
		where $\cos{\theta_{\rm lb,st}}$ is the maximum value of $\cos{\theta_{\rm lb,0}}$ for which a system can reach $r_{\rm p, ST}$, given in equation (\ref{eq:rpmig}). We can integrate $f^{\; \rm st+ct}_{\rm mig}$ over orbital parameters $a_0$, $a_{\rm b}$, and $e_{\rm b}$ (as before) to find $\mathcal{F}^{\; \rm st+ct}_{\rm mig}$. Because $r_{\rm p,ST}$ is generally less than $r_{\rm p,crit}$, the percentage of systems that are not chaotic but become HJs due to weak tidal friction is much less than $1\%$.
		
\begin{figure}
			\begin{center}
				\includegraphics[width = \columnwidth]{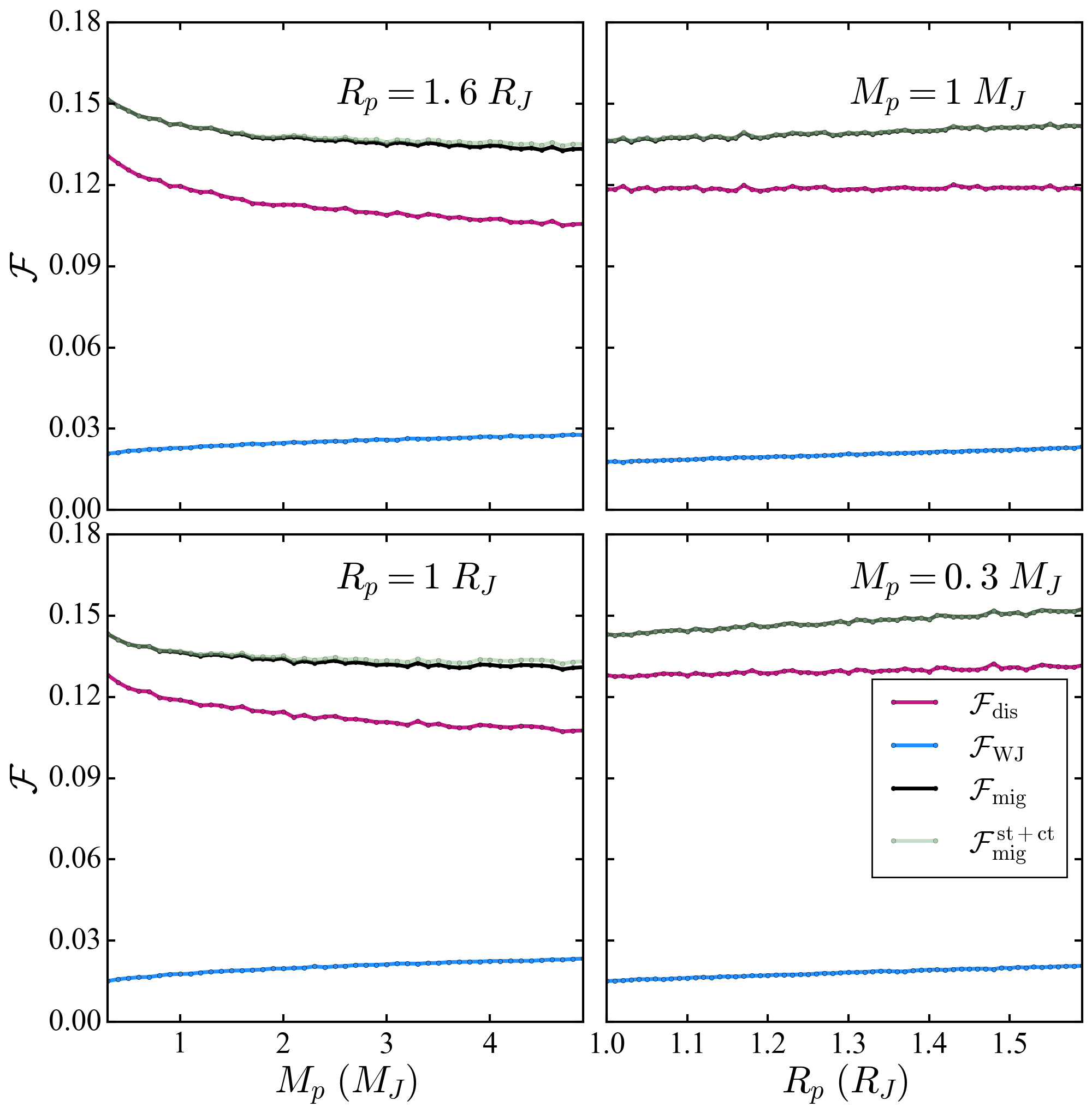}
				\caption{Analytical results for the planet migration/disruption fractions for various planet masses and radii (see Section~\ref{sec:Rates}). $\mathcal{F}_{\rm mig}$ gives the fraction of systems that experience chaotic tidal migration, $\mathcal{F}_{\rm dis}$ gives the upper bound on the fraction of sytems that are tidally disrupted, and $\mathcal{F}_{\rm WJ} = \mathcal{F}_{\rm mig} - \mathcal{F}_{\rm dis}$ gives the lower bound on the fraction of eccentric WJs formed in this scenario. 
				$\mathcal{F}^{\;\rm st+ct}_{\rm mig}$ gives the fraction of systems that experience chaotic tidal migration or standard LK migration with static tides. Note that $\mathcal{F}^{\;\rm st+ct}_{\rm mig} \approx \mathcal{F}_{\rm mig}$ (the green line rests nearly on top of the black line). This suggests that almost all planets (with $\Delta t_L = 1$ s) that could migrate via standard LK migration within 1 Gry will experience chaotic tides.}
				\label{fig:Rates}
			\end{center}
		\end{figure}

		In our population synthesis calculations (Section~\ref{sec:popSynthesis}) we have only considered planets with $M_p=M_J$ and $0.3 M_J$ and radius $R_p = R_J$ or $1.6\;R_J$. With analytic calculations, we can easily predict the fractions of systems that survive chaotic tides or suffer disruption as a function of $M_p$ and $R_p$. The results are shown in Fig.~\ref{fig:Rates}. The fraction of systems that undergo chaotic tidal migration is nearly independent of $M_p$, $R_p$ ($\mathcal{F}_{\rm mig}\sim 13- 15$\%). The WJ formation fraction from LK chaotic tides is also insensitive to $M_p$ and $R_p$ because $r_{\rm p, crit}$ and $r_{\rm p,dis}$ have roughly the same scaling with $M_p$ and $R_p$. This result differs from the standard picture of LK migration, where Saturn-mass planets that migrate are destined for tidal disruption (see ASL16).
		
\section{Summary and Discussion}\label{sec:Summary}

\subsection{Summary of Key Results}

In this paper, we have systematically studied the role of dynamical
tides in high-eccentricity gas giant migration via the LK effect. Our
investigation has revealed (i) the conditions under which dynamical
tides in the planet have a significant effect on migration, (ii)
the characteristics of the planet population that results from LK
migration with dynamical tides, and (iii) the expected HJ fraction from
this formation path. The overall summary of the paper is already given 
in the abstract. Here we provide a guide to the key results
of each section.

In Section~\ref{sec:Chaos}, we explored the orbital parameters
necessary for a gas giant to undergo chaotic tidal migration. When a
planet's orbit is very eccentric and its pericentre distance is
sufficiently small, the tidal force from the host star can excite
planetary oscillations at pericentre in such a way that these
oscillations chaotically climb in amplitude over repeated close
passages (an example is shown in the bottom panel of
Fig.~\ref{fig:3Behaviors}). We found that, for the $\gamma=2$
polytrope planetary model, the fundamental mode (f-mode) is more strongly excited 
than inertial modes, and is the most likely to experience
chaotic growth over multiple orbits. The condition for these
``chaotic tides'' to operate is given by equation~(\ref{eq:echaos}) and
plotted in Fig.~\ref{fig:HJParams}. In general, chaotic tides can occur
in planets on highly eccentric orbits with small pericentre
distances. We also explored (in
Sections~\ref{sec:extendChaosCondition} and
\ref{sec:DissipativeChaosCondition}) how the conditions for chaotic
tides become more generous when the f-mode already has some non-zero
energy.

In Section \ref{sec:TidesandKozai}, we developed a model to couple the LK
migration with orbital changes due to dynamical tides.
The model is fully described in Section~\ref{sec:FullModel}, and can
be readily adapted to other high-eccentricity migration mechanisms.
Although the model is largely {\it ab initio}, there are two free
parameters that characterize the (uncertain) nonlinear mode
dissipation: $E_{\rm max}$ (the maximum energy the mode can reach
before nonlinear dissipation sets in) and $E_{\rm resid}$ (the
residual energy in the mode after an episode of nonlinear
dissipation).  We found that these parameters (with reasonable values)
do not change the general features of evolution due to chaotic
tides, but can alter the details (see Figs.~\ref{fig:EmaxComparison}-\ref{fig:EresComparison}).

In Sections \ref{sec:Results} and \ref{sec:Features}, we 
presented example calculations of the LK chaotic tidal migration, 
and explained some key features of this migration mechanism:
\begin{itemize}
	\item
	Chaotic tidal migration leads to rapid formation of WJs (see
	Fig.~\ref{fig:LongTermEvolution} for an example)
	on timescales a few to 100 Myrs (see Fig. \ref{fig:Histograms}). 
	These WJs have eccentricities in
	the range $0.85 \lesssim e_{\rm WJ} \lesssim 0.98$ and semi-major axes between 
	0.1 and 2 au (with lower values more likely), depending on the
	giant planet mass and radius (see Fig.~\ref{fig:scatterPlots}).
	\item
	These eccentric WJs efficiently circularize to HJs via dissipation of static tides. 
	This process is faster than standard LK migration because
	a planet that experiences chaotic tides quickly becomes decoupled from
	the influence of the perturber and the eccentricity oscillations are frozen
	(see Figs.~\ref{fig:LongTermEvolution}, \ref{fig:LongTermEvolutionLargerEta} and 
	\ref{fig:SavedSystemExample}).
	\item
	Some giant planets that are otherwise fated for tidal disruption can
	be saved by chaotic tides (see Fig.~\ref{fig:SavedSystemExample}). This occurs when the orbital decay timescale associated with chaotic tides is shorter than the timescale for the perturber to change the planet's eccentricity (see Section \ref{sec:Saved}).
\end{itemize}

In Section~\ref{sec:popSynthesis}, we conducted a population synthesis study
to determine the formation fractions and properties of eccentric WJs
and HJs for a few combinations of planet mass and radius. Our study sampled
the same range of initial orbital parameters as in Anderson et al.~(2016) 
for the standard LK migration scenario.
The calculated fractions for chaotic tidal migration
($\mathcal{ F}_{\rm mig}$), tidal disruption ($\mathcal{ F}_{\rm dis}$), 
WJ formation ($\mathcal{ F}_{\rm WJ}$; note that $\mathcal{ F}_{\rm mig} = 
\mathcal{ F}_{\rm dis} + \mathcal{ F}_{\rm WJ}$) and HJ formation within 1 Gyr 
($\mathcal{ F}_{\rm HJ}$), are provided in Table \ref{tab:PopSynthResults}. 
For an $M_p=1 M_J, R_p=1 R_J$ planet, LK chaotic tidal migration
produces more HJs than standard LK migration ($\mathcal{ F}_{\rm HJ}$
increases from $2.4\%$ to $3.9\%$; see Table 3 in ASL16).
More importantly, chaotic tides can produce hot giant planets with a broad range of masses and radii. This is in stark contrast with the standard LK
migration, which produces very few hot Saturns because of the
severe tidal disruption experienced by such low-mass giants (ASL16; Munoz et al.~2016).

The orbital properties of planets from our population synthesis that
survived chaotic tidal migration are shown in
Figs.~\ref{fig:Histograms} and \ref{fig:HJPeriods}. The WJs that are
produced directly by chaotic tidal dissipation have
pericentre distributions that peak near $r_{\rm p, WJ} \sim 3 r_{\rm tide} 
= 3 R_p (M_\star/M_p)^{1/3}$. 
These pericentre values are smaller than those of observed high-eccentricity WJs 
such as HD80606 b \citep{Hebrard10}. 
This is expected as the eccentric WJs formed by chaotic tides are ``transient'' and should move quickly 
through the high-eccentricity phase.
The HJs that form via LK chaotic tidal migration exhibit a pile-up
around a 3~day orbital period, depending on the planet's mass-radius relation
and the assumed tidal disruption criterion. This is a feature of all
tidal migration mechanisms. However, our predicted HJ period
distribution for LK chaotic tidal migration differs from that for the
standard LK migration, particularly for low-mass giant planets
(compare Fig.~\ref{fig:HJPeriods} with the middle row of Fig.~23 from
ASL16). Chaotic tidal migration can produce Saturn-mass planets at
periods longer than 5 days and generates a wider period distribution
for such planets. Finally, we found that chaotic tidal migration
yields similar spin-orbit misalignments as the standard LK migration
(compare Fig.~\ref{fig:HJPeriods} with Fig.~24 of ASL16). For all
three planet models we considered, the distribution of final
spin-orbit misalignments is bimodal with peaks at
$\theta_{\rm sl}^{\rm F}\sim 30^\circ$ and $\theta_{\rm sl}^{\rm F} \sim 130^\circ$.

In Section~\ref{sec:Rates}, we used an analytical method, developed in
\citet{Munoz16}, to understand how the HJ formation rate varies with
planet mass and radius. We calculated a lower-bound on the eccentric
WJ formation rate (not accounting for planets that are spared from
tidal disruption by chaotic tides). The predicted (analytical) rates
are in agreement with the results from the population synthesis study
(see Section~\ref{sec:popSynthesis}) and are shown in
Fig.~\ref{fig:Rates}. We confirmed that the WJ formation rate from LK
chaotic tidal migration is constant over a reasonable range of giant
planet masses and radii, as suggested by, e.g.,
Fig.~\ref{fig:phaseSpaceScan} and Table \ref{tab:PopSynthResults}.

\subsection{Discussion}

The results presented in this paper show that chaotic tides endow the LK migration scenario with a number of ``favorable'' features \citep[see also][]{Wu18}. These not only reduce the theoretical uncertainties
regarding tidal dissipation that are inherent in the theory, but also
may help reconcile some of the discrepancies between observations and
predictions of LK migration.
Chaotic tides drastically reduce the amount of time that a gas giant
spends at high eccentricity. This could explain the lack of
observations of super-eccentric gas giants \citep{Dawson15}. In
addition, chaotic tides quickly decouple a gas giant experiencing LK
oscillations from the stellar perturber. This allows planets at larger
pericentre distances to migrate within the lifetime of their host stars.
In this way, chaotic tidal migration naturally produces a period
distribution with a longer tail. Indeed, HJs beyond the 3-day pile-up
are observed but difficult to explain with standard high-e migration scenarios.

Although chaotic tides increase the HJ yield from LK migration in
stellar binaries, particularly for low-mass planets, we should not
expect this particular formation channel to account for all HJs.
The occurrence rate of HJs produced by this channel can be computed from
\begin{equation}
\mathcal{R}_{\rm HJ} = \mathcal{F}_b \times \mathcal{F}_p \times \mathcal{F}_{\rm HJ},
\end{equation} 
where $\mathcal{F}_{b}$ is the fraction of stars with a binary
companion and $\mathcal{F}_{p}$ is the
fraction of solar-type stars with a giant planet at a few au. Assuming
$\mathcal{F}_{b}\sim 50\%$ \citep{Raghavan10, Ngo15}, $\mathcal{F}_{p}\sim 10\%$ (as in ASL16), and $\mathcal{F}_{\rm HJ}\sim 4\%$ from our
population synthesis calculation, we obtain an estimate of $\mathcal{R}_{\rm HJ} \sim
0.2\%$, which is nearly an order of magnitude smaller than the
observed occurrence rate of $1\%$ \citep{Marcy05,Wright12,Fressin13}. 
Thus, LK chaotic tidal migration in stellar binaries can
roughly account for $20-30\%$ 
of the observed HJ population.

However, we expect that many of the ``nice'' features of chaotic tides
may also apply to other flavours of high-$e$ migration scenarios, such
as LK migration induced by planetary companions and secular chaos in
multi-planet systems.  The eccentricity and pericentre ranges of
planets that are susceptible to chaotic tides are set by the boundary
for chaotic f-mode behaviour (see Section \ref{sec:Chaos}), and do not depend on a
specific high-$e$ migration scenario.
Chaotic tides can save a planet from tidal disruption when the
time-scale for energy transfer to the planet's oscillation mode is
shorter than the time-scale for driving and maintaining the planet's
high-eccentricity (see Section \ref{sec:Saved}) -- this condition can be satisfied
by all secular eccentricity excitation mechanisms. Indeed, 
the recent work by \citet{Teyssandier18} showed that chaotic tides
significantly increase the HJ formation fraction in the secular-chaos
high-$e$ migration scenario. Overall, chaotic tides boost the importance
of high-$e$ migration for the formation of HJs.
 
The story of chaotic tidal migration hinges upon the planet's ability
to survive rapid tidal heating. There is reason to expect that the
planet interior could survive this process \citep{Wu18} if most of the
dissipated energy goes into the outer layers of the planet, which can
quickly radiate heat. However, if the tidal energy is deposited at a
larger depth, the planet's envelope may expand. Such changes in the
structure of the planet would affect the f-mode frequency and could
have a larger influence on the evolution of the planet's orbit. A
sudden expansion of the planet's radius may also put the planet in
danger of tidal disruption. Many of the planets that survive chaotic
tidal migration come close to the tidal disruption radius. Expansion
of the planet could easily lead to stripping of the outer layers and
initiate mass transfer or mass loss. The effect of mode energy
dissipation on the planetary structure is a very important problem for
future study.

\section*{Acknowledgements} 
This work is supported in part by the NSF grant AST1715246
and NASA grant NNX14AP31G. MV is supported by a NASA Earth and Space Sciences Fellowship in Astrophysic. KRA is supported by a NASA Earth and Space Sciences Fellowship in Planetary Science.

\bibliographystyle{mnras}	
\bibliography{References}

\bsp
\label{lastpage}
\end{document}